\documentclass[twocolumn]{aastex61}

\usepackage{natbib,graphicx,longtable,afterpage,amsmath,rotating,adjustbox,tabularx}

\makeatletter

\newcommand{\Rmnum}[1]{\expandafter\@slowromancap\romannumeral #1@}
\makeatother 

\shorttitle{Black hole mass and Eddington ratio distributions of the $z\sim4$ HSC quasars}
\shortauthors{He et al.}

\begin{document}

\title{Black hole mass and Eddington ratio distributions of less-luminous quasars at $z\sim4$ in the Subaru Hyper Suprime-Cam Wide field}

\author[0000-0001-7759-6410]{Wanqiu He}
\email{wanqiu.he@nao.ac.jp}
\affil{National Astronomical Observatory of Japan, 2-21-1 Osawa, Mitaka, Tokyo 181-8588, Japan}

\author[0000-0002-2651-1701]{Masayuki Akiyama}
\affiliation{Astronomical Institute, Tohoku University, 6-3, Aramaki, Aoba-ku, Sendai, Miyagi, 980-8578, Japan}

\author[0000-0003-0150-8845]{Motohiro Enoki}
\affiliation{Center for General Education, Tokyo Keizai University, 1-7-34, Minami-cho, Kokubunji, Tokyo, 185-8502, Japan}

\author[0000-0002-4377-903X]{Kohei Ichikawa}
\affiliation{{Global Center for Science and Engineering, Faculty of
Science and Engineering, Waseda University, 3-4-1,
Okubo, Shinjuku, Tokyo 169-8555, Japan}}
\affiliation{
{Department of Physics, School of Advanced Science and Engineering,
Faculty of Science and Engineering, Waseda University, 3-4-1,
Okubo, Shinjuku, Tokyo 169-8555, Japan}}

\author[0000-0001-9840-4959]{Kohei Inayoshi}
\affiliation{Kavli Institute for Astronomy and Astrophysics, Peking University, Beijing 100871, China}

\author[0000-0003-3954-4219]{Nobunari Kashikawa}
\affiliation{Department of Astronomy, Graduate School of Science, The University of Tokyo, 7-3-1 Hongo, Bunkyo, Tokyo 113-0033, Japan}
\affiliation{Research Center for the Early Universe, The University of Tokyo, 7-3-1 Hongo, Bunkyo-ku, Tokyo 113-0033, Japan}

\author[0000-0002-3866-9645]{Toshihiro Kawaguchi}
\affiliation{Department of Economics, Management and Information Science, Onomichi City University, 1600-2, Hisayamada, Onomichi, Hiroshima, 722-850, Japan}

\author[0000-0001-5063-0340]{Yoshiki Matsuoka}
\affiliation{Research Center for Space and Cosmic Evolution, Ehime University, 2-5 Bunkyo-cho, Matsuyama, Ehime 790-8577, Japan}

\author[0000-0002-7402-5441]{Tohru Nagao}
\affiliation{Research Center for Space and Cosmic Evolution, Ehime University, 2-5 Bunkyo-cho, Matsuyama, Ehime 790-8577, Japan}

\author[0000-0003-2984-6803]{Masafusa Onoue}
\affiliation{Kavli Institute for Astronomy and Astrophysics, Peking University, Beijing 100871, China}
\affiliation{Kavli Institute for the Physics and Mathematics of the Universe (Kavli IPMU, WPI), The University of Tokyo, Chiba 277-8583, Japan}

\author{Taira Oogi}
\affiliation{Research Center for Space and Cosmic Evolution, Ehime University, 2-5 Bunkyo-cho, Matsuyama, Ehime 790-8577, Japan}

\author[0000-0002-6660-6131]{Andreas Schulze}
\affiliation{OmegaLambdaTec, Lichtenbergstrabe 8, D-85748, Garching, Germany}
\affil{National Astronomical Observatory of Japan, 2-21-1 Osawa, Mitaka, Tokyo 181-8588, Japan}

\author[0000-0002-3531-7863]{Yoshiki Toba}
\affiliation{National Astronomical Observatory of Japan, 2-21-1 Osawa, Mitaka, Tokyo 181-8588, Japan}
\affiliation{Research Center for Space and Cosmic Evolution, Ehime University, 2-5 Bunkyo-cho, Matsuyama, Ehime 790-8577, Japan}
\affiliation{Academia Sinica Institute of Astronomy and Astrophysics, 11F of Astronomy-Mathematics Building, AS/NTU, No.1, Section 4, Roosevelt Road, Taipei 10617, Taiwan}

\author[0000-0001-7821-6715]{Yoshihiro Ueda}
\affiliation{Department of Astronomy, Kyoto University, Kitashirakawa-Oiwake-Cho, Sakyo, Kyoto 606-8502, Kinki, Japan}


\begin{abstract}
We {investigate} the black hole mass function (BHMF) and Eddington ratio distribution function (ERDF) of broad-line AGNs at $z=4$, based on a sample of 52 quasars with $i<23.2$ at $3.50\leq z \leq4.25$ from the Hyper Suprime-Cam Subaru Strategic Program (HSC-SSP) S16A-Wide2 dataset, and 1,462 quasars with $i<20.2$ in the same redshift range from the Sloan Digital Sky Survey (SDSS) DR7 quasar catalog. Virial BH masses of quasars are estimated using the width of the ${\rm C_{\Rmnum{4}}}$ 1549{\AA} line and the continuum luminosity at 1350{\AA}. To obtain the \textit{intrinsic} broad-line AGN BHMF and ERDF, we {correct} for the incompleteness in the low-mass and/or low-Eddington-ratio ranges caused by the flux-limited selection. The resulting BHMF is constrained down to $\log M_{\rm BH}/M_{\odot}\sim7.5$. In comparison with broad-line AGN BHMFs at $z\sim2$ in {the} literature, we {find} that the number density of massive SMBHs peaks at higher redshifts, consistent with the "down-sizing" evolutionary scenario. Additionally, the resulting ERDF shows a negative dependence on BH mass, suggesting more massive SMBHs tend to accrete at lower Eddington ratios at $z=4$. With the derived \textit{intrinsic} broad-line AGN BHMF, we also {evaluate} the active fraction of broad-line AGNs among the entire SMBH population at $z=4$. The resulting active fraction may suggest a positive dependence on BH mass. Finally, we {examine} the time evolution of broad-line AGN BHMF between $z=4$ and 6 through solving the continuity equation. The results suggest that the broad-line AGN BHMFs at $z=4\sim6$ only show evolution in their normalization, but with no significant changes in {their} shape.
 
\end{abstract}


\section{Introduction}
\label{sec:intro}

Understanding the formation and evolution of supermassive black holes (SMBHs), which are found to be ubiquitous at the centers of local massive galaxies \citep[e.g.,][]{kr1995}, is one of the important goals in observational cosmology. The growth processes of SMBHs may be physically associated with the galaxy evolution processes, as suggested by a series of tight scaling relationships observed between the mass of SMBHs and the properties of the host spheroids, such as their stellar velocity dispersion \citep[e.g.,][]{Gebhardt2000, MF2001}, bulge mass \citep[e.g.,][]{Magorrian1998,MH2003}. Those empirical correlations imply a physical connection between the evolution of SMBHs and host galaxies, i.e., the ``co-evolution''. But its physical mechanism is still under debate: in one scenario, galaxy minor mergers are proposed to naturally invoke the hierarchical assembly of SMBHs and stellar mass of host galaxies \citep[e.g.,][]{Peng07,Jahnke}; in another scenario, galaxy major mergers and the following AGN feedback are thought to be the mechanism to regulate the growth of SMBHs together with star formation in the central region of their host galaxies \citep[e.g.,][]{Fabian1999,Di2005,Springel2005,Hopkins06}.

Luminosity functions of AGNs can reflect the SMBH growth under the active accretion phase, and they are already intensively investigated using large samples of AGNs across cosmic time {\citep[e.g.,][]{Ueda03,Fan04,Richards06b,Silverman08,Croom09,Ross13,Toba13,Toba14,Ueda14,akiyama2018,McGreer18,Matsuoka18,Matsuoka23}}. Following the idea proposed by \citet{Soltan1982}, the total mass accreted towards SMBHs, which is obtained by integrating the AGN luminosity function through luminosity and redshift, is compared to the mass density of SMBHs in the local universe \citep[e.g.,][]{YT2002,Shankar09}. Consistency between the two quantities indicates the local SMBH mass density is mostly accumulated by mass accretion in the AGN phase, and requires the average radiative efficiency to be in the order of 10\%.

Based on luminosity functions, the cosmological evolution of the comoving number density of AGNs has been examined. The number density of luminous AGNs, e.g., quasars, is found to peak at higher redshifts than that of less-luminous AGNs \citep[e.g.,][]{Ueda14,akiyama2018,Matsuoka18,niida2020}. Such anti-hierarchical trend is called ``down-sizing'' evolution. At high redshifts, there are also observational results suggesting that the abundance of less-luminous AGNs tends to decline more mildly towards higher redshifts than that of luminous AGNs \citep[e.g.,][]{Ueda14}. Such trend is called ``up-sizing'' evolution, which follows the direct expectation of hierarchical structure formation.

However, the luminosity of AGNs does not directly reflect the mass assembly history of SMBHs. The luminosity function of AGNs is the convolution of the black hole mass function (BHMF) and the Eddington-ratio distribution function (ERDF). {Some studies find that the Eddington ratio, a ratio between the observed bolometric and the Eddington-limited luminosity, is distributed over a wide range \citep[e.g.,][]{AS2010,Suh2015,Jones2016}. Thus, constraining BHMF and ERDF along the cosmic time becomes important in studying the growth history of SMBHs quantitatively.} In the local universe, by applying the scaling relationships between the mass of SMBHs and properties of their host spheroids for normal galaxies, the BHMF is evaluated in many studies \citep[e.g.,][]{YT2002,Marconi,Shankar09}. Beyond the local universe, since it remains controversial how the scaling relationships evolve in the early universe, determining the BHMF by using the scaling relationships is not applicable.

For broad-line AGNs, an alternative method is available to estimate their virial BH masses, and then determine the distribution functions. The virial BH mass can be estimated from a single-epoch spectrum of broad-line AGNs with broad emission lines in their spectra. Multiple calibrations are made for the broad H$\alpha$, H$\beta$, ${\rm Mg_{\Rmnum{2}}}$ and ${\rm C_{\Rmnum{4}}}$ emission lines \citep[e.g.,][]{VP06,wang2009,SL2012,TN12,Denney13,park13,park17}. There are already a large number of broad-line AGNs with virial BH mass estimated \citep[e.g.,][]{Kollmeier06,Shemmer,Netzer07a,Zuo,Trakhtenbrot16,S2018,Banados18,Onoue19,wang2021,Eilers23,Harikane23,Onoue23}, including {the second highest-redshift AGN, CEERS\_1019 at $z=8.679$, with a BH mass of $ 10^{6.95} M_\odot$ \citep{Larson23}}. 

With the BH mass estimates, some studies jointly determine the BHMF and ERDF by fitting bivariate distribution models of BH mass and Eddington ratio with observed distributions, either using the Bayesian method \citep[e.g.,][]{SK2012,ks2013}, or the Maximum-likelihood method \citep[e.g.,][]{AS2010,nobuta,AS2015,Ananna22}. While different methods and quasar samples are used, those studies result in a common trend of ``down-sizing'' in the growth history of SMBHs: the number density of quasars with less-massive SMBHs mildly increases from $z = 2-3$ to the local universe, while that of quasars with massive SMBHs sharply declines towards $z = 0$, i.e., the massive SMBHs stop growing first while the less-massive ones grow continuously. 

The evolutionary trend remains uncertain towards higher redshifts. Most of the current statistical studies only rely on the luminous quasars brighter than $i=20$ at $z\gtrsim3$. Due to the flux limit, a large fraction of AGNs with low Eddington ratios of $\log \lambda_{\rm Edd}<-0.5$ or small BH masses of $\log M_{\rm BH}/M_\odot<9$ can be missed, and the constraints on the less-massive end of BHMF and low-Eddington-ratio end of ERDF are not tight. At $z>3.2$, it is found that the fraction of observable quasars above the SDSS detection limit decreases to less than $10$\% at $M_{\rm BH} < 10^9 M_{\odot}$, indicating the constraints can have large uncertainty in the less-massive ends \citep{SK2012}. \citet{AS2015} also demonstrate a sample of AGNs covering a wide luminosity range is crucial to constrain the BHMFs over a wide BH mass range.

In this work, in order to trace the early growth of SMBHs, we constrain the BHMF and ERDF of broad-line AGNs at $z = 4$ using a sample of less-luminous quasars selected from the Hyper Suprime-Cam Subaru Strategic Program \citep[HSC-SSP;][]{aihara2018a}. Here, HSC-SSP is an imaging survey utilizing Hyper Suprime-Cam (HSC), the wide-field CCD camera attached to the prime focus of the Subaru 8.2m telescope \citep[][]{miyazaki18}. In the most recent data release PDR3, the Wide layer of HSC-SSP covers $\sim$1200 deg$^2$ in the $grizy$-bands with the 5$\sigma$ detection limits of 26.5, 26.5, 26.2, 25.2, and 24.4 mag, respectively \citep{aihara22}. Using the wide coverage and deep detection limit of the HSC-SSP, \citet{akiyama2018} construct a sample of 1666 $z\sim4$ less-luminous quasars from the S16A-Wide2 data release, which has a sky coverage of $\sim170$ deg$^2$ \citep{aihara2018b}. The sample covers $\sim 2-3$ mag fainter magnitude range than the SDSS quasars at $z=4$, with the sample size a few times larger than other $z\sim3-4$ quasar samples in the similar luminosity ranges \citep[e.g.,][]{Netzer07a,Trakhtenbrot16,S2018}.

The outline of this paper is as follows. We describe the quasar samples and spectroscopic observations in section 2. Line fitting of the ${\rm C_{\Rmnum{4}}}$ broad emission line, and estimates of the virial BH mass and Eddington ratio are described in section 3. Section 4 introduces the $V_{\rm max}$ and Maximum-likelihood method used to constrain the $z=4$ broad-line AGN BHMF and ERDF. The final section 5 discusses the cosmological evolution of the broad-line AGN BHMFs and ERDFs. In addition, we investigate the active fraction of broad-line AGNs at $z=4$ based on the derived BHMFs and ERDFs. We also examine the evolution of broad-line AGN BHMFs and ERDFs between $z=4$ and 6 through solving the continuity equation. Throughout this paper we assume a $\Lambda$CDM cosmology with $\Omega_{m}=0.3,~\Omega_{\Lambda}=0.7$, and $H_{0}=70~{\rm km~s^{-1}~Mpc^{-1}}$. {The PSF magnitudes of quasars, which are determined by fitting a model PSF to the image of object, are adopted.} All the magnitudes are given in the AB magnitude system.


\section{Sample selection, observations and redshift determination}
\label{sec:section2}

\subsection{Target selections of $z\sim4$ quasar candidates for spectroscopic observations}
\label{sec:s2sample}

We use the sample of less-luminous $z\sim4$ quasar candidates constructed with the HSC-SSP S16A-Wide2 dataset \citep[for details see][]{akiyama2018}. To build the $z\sim4$ quasar catalog, \citet{akiyama2018} adopt the following $g$-band dropout color criteria:
\begin{eqnarray}
  0.65(g-r)-0.30&>&(r-z),  \label{eq:c1_grz1}\\
  3.50(g-r)-2.90&>&(r-z), \label{eq:c1_grz2} \\
  (g-r)&<&1.5.  \label{eq:c1_grz3} 
\end{eqnarray}
Additional color criteria are set to remove contamination from Galactic stars as follows:
\begin{eqnarray}
 -2.25(i-z)+0.40&>&(z-y), \label{eq:c1_izy1}\\
  (i-z)&>&-0.3. \label{eq:c1_izy2}
\end{eqnarray}
We refer to the color criteria as ``c1''. By applying the c1 criteria for objects with stellar morphology (for the criteria see equation 1-2 in \citealt{akiyama2018}) on the HSC-SSP S16A-Wide2 dataset, a sample of 1,666 $z\sim4$ quasar candidates at $20<i<24$ is constructed in an effective area of 172 deg$^{2}$. Their photometric redshifts are estimated to distribute between $z_{\rm phot}=3.4$ and $4.6$, with a mean and standard deviation at $z_{\rm phot}=3.9$ and 0.2, respectively \citep{akiyama2018}. 

Additionally, in order to examine the completeness of the c1 criteria for selecting quasars at $3.5<z<4.5$, we include supplemental $z\sim4$ quasar candidates in the spectroscopic observations. Since the c1 criteria are determined to minimize the contamination from compact galaxies, Galactic stars, and quasars at other redshifts than $3.5<z<4.5$, they can suffer from a relatively low completeness. Thus, we expand the color selection window on the $g-r$ vs. $r-z$ plane to recover the $z\sim4$ less-luminous quasars that are missed by the c1 criteria. 

\begin{figure*} [!ht]
\centering
\includegraphics[keepaspectratio,width=.9\textwidth]{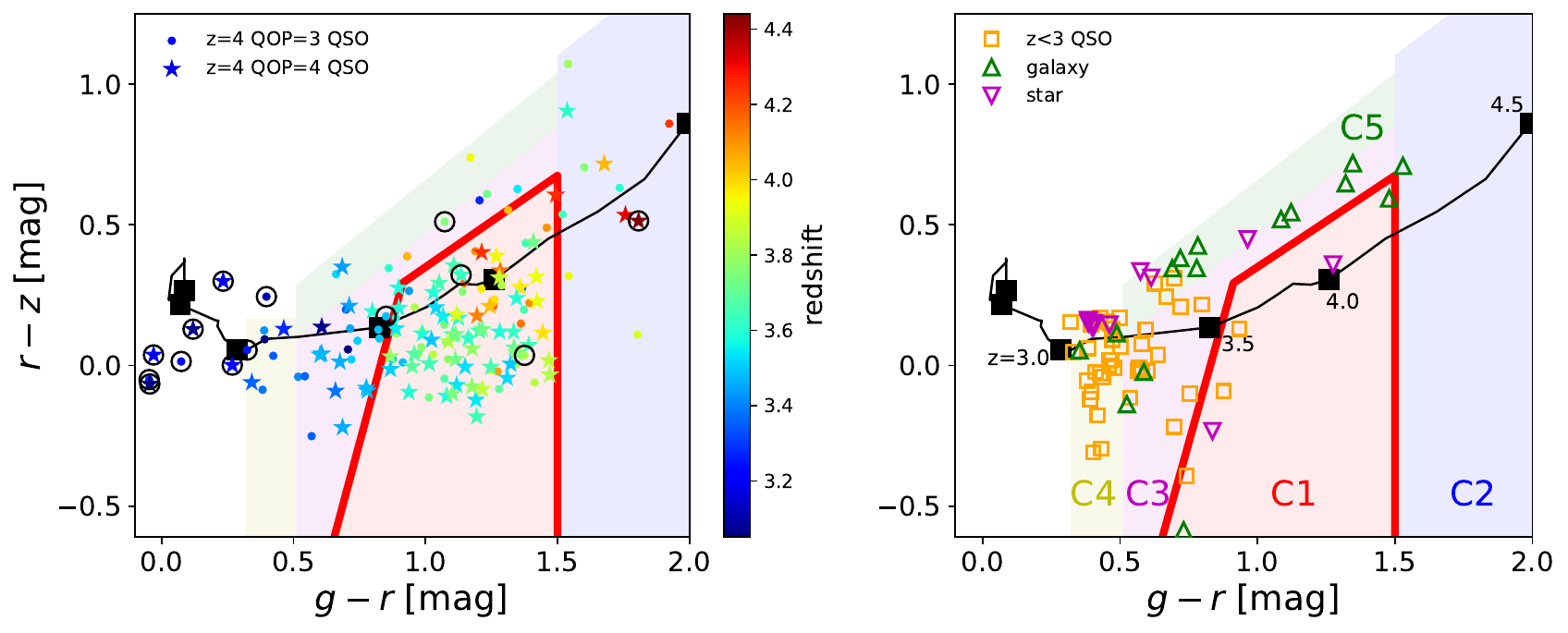}
\caption{Left) the $g-r$ vs. $r-z$ color-color diagram of the identified quasars at $3<z<4.5$. Red (c1), blue (c2), magenta (c3), yellow (c4) and green (c5) shaded areas denote the c1-5 selection criteria. Black solid line shows the mean color track of the model quasars in \citet{akiyama2018}, with black squares implying its colors at $z=2$, $2.5$, $3$, $3.5$, $4$, and $4.5$ from left to right. Stars and dots show the identified quasars with QOP=4 (good quality) and QOP=3 (low quality), respectively. Symbols are color-coded with their spectroscopic redshift, as suggested by the color bar on the right side. Black open circles mark quasars selected by the Lyman break criteria in \citet{z4lbg} or radio detection in \citet{wergs1}. Right) the $g-r$ vs. $r-z$ color-color diagram of the identified contaminants. Orange squares, green triangles and purple inverted triangles represent the $z<3$ quasars, galaxies and Galactic stars identified in the spectroscopic observations, respectively.}
\label{fig:CCD}
\end{figure*}

The additional c2-5 selection criteria are summarized in Table~\ref{tab:obs_c25}. 
\begin{table}[!ht]
\caption{The c1-5 color selection criteria and number of the observed candidates in the spectroscopic observations}\label{tab:obs_c25}
\begin{adjustbox}{width=0.5\textwidth,center=7.5cm}
\centering
\begin{tabular}{cccccc}
\hline\hline
         & selection criteria  & Wide2  & Udeep\&Deep & total  \\
\hline
c1 & Equation~\ref{eq:c1_grz1}-\ref{eq:c1_izy2} &  141 & 93 & 234 \\
\hline
 & $1.50<(g-r)<2.05$ &  &  & \\
c2 & $0.73(g-r)+0.009>(r-z)$ & 216 & -- & 216\\ 
 &  $(i-z)>-0.3$ & & & \\
\hline
 & $0.51<(g-r)<1.50$ &  & &\\
c3 & $0.76(g-r)-0.29>(r-z)$ & 272 & 52 & 324\\
 &  $(i-z)>-0.3$ & & &\\
 & exclude c1 & & &\\
\hline
c4 & $0.32<(g-r)<0.51$ & 794 & -- & 794 \\
 & $(r-z)<0.17$ &  &  &\\
\hline
 & $0.51<(g-r)<1.50$ &  & &\\
c5 & $0.76(g-r)-0.10>(r-z)$ & 793 & -- & 793\\
 & $0.76(g-r)-0.29<(r-z)$ & & &\\
 & Equation~\ref{eq:c1_izy1}-\ref{eq:c1_izy2} &  & \\
\hline\\
\end{tabular}
\end{adjustbox}
\end{table}
In Figure~\ref{fig:CCD}, the red, blue, magenta, yellow and green shaded areas show the c1-5 selection criteria, respectively. 
Here, the selection windows are designed according to the mean color track of the model quasars in \citet{akiyama2018}, which is plotted as a function of redshift by solid black line. Since the model quasars do not consider the dust reddening to the nucleus, we set the c2 criteria to include quasars at the same redshifts with c1 but reddened by dust. The region also contains quasars in the redshift range of $4.0<z<4.5$, where the completeness of the c1 criteria significantly drops. The c3 criteria extend to the area where contaminations by quasars at $z<3.5$ and Galactic stars dramatically increase. The c4 and c5 criteria further expand the selection area to cover those $z\sim4$ quasars with large offsets away from the model colors. By applying the supplemental c2-5 color criteria for objects with stellar morphology in the HSC-SSP S16A-Wide2 dataset, we select additional 11009 c2, 10060 c3, 84412 c4 and 51525 c5 targets with $20<i<24$. They are included as fillers for the multi-object spectroscopic observations.

\subsection{Spectroscopic observations and data processing}
\label{sec:s2obs}

We carried out spectroscopic observations of the $z\sim4$ quasar candidates using the two-degree field (2dF) fibre positioner \citep{2df} with the AAOmega spectrograph \citep{aao} mounted on the 3.9m Anglo-Australian Telescope (AAT) during three observing runs S17A/08, S18A/16, and S18B/03 between 2017 to 2019, and the DEep Imaging Multi-Object Spectrograph (DEIMOS) \citep{deimos} mounted on the Keck II telescope during the observing run S356 in 2019.  

\subsubsection{Observations with AAT/AAOmega}

The 2dF fibre positioner is able to place up to 400 fibres (8 fibre bundles for guiding, 25 fibres for sky subtraction, and remaining fibres for science objects) within a 2 deg diameter FoV, and AAOmega is a dual-beam spectrograph. In the AAT S17A/08 observing run, to fully cover both of the Ly$\alpha$ 1215{\AA} and ${\rm C_{\Rmnum{4}}}$ 1549{\AA} emission lines of quasars at $3<z<4.5$, we {choose} the following setup: the X6700 dichroic beam splitter, the 580V grating with central wavelength at 5750{\AA} in the blue channel, and the 385R grating with central wavelength at 8200{\AA} in the red channel. The configuration covers a wavelength range of 4800-9800{\AA}, with a resolution of R$\sim$1600. Target FoVs are determined to maximize the number of c1 quasar candidates, and c2-5 targets are added as fillers in each field. In total, we pick up 7 target fields, which contain 84 c1 and 1695 c2-5 candidates with $20<i_{\rm psf}<23$. During the observing run, the seeing {is} $\sim1.4^{\prime\prime}-1.8^{\prime\prime}$. Each field {is} observed with a net exposure time of 160-300 minutes, typically divided into multiple exposures of 20-30 minutes. Details are summarized in Table~\ref{tab:obs_log}.
\begin{table*}[!ht]
\caption{Target fields of the $z\sim4$ quasars in the HSC-SSP S16A-Wide2 dataset}\label{tab:obs_log}
\begin{adjustbox}{width=0.95\textwidth,center=15cm}
\centering
\begin{tabular}{lccccccccc}
\hline\hline
\multicolumn{10}{c}{Jul. 2017, AAT/2dF-AAOmega, PI: Akiyama, 17A/08}\\
\hline
Field         & RA  & DEC  & $i_{\rm psf}-$mag limit   & Exp. time [min] &c1 &c2 &c3 &c4 &c5 \\
\hline
Wide-XMM & 02:14:38.06  & $-$05:56:36.1 & 23.0 & 180 &11&9&15&163&88  \\
Wide-GAMA15H & 14:08:27.76 & $-$01:10:55.8 & 23.0 &250&13&15&28&98&94 \\
Wide-GAMA15H & 14:19:16.23  & $-$00:26:27.3 & 23.0 &200&13&21&26&90&117  \\
Wide-GAMA15H & 14:40:59.68  & $+$01:09:02.5 & 23.0 & 260&14&22&32&97&125  \\
Wide-VVDS & 22:06:09.46  & $+$01:24:38.4 & 23.0 & 300&9&20&32&80&90\\
Wide-VVDS & 22:36:44.23  & $+$00:42:45.1 & 23.0 & 160&7&28&33&79&97  \\
Wide-VVDS & 22:46:38.19  & $+$01:11:16.8 & 23.0 & 220&17&23&32&75&66 \\
\hline
\multicolumn{10}{c}{Jun. 2018, AAT/2dF-AAOmega, PI: He, 18A/16}\\
\hline
Wide-GAMA15H & 14:43:30.48  & $-$00:17:42.1 & 23.5 &180&16&66&60&50&60 \\
\hline
\multicolumn{10}{c}{Apr. 2019, KeckII/DEIMOS, PI: He, S356}\\
\hline
Wide-GAMA09H & 09:14:08.05  & $-$01:27:22.2 & 23.5 & 90&3&2&0&2&5  \\
Wide-GAMA09H & 09:25:39.79  & $+$02:33:47.7 & 23.5 &60&3&1&2&1&4  \\
Wide-WIDE12H & 11:49:23.68    & $-$01:47:25.6 & 23.5 & 60&3&2&2&4&8 \\
Wide-WIDE12H & 11:56:13.07    & $-$01:01:25.8 & 23.5 & 60&4&0&0&1&5  \\
Wide-WIDE12H & 12:04:10.30     & $+$00:33:30.9 & 23.5 & 60&3&1&1&5&7  \\
Wide-GAMA15H & 14:15:45.54     & $+$01:04:05.0 & 23.5 & 60&3&0&2&8&4  \\
Wide-GAMA15H & 14:18:34.62   & $+$01:06:39.8 & 23.5 &60&3&1&0&14&2  \\
Wide-GAMA15H & 14:47:50.67    & $-$01:38:37.1 & 23.5 & 70&4&1&1&6&2  \\
Wide-GAMA15H & 14:54:09.40    & $-$00:12:19.1 & 23.5 & 60&4&2&0&10&2 \\
Wide-HECTOMAP & 15:57:10.24    & $+$42:09:33.5 & 23.5 & 90&4&1&1&4&7  \\
Wide-HECTOMAP & 16:14:29.27    & $+$42:07:03.6 & 23.5 & 40&3&1&2&4&5  \\
Wide-HECTOMAP & 16:34:36.23    & $+$42:54:53.8 & 23.5 & 40&4&0&3&3&5  \\
\hline\\
\end{tabular}
\end{adjustbox}
\end{table*}

In the AAT S18A/16 and S18B/03 observing runs, since a spectral break between the red and blue channels after the splicing process is found in some spectra with the setup using the dichroic beam splitter at 6700{\AA}, we {choose} the standard setups: the X5700 dichroic beam splitter, the 580V grating with central wavelength at 4821{\AA} in the blue channel, and the 385R grating with central wavelength at 7251{\AA} in the red channel. The configuration covers a wavelength range of 3800-8800{\AA}, with a resolution of R$\sim$1400. In the S18A/16 run, we {observe} one FoV containing 16 c1 and 236 c2-5 candidates with $20<i_{\rm psf}<23.5$. In the S18B/03 observing run, we {observe} three FoVs including 93 c1 and 52 c3 candidates with $20<i_{\rm psf}<24$ as fillers of a program targeting $z>4$ Lyman break galaxies. Both of the observing runs {have} bad weather conditions, and the effective exposure time is severely reduced, resulting in low SNRs in most of the spectra. Details of the S18A/16 and S18B/03 observing runs are summarized in Table~\ref{tab:obs_log} and \ref{tab:obs_log2}, respectively.
\begin{table*}[!t]
\caption{Target fields of the $z\sim4$ quasars in the HSC-SSP S16A Udeep/Deep dataset}\label{tab:obs_log2}
\begin{adjustbox}{width=0.95\textwidth,center=15cm}
\centering
\begin{tabular}{lccccccccc}
\hline\hline
\multicolumn{10}{c}{Jan. 2019, AAT/2dF-AAOmega, PI: Ono, 18B/03}\\
\hline
Field         & RA  & DEC  & $i_{\rm psf}$-mag limit   & Exp. time [min] &c1 &c2 &c3 &c4 &c5 \\
\hline
Udeep-SXDS & 02:19:30.00  & $-$04:52:00.0 & 24 &30&26&-&8&-&- \\
Udeep-SXDS &02:22:00.00 &$-$04:36:00.0&24&60&29&-&24&-&-\\
Deep-COSMOS &10:00:48.80 &$+$02:13:12.0&24&130&18&-&8&-&-\\
Deep-COSMOS\footnote{It adopts a different fibre configuration for the same field due to conflicts of fibre in the FoV.} &10:00:55.80 &$+$02:13:12.0&24&130&20&-&12&-&-\\
\hline\\
\end{tabular}
\end{adjustbox}
\end{table*}

The raw data is processed by the \texttt{OzDES} pipeline, which uses \texttt{python} scripts to reduce the individual science frame with the \texttt{v6.46 2dfdr} pipeline, and then to combine the spectra from multiple exposures (private communication with Lidman, C.). Wavelength dependence of the sensitivity is corrected using the default sensitivity function provided in the pipeline. \texttt{PyCosmic} \citep{pcosmic} is used to remove pixels affected by cosmic rays. {There are 5 spectra affected by a severe splicing break at 6700{\AA} between the red and blue channels, thus we manually re-normalize} the blue channel by matching the mean count in 6600-6700{\AA} in the blue channel to that in 6800-6900{\AA} in the red channel. For each spectrum, flux calibration is applied by adjusting its normalization to minimize the average difference between the $riz$-band PSF magnitudes cataloged in the HSC-SSP S16A-Wide2 dataset, and those directly calculated from the reduced spectrum. The $riz$-bands are set since they are mostly covered by the observed wavelength range. The cataloged magnitudes {have been corrected} for the galactic extinction. 

\subsubsection{Observation with Keck/DEIMOS}

DEIMOS is a multi-slit imaging spectrograph, which covers up to $\sim100$ objects within a FoV of $16^{\prime}\times4^{\prime}$. In the S19A/S356 observing run, we {adopt} a setup with a slit width of $1^{\prime\prime}$, the 600 line mm$^{-1}$ (600ZD) grating with central wavelength at 7500{\AA}, and the GG495 blue blocking filter to cover both of the Ly$\alpha$ 1215{\AA} and ${\rm C_{\Rmnum{4}}}$ 1549{\AA} emission lines at $3<z<4.5$. The configuration covers a typical wavelength range of $\sim$5000-10000{\AA} with a resolution of $R\sim$1600, but the range depends on the target position on the mask. We {observe} 12 FoVs, which contain in total 41 c1 and 144 c2-5 candidates with $20<i_{\rm psf}<23.5$. The seeing condition {is} $\sim0.5^{\prime\prime}-0.7^{\prime\prime}$. Each field {is} observed with a total exposure time of 40-90 minutes, typically divided into 2-3 exposures of 20-30 minutes. Details are summarized in Table~\ref{tab:obs_log}.

The raw data is processed by the \texttt{DEEP2} pipeline \citep{cooper, newman}. At first bias subtraction, cosmic ray removal, and flat-fielding {are} applied in a standard manner, then wavelength calibration and slit-tilt correction are determined. Sky subtraction is applied to the tilt-corrected data, and extractions of spectra are conducted. The extracted spectra from multiple frames are combined. Two CCD chips 2 and 6 of the J091408$-$012722 field, covering one c1 and one c4 candidates, {fail} in the reduction with the pipeline. We then use the \texttt{IRAF longslit} and \texttt{twodspec} packages to reduce their spectra by cutting the data into longslit spectra. Feige 34 and BD$+$25d3941 are observed to obtain the throughput as a function of wavelength, and the flux calibration is applied using the same method described in the AAOmega data processing.

In total, we cover 234 c1, 216 c2, 324 c3, 794 c4, and 793 c5 candidates, as summarized in Table~\ref{tab:obs_c25}. We {observe} 141 c1 quasars on the HSC Wide layer, which account for $141/1666=8.5$\% of the full quasar sample in \citet{akiyama2018}. The photometric redshift of the 141 quasars spans between $z_{\rm phot}=3.4$ and $4.5$, with the mean at $z=3.8$ and the standard deviation of 0.2. This distribution is consistent with that of the full quasar sample in \citet{akiyama2018}, suggesting we do not introduce bias in the distribution of photometric redshift when selecting targets for spectroscopic observations.  

\subsection{Classification and redshift determination}
\label{sec:s2iden}

\begin{figure*} [!ht]
\centering
\includegraphics[keepaspectratio,width=.9\textwidth]{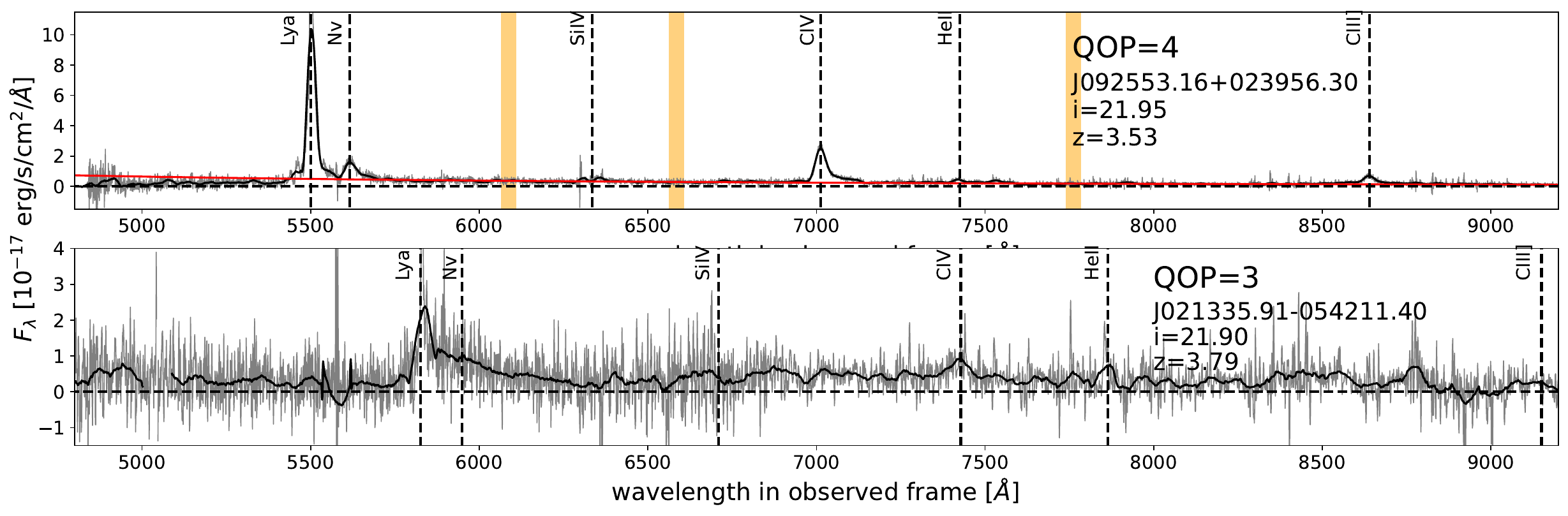}
\caption{Examples of spectra assigned with a quality flag of QOP=4 (top) and QOP=3 (bottom). Gray line shows the obtained spectrum, and black solid line displays the spectrum smoothed by a Savitzky-Golay filter. Orange shades highlight the wavelength ranges used to determine the best-fit power-law model of the spectrum continuum, and red solid line shows that best-fit result. Central wavelength of strong emission lines expected from the systemic redshift is denoted by vertical dashed lines. }
\label{fig:spec_ex}
\end{figure*}

We use the web-based program \texttt{MARZ} \citep{marz} to manually inspect each spectrum and determine its redshift. \texttt{MARZ} can automatically determine the best-fit object type and redshift. It matches input spectra against a variety of spectral templates of stars and galaxies using a modified version of \texttt{AUTOZ} cross-correlation algorithm implemented by \citet{autoz}. For {an} incorrect matching, users are allowed to manually re-fit the spectrum by iterating the automatic results, modifying the choice of a template, or marking specific spectral features. \texttt{MARZ} is originally developed for the Australian Dark Energy Survey (OzDES) carried out with the AAT/2dF-AAOmega spectrograph, while it can also handle spectrum taken with other instruments and pipelines if the spectrum is in the standard FITS format. In the course of inspection, we assign a quality flag to each spectrum with the following criteria: 
\begin{itemize}
\item QOP=4 - quasars at $3<z<4.5$ with both the Ly$\alpha$ and ${\rm C_{\Rmnum{4}}}$ emission lines detected, and the ${\rm C_{\Rmnum{4}}}$ line profile detected with sufficient SNRs; 
\item QOP=3 - quasars at $3<z<4.5$ with only the Ly$\alpha$ emission line detected, but the ${\rm C_{\Rmnum{4}}}$ emission line highly affected by the absorption lines, or barely detected that its profile can not be well constrained; 
\item QOP=2 - quasars at $z<3$ with other multiple emission lines detected; 
\item QOP=1 - galaxies; 
\item QOP=6 - stars; 
\item QOP=0 - unknown or non-detection.
\end{itemize}
Two example spectra assigned with the quality flag of QOP=4 and 3 are shown in the top and bottom panels of Figure~\ref{fig:spec_ex}, respectively.

There are 125 out of the 234 c1 candidates observed under good observing conditions, i.e., during the S17A/08 and S19A/S356 runs. Among them, we identify 78 quasars at $3<z<4.5$ with 51 QOP=4 and 27 QOP=3 quality flags. The resulting success rate of c1 candidates to be a $z\sim4$ quasar is $78/125=62.4$\%. There are 1 and 4 c1 candidates identified as a Galactic star and $z<3$ quasars, respectively. The remaining objects show no clear features to be identified with any of the templates due to either low SNRs or failure in the reduction of their spectra. Among the 150 c2, 212 c3, 744 c4 and 733 c5 candidates observed in the two observing runs, 9, 18, 6 and 3 of them are identified with a $3<z<4.5$ quasar, resulting in the identification rates of 6\%, 8.5\%, 0.8\% and 0.4\%, respectively. 

Under observing runs with poor weather conditions, i.e., the S18A/16 and S18B/03 runs, majority of the observed candidates have too low SNR to be identified, and we identify 14, 1 and 8 quasars at $3<z<4.5$ from the c1, c2, and c3 candidates, respectively. Moreover, we also identify 14 $3<z<4.5$ quasars, including 7 with QOP=4, among fillers selected from $z\sim4$ Lyman-break galaxy candidates \citep{z4lbg} and radio-detected candidates \citep{wergs1}. Details are summarized in Table~\ref{tab:obs_det}.
\begin{table*}[!ht]
\caption{Identification of the $z\sim4$ quasar candidates}\label{tab:obs_det}
\begin{adjustbox}{width=0.9\textwidth,center=16cm}
\centering
\begin{tabular}{cccccccc}
\hline\hline
\multicolumn{8}{c}{Targets observed under good conditions}\\
\hline
criteria  & total & QOP=4 quasar & QOP=3 quasar & success rate & $z<3$ quasar &galaxy &star\\
\hline
c1 & 125 & 51 & 27 (3)\footnote{Numbers in the brackets are for the BAL quasar identified in each criteria.}  & 62.4\%& 4 & 0 & 1 \\
c2 & 150 & 3 & 6 (1) & 6.0\%& 0 & 1 & 0 \\
c3 & 212 & 8 & 10 (2)  & 8.5\%& 12 & 5 & 0 \\
c4 & 744 & 2 & 4 (1) & 0.8\%& 26 & 1 & 8 \\
c5 & 733 & 1 & 2 (-)& 0.4\%& 3 & 4 & 3 \\
\hline
\multicolumn{8}{c}{Targets observed under poor conditions}\\
\hline
c1 & 109 & 5& 9 (1) &-& 0 & 1 & 1 \\
c2 & 66 & 0 & 1 (1)&-& 0 & 0& 0 \\
c3 & 112 & 3 & 5 (1) & -& 0 & 1 & 0\\
\hline
collaborator & - & 7 &7 (1)&-&-&-&- \\
\hline
total & - & 80 & 71 (11) &-& 45 & 13 & 13\\
\hline\\
\end{tabular}
\end{adjustbox}
\end{table*}

In total, we identify 151 quasars at $3<z<4.5$, containing 80 with QOP=4, and 71 with QOP=3 quality flags. Among them, 11 show strong and broad absorption feature of the ${\rm C_{\Rmnum{4}}}$ line, and they are identified as broad absorption line (BAL) quasars. Details are summarized in Table~\ref{tab:z4qsoIden}. 
\begin{table}[!ht]
\caption{Identified $3<z<4.5$ quasars}\label{tab:z4qsoIden}
\begin{adjustbox}{width=0.55\textwidth,center=7.5cm}
\centering
\begin{tabular}{crrccccc}
\hline\hline
Object & ra & dec& criteria & QOP & type &  $i-$mag & $z_{\rm spec}$  \\
\hline
J095825.79$+$012628.64&149.61&1.44&collaborator\footnote{Quasars selected by the Lyman break color or radio detection}&4&	&20.57&3.167\\
J100142.97$+$030118.85&150.43&3.02&c3&4&	&21.55&3.451\\
J021917.34$-$040443.21&34.82&$-$4.08&c3&4&	&21.13&3.483\\
J095931.01$+$021332.89&149.88&2.23&c1&4&	&22.73&3.638\\
J100149.65$+$030657.10&150.46&3.12&c1&4&	&21.03&3.506\\
J095906.46$+$022639.41&149.78&2.44&c1&3&	&22.38&4.161\\
J100334.35$+$015649.98&150.89&1.95&c1&3&	&22.52&3.899\\
J224816.00$+$020609.59&342.07&2.10&c1&3&BAL&20.11&3.638\\
\hline\
\end{tabular}
\end{adjustbox}
\tablecomments{Table~\ref{tab:z4qsoIden} is published in its entirety in the machine-readable format. A portion is shown here for guidance regarding its form and content.}
\end{table}

In Figure~\ref{fig:CCD}, we plot the $g-r$ and $r-z$ color distribution of the identified sources. The identified $3<z<4.5$ quasars are shown by stars (QOP=4) and dots (QOP=3) in the left panel. Their distribution follows the color track of model quasars between $z=3$ and 4.5. There are 9 quasars, marked by open black circles, locating outside of the c1-5 windows {due to different selection criteria of the Lyman break color or radio detection}. Additional 5 quasars fall in the c1-3 selection windows, but they are removed from our samples by the $i-z$ and $z-y$ color criteria. In the right panel, color distributions of the contaminating $z<3$ quasars, galaxies and Galactic stars are plotted by orange squares, green triangles and purple inverted triangles, respectively. Majority of them locate in the c3-5 selection windows as expected, with a few contaminants locating at the edges of the c1 window. The 4000{\AA} break in $z\sim0.5$ galaxies can mimic the Lyman break at $z\sim4$. Also, the broad ${\rm Mg_{\Rmnum{2}}}$ 2798{\AA} emission line in some quasar spectra mimics the Ly$\alpha$ 1215{\AA} emission line.

\begin{figure*} [!ht]
\centering
\includegraphics[keepaspectratio,width=0.7\textwidth]{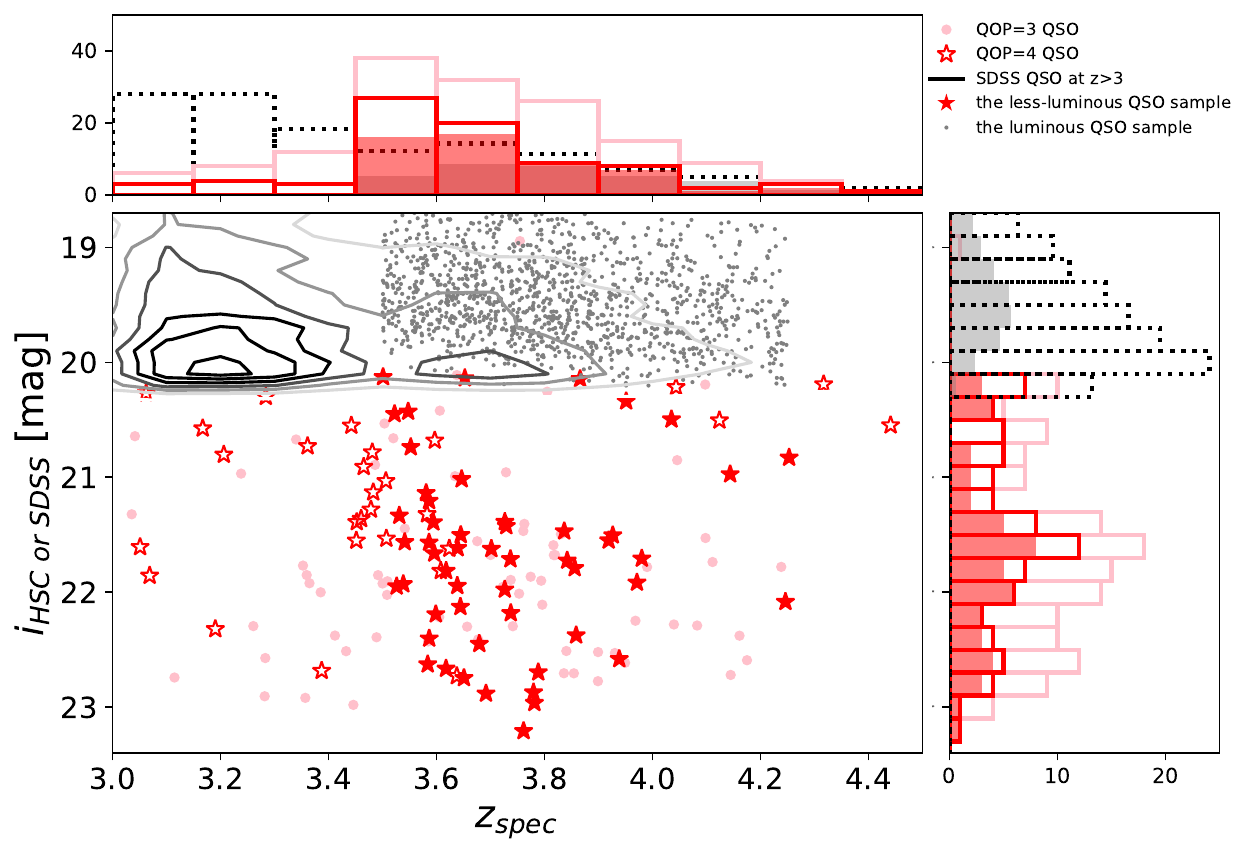}
\caption{Distribution of the $i$-band magnitude and spectroscopic redshift, $z_{\rm spec}$, of the identified quasars. Red stars and pink dots indicate the identified $3<z<4.5$ quasars with QOP=4 and QOP=3 quality flags, respectively. Black contours show the distributions of the luminous SDSS DR7 quasars at $3<z<4.5$. Filled red stars and grey dots represent the 52 QOP=4 quasars and 1,462 SDSS quasars at $3.5<z<4.25$ used in the determination of the $z=4$ BHMF, respectively. Horizontal and vertical histograms plot the distributions of redshift and $i$-band magnitude of the quasar samples. Red open histogram, red filled histogram, pink open histogram, black open histogram and grey filled histogram represent the distributions of the 80 QOP=4 quasars, the 52 QOP=4 quasars for determining the $z=4$ BHMF, the entire 151 QOP=4 plus QOP=3 quasars, the SDSS DR7 quasars at $3<z<4.5$, and the 1,462 SDSS quasars for determining the $z=4$ BHMF, respectively. Histograms of the SDSS quasars are scaled down by a factor of 45.}
\label{fig:magz}
\end{figure*}

In Figure~\ref{fig:magz}, 
we plot the $i$-band magnitude vs. redshift distribution of the identified quasars at $3<z<4.5$. The pink open histogram in the horizontal panel shows that the 151 identified quasars cover a redshift range of $3.04\leq z_{\rm spec} \leq 4.44$, with a mean and standard deviation at 3.67 and 0.28, respectively. The vertical panel displays the $i$-band magnitude distribution of the identified quasars. Pink open histogram represents the entire 151 identified quasars, and they cover a magnitude range of $18.9<i_{\rm psf}<23.2$, with a median at $i=21.7$. If only considering the QOP=4 quasars, whose distribution is plotted by red open histogram, we see they cover a similar magnitude range, but with a slightly brighter median at $i=21.5$, indicating no or weak detection of the ${\rm C_{\Rmnum{4}}}$ emission line in the QOP=3 quasars is partly due to their faintness. 

Here, we examine the completeness of the c1 criteria in selecting quasars at $3.5\leq z \leq4.5$. Since the c1-5 criteria are determined following the two-color distributions of model quasars at $3.5<z<4.5$, we assume the combined c1-5 criteria are wide enough to cover the typical quasars in the redshift range. We use the data taken in the observing run S17A/08 {as it covers a large number of c1-5 candidates under good conditions}. There are 84 c1, 138 c2, 198 c3, 682 c4 and 677 c5 candidates observed in total, which occupy 13.2\%, 25.0\%, 9.4\%, 1.3\% and 2.0\% of the full c1-5 samples down to $i=23$, respectively. Among the observed candidates, we identify 60 c1, 9 c2, 7 c3 and 1 c5 candidates to be quasars at $3.5\leq z \leq4.5$. As discussed in section~\ref{sec:s2obs}, we do not introduce bias in photometric redshift distribution when selecting the c1 candidates for spectroscopic observations, and c2-5 candidates are randomly selected over the target fields as fillers. Based on the statistics, the completeness of c1 criteria in selecting quasars at $3.5\leq z \leq4.5$ is (60/0.132)/(60/0.132+9/0.250+7/0.094+1/0.020)=74\%, which is broadly consistent with that evaluated with the SDSS $z\sim4$ quasars \citep[66\%; for details see][]{akiyama2018}. Except for the c1 criteria, the c3 criteria show the highest completeness of 12\%. However, the c3 criteria also show much higher contamination rate of 9\% than that of the c1 criteria of 1\% in the magnitude range of $20<i<23$. We thus confirm the c1 criteria established in \citet{akiyama2018} are able to select quasars at $3.5\leq z \leq4.5$ with high completeness and the minimal contamination.

\begin{figure} [!ht]
\centering
\includegraphics[keepaspectratio,width=0.45\textwidth]{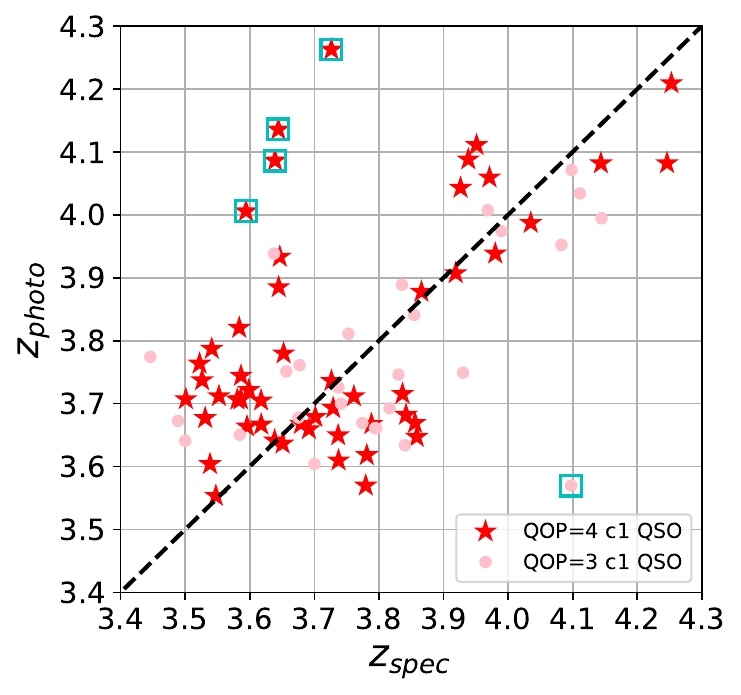}
\caption{Comparison of the photometric redshifts of the c1 quasars with their spectroscopic redshifts. Symbols have the same meaning with Figure~\ref{fig:magz}. Cyan open squares mark the quasars with $|z_{\rm phot}-z_{\rm spec}|>0.35$. Black dashed line indicates the one-to-one relation.}
\label{fig:zp_zs}
\end{figure}

The photometric redshift of c1 quasar candidates are estimated in \citet{akiyama2018}. {We identify 93 among them in the spectroscopic observations. The comparison between their photometric and spectroscopic redshifts is shown} in Figure~\ref{fig:zp_zs}.
The identified quasars with QOP=4 and QOP=3 quality flags are plotted by red stars and pink dots, respectively. Overall, their photometric redshifts are consistent with the spectroscopic redshifts. The accuracy of photometric redshift, which is defined as $1.48\times{\rm median}(|z_{\rm phot}-z_{\rm spec}|/(1+z_{\rm spec}))$ \citep[e.g.,][]{Ilbert06}, is 0.04. No quasars fall in the outlier of $|z_{\rm phot}-z_{\rm spec}|/(1+z_{\rm spec})>0.15$. We note that there are five quasars with the discrepancy between their photometric and spectroscopic redshifts larger than 0.35. After inspecting their spectra, we {find} the one in the lower right corner is a BAL quasar, and the remaining 4 quasars in the upper left corner show flatter continuum shape in their spectra than those expected from their broad-band photometries. The systematic offsets between the flux-calibrated and cataloged magnitudes of the 4 quasars reach 0.5-2 mag, resulting in a large discrepancy between the photometric and spectroscopic redshift. 
 
Moreover, we examine the distribution of spectroscopic redshift of the identified $3<z<4.5$ quasars in each FoV. In the 20 FoVs within the Wide2 dataset, the standard deviation of redshift distribution of each FoV is $\Delta z_{\rm spec}\sim$0.1-0.3, which corresponds to a velocity offset of $\Delta$v=6600-20000 km s$^{-1}$. Thus, most of the identified quasars in the target fields are just clustered in the projected plane by chance. There are 2 quasar pairs, which have $\Delta$v$<$2000 km s$^{-1}$ and the angular projected distance $r_{\perp}<4$ proper Mpc (pMpc). They are listed in Table~\ref{tab:qsopair}. Following the definition of a quasar pair in \citet{Onoue18}, i.e., $\Delta$v$<$3000 km s$^{-1}$ and $r_{\perp}<4$ pMpc, the 2 quasar pairs can have physical association, but none of them meet the tighter criterion {for physical association} in \citet{chiang13}, which is determined by the size of simulated proto-clusters at $z\sim4$. 
\begin{table*}[!ht]
\caption{Quasar pairs identified at $z\sim4$ in this work}\label{tab:qsopair}
\begin{adjustbox}{width=0.95\textwidth,center=16cm}
\centering
\begin{tabular}{ccccccccc}
\hline\hline
object & criteria & class & $z_{\rm spec}$ & $i-$mag & $\Delta z$ & $\Delta$V [km s$^{-1}$] & $\Delta \theta$ [arcsec] & $r_\perp$ [pMpc]\\
\hline
J021335.91$-$054211.40 &c1& QOP=3 & 3.794& 21.90 & 0.002 & 127 & 543.2 & 3.9\\
J021330.99$-$053313.19 &c1& QOP=3 & 3.796 & 22.11 & & & &\\
\hline
J115616.98$-$005937.06&c1 &QOP=4 & 3.761 & 23.21& 0.018 & 1157 & 154.3 & 1.1\\
J115626.34$-$005832.99 & c1 &QOP=4 &3.780& 22.87  & & & &\\
\hline\\
\end{tabular}
\end{adjustbox}
\end{table*}


\section{Black hole masses and Eddington ratios}
\label{sec:section3}

\subsection{Single-epoch virial BH mass estimator}
\label{sec:s3_intro}

Under the assumption that the broad-line region (BLR) is virialized, the virial BH mass can be estimated from the motion of the BLR through $M_{\rm BH}= fr_{\rm BLR} v^{2}/G$, where $f$ is a scaling factor accounting for the kinematic structure of the BLR, $v$ is the velocity of the BLR gas that can be derived from line width of broad emission lines, and $r_{\rm BLR}$ is the BLR radius. Based on the reverberation mapping of local broad-line AGNs, a tight correlation between the size of broad H$\beta$-emitting region ($r_{\rm BLR,H\beta}$) and the continuum luminosity around the H$\beta$ emission line at 5100{\AA} is found \citep[the $r-L$ relationship;][]{Kaspi}. The relationship enables us to estimate the virial BH mass of broad line AGNs simply with the H$\beta$ line width ($\Delta v_{\rm H\beta}$) and the 5100{\AA} continuum luminosity ($L_{\rm 5100}$), both of which can be derived from a single-epoch spectrum. 

The virial BH mass estimates have been extended to other broad emission lines, such as H$\alpha$, ${\rm Mg_{\Rmnum{2}}}$ and ${\rm C_{\Rmnum{4}}}$ emission lines \citep[e.g.,][]{VP06,SL2012,Denney13,park13,park17}. In this work, since only the ${\rm C_{\Rmnum{4}}}$ 1549{\AA} emission line of the $z\sim4$ quasars is covered among the calibrated broad emission lines in the optical wavelength range, we adopt the ${\rm C_{\Rmnum{4}}}$ virial BH mass estimate in \citet{VP06} as follows:
\begin{align}
\log \left( \frac{M_{\rm BH, FWHM}}{M_{\odot}} \right)  =&  6.66 + \log \left[ \left( \frac{\lambda L_{\lambda} (1350)}{10^{44}\ {\rm erg\ s^{-1}}} \right)^{0.53} \right] \notag \\ 
 &+ \log \left[ \left( {\rm \frac{FWHM_{\rm C_{\Rmnum{4}}}}{1000 km\ s^{-1}}} \right)^{2} \right], 
\label{eq:mbh_civ_fwhm}
\end{align}
where FWHM represents the line width of ${\rm C_{\Rmnum{4}}}$ 1549{\AA} emission line, and $L_{\rm 1350}$ is the monochromic UV luminosity at 1350{\AA}. The systematic uncertainty of the calibration is evaluated to be 0.36 dex \citep{VP06}. 

We note that there are also other studies providing calibrations of the ${\rm C_{\Rmnum{4}}}$-based single-epoch virial BH mass estimate \citep[e.g.,][]{SL2012,park13,park17}. For example, \citet{park13} use a sample of 26 local AGNs with both the H$\beta$ reverberation masses and UV archival spectra available to re-calibrate the ${\rm C_{\Rmnum{4}}}$-based virial BH mass estimator. Compared to the method used in \citet{VP06}, \citet{park13} perform the multi-component fitting on the ${\rm C_{\Rmnum{4}}}$ complex region to precisely deblend ${\rm C_{\Rmnum{4}}}$ from other contaminating lines, and they relax the constraint of the slope parameter in the calibration. The virial BH masses estimated with the calibration suggest those estimated with the calibration in \citet{VP06} (i.e., equation~\ref{eq:mbh_civ_fwhm}) can be underestimated in the low-mass range of $\log M_{\rm BH}/M_\odot<8$ and overestimated in the high-mass range of $\log M_{\rm BH}/M_\odot\sim9$. \citet{park17} extend the work by including additional six reverberation-mapped AGNs with BH masses down to $10^{6.5} M_\odot$. The calibration remains consistent with the one in \citet{park13}. Here, since the calibration in \citet{VP06} is commonly used in literature, we firstly adopt it to estimate the virial BH mass for comparison with literature, and we discuss how the other calibrations can affect the estimates in section~\ref{sec:s5_CIV_blueshift}. 

\subsection{Line width and 1350{\AA} monochromic luminosity}
\label{sec:s3_fit}

We measure the continuum flux and emission line width for the 80 QOP=4 quasars, whose ${\rm C_{\Rmnum{4}}}$ 1549{\AA} emission line has sufficient SNRs. Firstly, we fit the continuum of the obtained spectra by a single power-law model $F_{\lambda}\propto \lambda^{\alpha}$. {The fitting is carried out in the rest frame} wavelength ranges of 1340-1350{\AA}, 1450-1460{\AA}, 1700-1710{\AA}, 1800-1810{\AA} and 2000-2010{\AA}, since the contribution from emission and/or absorption lines in these wavelength ranges is small. We manually inspect the best-fit model of each spectrum, and shift the wavelength ranges for fitting by 10-100{\AA} to improve fitting results, especially when large systematic residuals in the wavelength range of 1300-1600{\AA} are found. In the top panel of Figure~\ref{fig:spec_ex}, we show the best-fit power-law model of {the} continuum of one example spectrum by a red line.

After subtracting the power-law component, the ${\rm C_{\Rmnum{4}}}$ emission line is fitted with Gaussian models over the wavelength range of 1500-1600{\AA}. We start from a single Gaussian model, and add an additional Gaussian if necessary. The fitting procedure utilizes the Levenberg-Marquardt optimization method provided by the {\tt PySpecKit} package \citep{pyspeckit}.
\begin{figure} [!ht]
\centering
\includegraphics[keepaspectratio,width=0.45\textwidth]{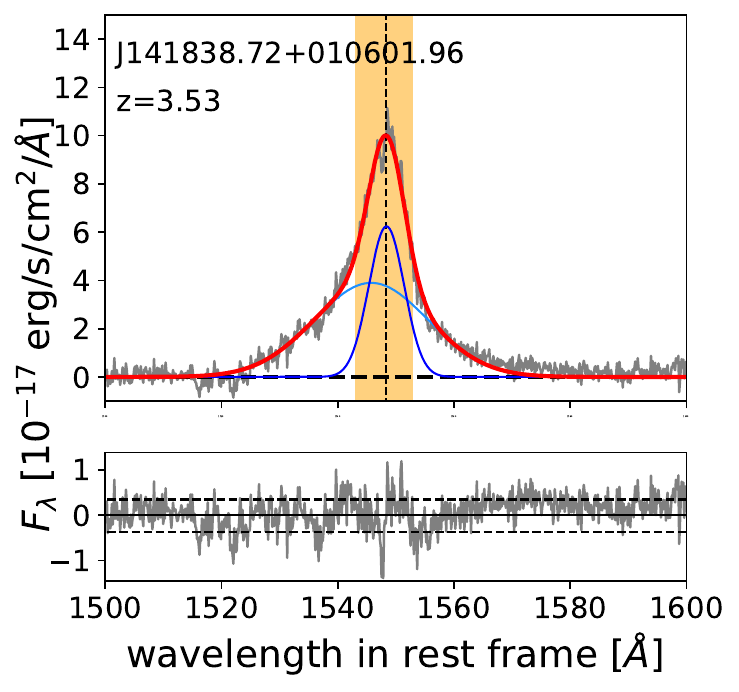}
\caption{Example of a fitting of the broad ${\rm C_{\Rmnum{4}}}$ emission line. Upper panel shows the continuum-subtracted spectrum (gray line), the best-fit Gaussian components (blue lines) and the best-fit model of the entire profile (red line) around ${\rm C_{\Rmnum{4}}}$ emission line. Lower panel displays residuals of the fitting by gray line. Horizontal dashed lines indicate the $1\sigma$ noise.}
\label{fig:CIV_fit}
\end{figure}
There are 13 ${\rm C_{\Rmnum{4}}}$ line profiles affected by narrow absorption lines. Since the SNRs of the spectra are not high enough to fit both the emission and absorption lines simultaneously, we fit the spectra by excluding the wavelength ranges affected by those absorption lines. In addition, we find large systematic residuals from the continuum subtraction at the edges of the ${\rm C_{\Rmnum{4}}}$ emission line in 15 spectra, and we exclude those wavelength ranges in fitting as well. Moreover, there is one ${\rm C_{\Rmnum{4}}}$ line profile dominated by spiky narrow component. To prevent the influence of the narrow-line region on estimating the virial BH mass, we remove the component from the best-fit model. There are 6 ${\rm C_{\Rmnum{4}}}$ emission lines well fitted with a single Gaussian, but for the remaining 74 spectra, two Gaussians are required to smoothly describe the observed profile. An example of {the} fitting is shown in Figure~\ref{fig:CIV_fit}.

With the best-fit profile constructed by combining all the Gaussian components, we measure the ${\rm C_{\Rmnum{4}}}$ line width. Both the FWHM and line dispersion can be used to describe the broad line width. While there are studies suggesting the line dispersion is more appropriate in measuring the broad line width \citep[e.g.,][]{Denney13,park17}, it requires high S/N data to accurately fit the line wings, especially for the line profile with extended wings like ${\rm C_{\Rmnum{4}}}$ line \citep[e.g.,][]{park17}. Here, since most of our spectra have limited SNR$\sim$2-5 {at the peak of ${\rm C_{\Rmnum{4}}}$ line}, and systematic residuals of the background and continuum subtraction are found in some spectra, especially in those taken with AAT/AAOmega, we only measure the FWHM as line width for virial BH mass estimates. The derived FWHMs are corrected for the instrumental broadening of the spectrograph, which is evaluated using the line width of the night sky emission lines. Table~\ref{tab:mass} summarizes the results.
\begin{table*}[!ht]
\caption{Continuum and emission line properties of the 80 QOP=4 quasars}\label{tab:mass}
\begin{adjustbox}{width=0.95\textwidth,center=16cm}
\centering
\begin{tabular}{lllccr}
\hline\hline
Object  &  $\log L_{1350 {\AA}}$ [erg/s]& FWHM(${\rm C_{\Rmnum{4}}}$) [km s$^{-1}$]&  $\log L_{\rm bol}$ [erg/s]  & log $M_{\rm BH}/M_{\odot}$ & log $\lambda_{\rm Edd}$ \\
\hline
J095825.79$+$012628.64&$42.69^{+0.04}_{-0.05}$&$7868^{+530}_{-1886}$&$46.40^{+0.04}_{-0.05}$&$9.42^{+0.06}_{-0.32}$&$-1.12^{+0.32}_{-0.06}$\\
J095819.36$+$013530.52\footnote{Narrow-line quasar}&$42.18^{+0.07}_{-0.12}$&$952^{+71}_{-53}$&$45.89^{+0.07}_{-0.12}$&$7.31^{+0.08}_{-0.10}$&$0.48^{+0.05}_{-0.07}$\\
J095848.37$+$013818.12&$41.99^{+0.00}_{-0.00}$\footnote{Measured from broad-band photometries}&$1743^{+130}_{-148}$&$45.70^{+0.00}_{-0.00}$&$7.74^{+0.06}_{-0.08}$&$-0.13^{+0.08}_{-0.06}$\\
J100142.97$+$030118.85&$42.09^{+0.06}_{-0.08}$&$1464^{+106}_{-100}$&$45.80^{+0.06}_{-0.08}$&$7.64^{+0.07}_{-0.06}$&$0.06^{+0.08}_{-0.07}$\\
J021917.34$-$040443.21&$42.35^{+0.08}_{-0.09}$&$1355^{+139}_{-97}$&$46.06^{+0.08}_{-0.09}$&$7.71^{+0.08}_{-0.08}$&$0.25^{+0.07}_{-0.10}$\\
J220522.42$+$021149.32&$42.52^{+0.00}_{-0.00}$$^{\rm b}$&$2355^{+207}_{-468}$&$46.23^{+0.00}_{-0.00}$&$8.28^{+0.07}_{-0.24}$&$-0.15^{+0.24}_{-0.07}$\\
\hline
\end{tabular}
\end{adjustbox}
\tablecomments{Table~\ref{tab:mass} is published in its entirety in the machine-readable format. A portion is shown here for guidance regarding its form and content.}
\end{table*}

The monochromic luminosity at 1350{\AA}, $L_{1350}$, is derived from the best-fit power-law model. Considering the stellar morphology of the identified quasars, we assume contribution of the host galaxy emission is negligible in the UV wavelength range. As described in section~\ref{sec:s2obs}, the spectra taken with the AAT/AAOmega can be affected by a splicing break between the red and blue arms. Thus, we also measure $L_{1350}$ using the $grizy$ 5-band PSF magnitudes in the HSC-SSP S16A-Wide2 dataset. The magnitudes have been corrected for the galactic extinction. Firstly we subtract the contribution of the emission lines, such as, Ly$\alpha$ and ${\rm C_{\Rmnum{4}}}$, to the corresponding broad-band magnitudes with the line flux measured in the continuum-subtracted spectra. Then, we fit the subtracted 5-band magnitudes with a power-law model $F_{\lambda}\propto \lambda^{\alpha}$, and derive $L_{1350}$ with the best-fit power-law model.

\begin{figure}
\centering
\includegraphics[keepaspectratio,width=0.45\textwidth]{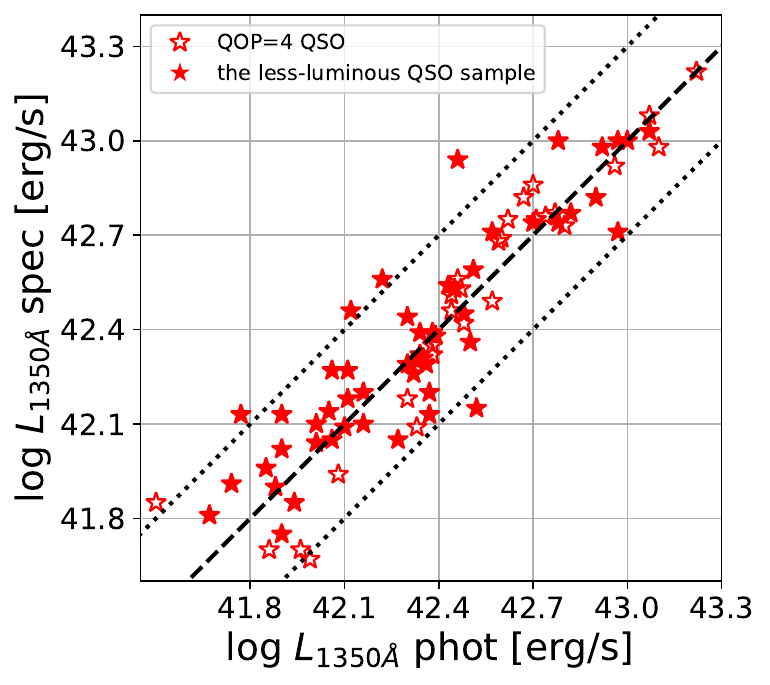}
\caption{Comparison of $L_{1350}$ of 80 QOP=4 quasars derived with the spectra and those derived with the broad-band photometries. Symbols have the same meaning with Figure~\ref{fig:magz}. The dashed line indicates the one-to-one relation, and the dotted lines delimit the area where $|\log L_{1350-\rm phot}-\log L_{1350-\rm spec}|<0.3$.}
\label{fig:L1350}
\end{figure}

In Figure~\ref{fig:L1350}, we compare $L_{1350}$ of quasars derived with their spectra with those derived from broad-band photometries.
For the majority of the objects, the two measurements show consistency, with a median discrepancy of $-0.01$ dex and a standard deviation of 0.15 dex. Therefore, we directly use their $L_{1350-\rm spec}$ as $L_{1350}$. We note that there are 7 objects having the discrepancy greater than 0.3 dex. By inspecting their spectra, we find that the two objects below the lower dashed line show weak detection of continuum and are affected by a splicing break. We thus adopt their $L_{1350-\rm phot}$ as $L_{1350}$ afterwards. For the five objects above the upper dashed line, their spectra show large systematic residuals of background subtraction in the wavelength range of $\lambda_{\rm obs}>6000${\AA}, resulting in over-subtraction in the $izy$-band photometries. We keep their $L_{1350-\rm spec}$ as $L_{1350}$. Table~\ref{tab:mass} summarizes the measurements of $L_{1350}$ of the 80 QOP=4 quasars.

\begin{figure} [!ht]
\centering
\includegraphics[keepaspectratio,width=0.45\textwidth]{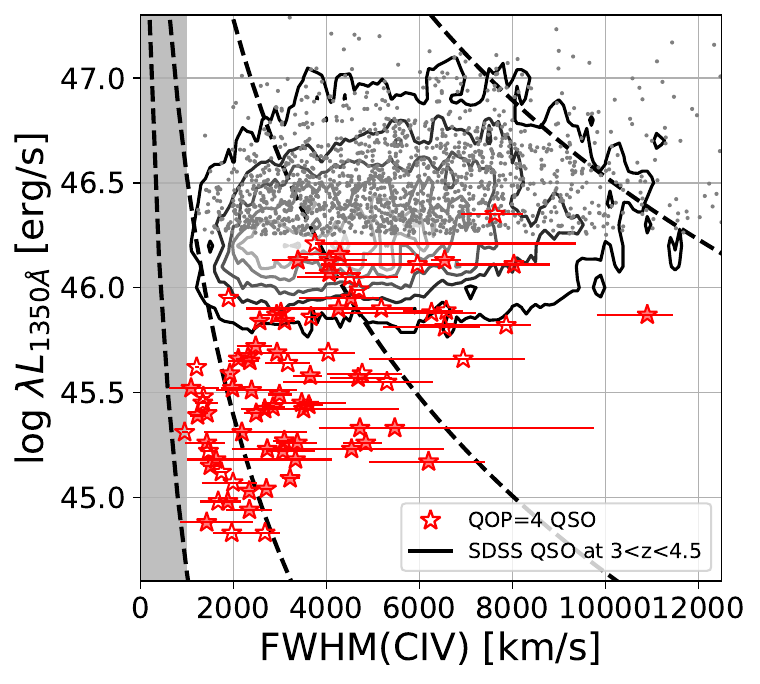}
\caption{Distribution of $L_{\rm 1350}$ of 80 QOP=4 quasars at $3<z<4.5$ as a function of FWHM of ${\rm C_{\Rmnum{4}}}$ 1549{\AA} emission line. Symbols have the same meaning with Figure~\ref{fig:magz}. Black contours show the same distributions for the luminous SDSS DR7 quasars at $3<z<4.5$. Black dashed lines show a constant BH mass of $10^{7},~10^{8},~10^{9},~10^{10}M_{\odot}$ from left to right. Grey shaded area suggests FWHM$<1000$ km s$^{-1}$.}
\label{fig:L1350_fwhm_sig}
\end{figure}

In Figure~\ref{fig:L1350_fwhm_sig}, 
the obtained continuum luminosity $L_{\rm 1350}$ of the 80 QOP=4 quasars at $3<z<4.5$ is plotted against FWHM(${\rm C_{\Rmnum{4}}}$). The quasars distribute in the luminosity range of $\log \lambda L_{\rm 1350}~({\rm erg}~{\rm s}^{-1})=44.83-46.35$. For the line width, the quasars cover a range of $952-10902$ km s$^{-1}$ in FWHM, with a median of 3081 km s$^{-1}$. Compared with the luminous SDSS DR7 quasars in the same redshift range, whose median FWHM is estimated to be 5266 km s$^{-1}$, the line width of the QOP=4 quasars is systematically smaller. Seen from the constant BH mass relation plotted by black dashed lines, the QOP=4 quasars are expected to harbor less-massive SMBHs than the SDSS quasars. We note that there is one quasar measured to have FWHM less than 1000 km s$^{-1}$. By visually inspecting the spectrum, we find that its ${\rm C_{\Rmnum{4}}}$ emission line can be fit with a single Gaussian model with FWHM$<1000$ km s$^{-1}$. Thus, we classify this object as a narrow-line AGN, and reject it in the broad-line quasar sample.

The uncertainty of the continuum and emission line measurements is determined by fitting mock spectra generated from the best-fit model of each quasar. We add Gaussian random noise, which is determined by the standard deviation of the residuals of the fitting, to the best-fit model spectra. Then, we apply the same fitting procedure to each mock spectrum, and measure its line width and continuum luminosity. For each quasar, we construct 50 mock spectra. {The $1\sigma$ uncertainty} in the ${\rm C_{\Rmnum{4}}}$ line width and $L_{\rm 1350}$ measurements are then evaluated by the rms scatter of those measured values from the mock spectra. Table~\ref{tab:mass} summarizes the uncertainties of the 80 QOP=4 quasars.

\subsection{BH mass, bolometric luminosity and Eddington ratio}
\label{sec:s3_mbh}

Based on the FWHM of ${\rm C_{\Rmnum{4}}}$ 1549{\AA} emission line and continuum luminosity $L_{\rm 1350}$ of the 80 QOP=4 quasars at $3<z<4.5$, we can estimate their virial BH mass following equation~(\ref{eq:mbh_civ_fwhm}). The derived BH mass covers a range of $7.4<\log M_{\rm BH}/M_{\odot}<9.7$, with a median of $\log M_{\rm BH}/M_{\odot}=8.5 $. For each quasar, {the $1\sigma$} error in the virial BH mass is calculated from the rms scatter of the 50 generated mock spectra as described in section~\ref{sec:s3_fit}. Table~\ref{tab:mass} summarizes the results.

There are quasars whose ${\rm C_{\Rmnum{4}}}$ emission line shows an extended blue wing, which is fitted by a blueshifted broad component. It is known that such blueshift correlates with spectral properties in quasar spectrum, such as equivalent width (EW) of ${\rm C_{\Rmnum{4}}}$ emission line, suggesting the blueshifted component may associate with a non-virial motion \citep[e.g.,][]{Richards11}. Thus, the necessity to correct for the contribution of the non-virial component to the ${\rm C_{\Rmnum{4}}}$ emission line profile in the virial BH mass estimates has been claimed \citep[e.g.,][]{SL2012}. \citet{Coatman} find the virial BH mass can be over-estimated by five times when the ${\rm C_{\Rmnum{4}}}$ emission line is strongly blueshifted by 3000 km s$^{-1}$. Here, since the wavelength range covered by our spectrum does not include other strong emission lines to directly determine the systemic redshift, we do not consider the blueshift correction in the virial BH mass estimates of the $z\sim4$ quasars in baseline. How the blueshift affects the virial BH mass estimates will be further discussed in section~\ref{sec:s5_CIV_blueshift}. 

We estimate the bolometric luminosity, $L_{\rm bol}$, from the monochromatic continuum luminosity at 1350{\AA} with $L_{\rm bol}=\kappa \lambda L_{\rm 1350}$, where $\kappa=3.81$ is the bolometric correction factor determined by the composite quasar SED \citep{Richards06}. Here, we note that the bolometric correction may depend on luminosity \citep[e.g.,][]{Netzer07b,lusso12,TN12,duras,shen2020}. But it becomes weak for the optical bolometric correction, especially in the low luminosity ranges \citep[e.g.,][]{lusso12,Runnoe12,shen2020}. Considering the relatively low luminosity range covered by our quasars, we neglect the luminosity dependence of the bolometric correction. The resulting bolometric luminosity of the 80 QOP=4 quasars spans a range of $45.41< \log L_{\rm bol}/ {\rm erg~s^{-1}}< 46.93$. 

The Eddington ratio is then calculated with $\lambda_{\rm Edd}=L_{\rm bol}/L_{\rm Edd}$, where $L_{\rm Edd}$ is the Eddington luminosity given by $L_{\rm Edd}=1.26\times10^{38}(M_{\rm BH}/M_{\odot})$ erg s$^{-1}$. For the 80 quasars, the estimated Eddington ratios distribute in a range of $-1.4<\log \lambda_{\rm Edd}<0.5$, with a median of $\log \lambda_{\rm Edd}=-0.4$. For each quasar, {the $1\sigma$} error in the Eddington ratio is calculated from the rms scatter of the 50 generated mock spectra described in section~\ref{sec:s3_fit}. Table~\ref{tab:mass} summarizes the results.

\subsection{Comparison with $z\sim4$ quasars in literature}
\label{sec:s3_comp}

We compare the distribution of the estimated virial BH mass of the 80 QOP=4 quasars at $3<z<4.5$ with literature. \citet{shen2011} compile similar measurements on the ${\rm C_{\Rmnum{4}}}$ 1549{\AA} emission line width and its continuum luminosity for quasars at $z>2$ in the SDSS DR7 quasar catalog. The virial BH masses are estimated using the same calibration with the ${\rm C_{\Rmnum{4}}}$ FWHM (i.e., equation~\ref{eq:mbh_civ_fwhm}). We use those estimates in the comparison. 

\begin{figure*} 
\centering
\includegraphics[keepaspectratio,width=\textwidth]{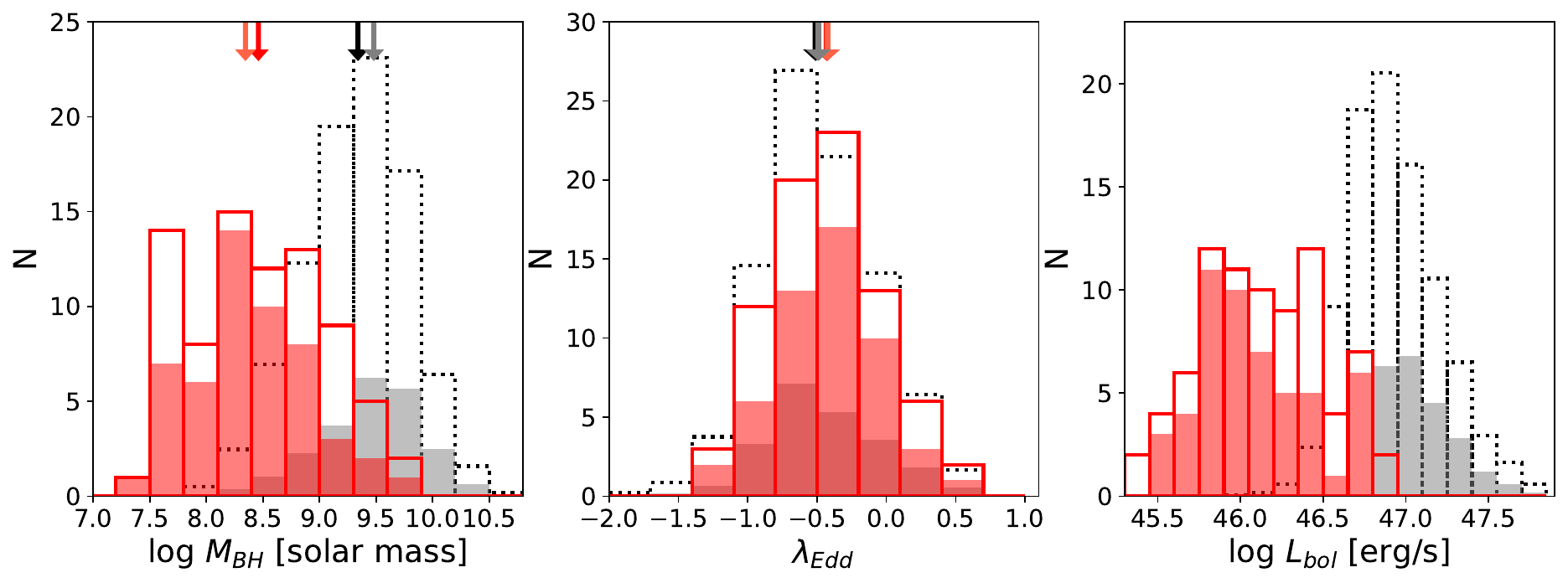}
\caption{Distributions of virial BH mass (left), Eddington ratio (middle) and bolometric luminosity (right) of the quasar samples at $z\sim4$. Red and black open histograms show the distributions of the 80 QOP=4 quasars and SDSS DR7 quasars at $3<z<4.5$, respectively, while red and grey filled histograms represent those of the 52 QOP=4 quasars and 1,462 SDSS DR7 quasars at $3.5<z<4.25$ used in the determination of the $z=4$ BHMF, respectively. Median value of each distribution is displayed with an arrow at the top. Histograms of the luminous quasars are scaled down by a factor of 65 for clarity.}
\label{fig:mbh-hist}
\end{figure*}

In Figure~\ref{fig:mbh-hist}, we compare the distributions of virial BH mass, Eddington ratio and bolometric luminosity of the QOP=4 quasars and SDSS DR7 quasars at $3<z<4.5$.
The QOP=4 quasars identified in this work covers the luminosity range down to $\log L_{\rm bol}/{\rm erg~s}^{-1}=45.4$, which is more than one order of magnitude less-luminous than the SDSS quasars in the same redshift range. Meanwhile, the QOP=4 quasars distribute in the BH mass range down to $10^{7.4} M_{\odot}$ with a median of $\log M_{\rm BH}/M_{\odot}=8.5$, which is around one order of magnitude less-massive than the SDSS quasars, whose median BH mass is estimated to be $\log M_{\rm BH}/M_{\odot}=9.3$. On the other hand, both of the quasar samples have the Eddington ratio similarly distributing around $\log \lambda_{\rm Edd}\sim-0.5$, with the QOP=4 quasars slightly extend towards the high Eddington ratios. The lower luminosities of the QOP=4 quasars are thus mainly driven by their smaller BH masses. In addition, we see a wider dispersion in virial BH mass for the QOP=4 quasars, which is consistent with previous studies focusing on deep AGN samples at $z<2$ \citep[e.g.,][]{Merloni10,AS2015}. 

\begin{figure*}
\centering
\includegraphics[keepaspectratio,width=.9\textwidth]{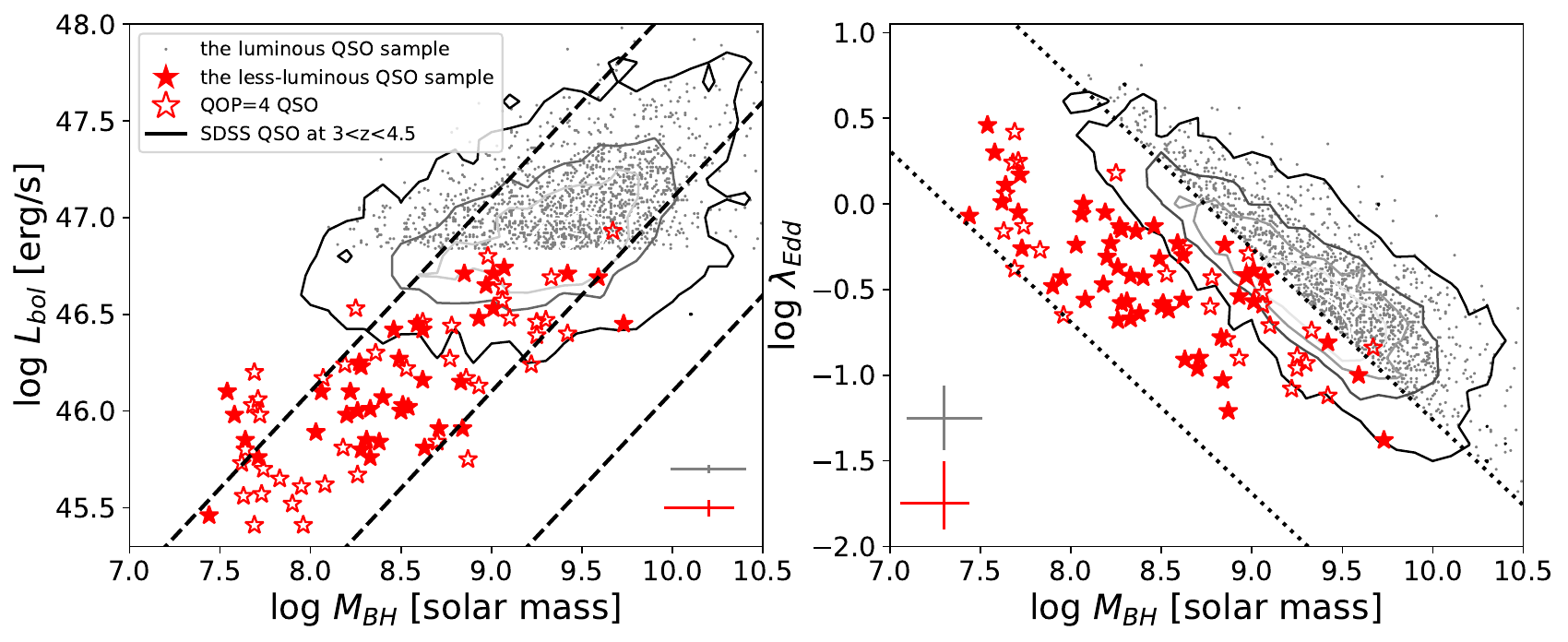}
\caption{$L_{\rm bol}$ vs. $M_{\rm BH}$ (left) and $\lambda_{\rm Edd}$ vs. $M_{\rm BH}$ (right) distributions of the 80 QOP=4 quasars at $3<z<4.5$. Symbols have the same meaning with Figure~\ref{fig:magz}. Black contours show the same distributions of the SDSS DR7 quasars in the same redshift range. Black dashed lines in the left panel indicate a constant Eddington ratio of 1, 0.1 and 0.01 from top to bottom. Black dotted lines in the right panel denote the luminosity limits of the luminous (upper) and less-luminous (lower) quasar samples used in the determination of the $z=4$ BHMF. {The average $1\sigma$ uncertainties of the estimates are displayed with the gray and red errorbars for the luminous and less-luminous quasar samples, respectively.}}
\label{fig:mbh-bol}
\end{figure*}

In Figure~\ref{fig:mbh-bol}, we show the observed bivariate distributions of virial BH mass, Eddington ratio and luminosity of the QOP=4 quasars and SDSS DR7 quasars at $3<z<4.5$. 
As plotted in the left panel, there is a positive correlation between the luminosity and BH mass. The ratio between the two quantities corresponds to the Eddington ratio, and the distribution suggests that these quasars are active SMBHs accreting at $\lambda_{\rm Edd}>0.01$. Meanwhile, we find there is a sharp decrease towards smaller Eddington ratios of $\lambda_{\rm Edd}<0.1$. It can be caused by the flux limit, especially in the less-massive range. As shown by the lower dotted line in the right panel, even the flux limit of less-luminous quasars is still too shallow to detect those population in the less-massive range of $<10^{9} M_{\odot}$. 

\begin{figure} [!ht]
\centering
\includegraphics[keepaspectratio,width=0.45\textwidth]{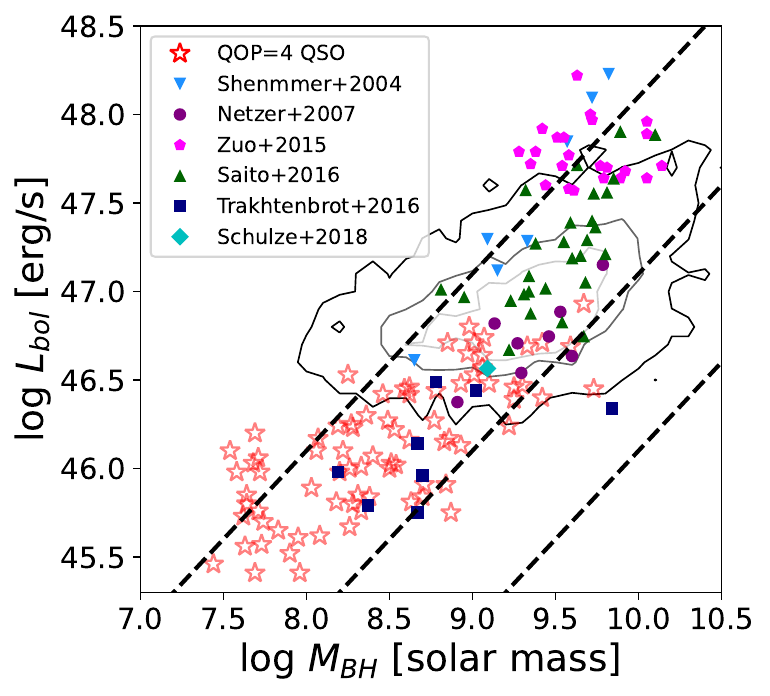}
\caption{$L_{\rm bol}$ vs. $M_{\rm BH}$ distribution of quasars at $3<z<4.5$. Red open stars show the 80 QOP=4 quasars in this work, and black contours display the same distribution of the SDSS DR7 quasars in the same redshift range. Blue inverted triangles, purple dots, magenta polygons, green triangles, navy squares and cyan diamonds represent the $z\sim3-4$ quasars with the virial BH mass estimated using the H$\beta$ or ${\rm Mg_{\Rmnum{2}}}$ emission line width and its corresponding continuum luminosity in \citet{Shemmer}, \citet{Netzer07a}, \citet{Zuo}, \citet{saito}, \citet{Trakhtenbrot16} and \citet{S2018}, respectively. Dashed lines have the same meaning with the left panel of Figure~\ref{fig:mbh-bol}.}
\label{fig:mbh-bol_mg2}
\end{figure}

While there are studies suggesting the ${\rm C_{\Rmnum{4}}}$ emission line profile can be used as a single-epoch virial BH mass estimator \citep[e.g.,][]{VP06}, some others indicate a large scatter between the line width of the ${\rm C_{\Rmnum{4}}}$ and H$\beta$ emission lines \citep[e.g.,][]{Netzer07a,SL2012}, and claim the H$\beta$ or ${\rm Mg_{\Rmnum{2}}}$ emission line width and its corresponding continuum luminosity can be more reliable estimators. Furthermore, the ${\rm C_{\Rmnum{4}}}$ emission line can be affected by the non-virial motion as well as the narrow or broad absorption lines \citep[for review see][]{shen2013}. Therefore, we further compare the virial BH mass of the 80 QOP=4 quasars to that of quasars estimated using the H$\beta$ or ${\rm Mg_{\Rmnum{2}}}$ emission line at $z\sim3.5$, which are collected from \citet{Shemmer}, \citet{Netzer07a}, \citet{Zuo}, \citet{saito}, \citet{Trakhtenbrot16} and \citet{S2018}. Results are plotted in Figure~\ref{fig:mbh-bol_mg2}.

The quasars identified in this work fall in a similar luminosity range as those in \citet{Netzer07a} and \citet{Trakhtenbrot16}. Both of the latter two studies adopt the H$\beta$ emission line as the virial BH mass estimator. We only pick up quasars at $3<z<4.25$ from their samples for the comparison. Overall, the QOP=4 quasars distribute in the similar ranges of virial BH mass and Eddington radio to their samples, while our quasars extend towards the low-mass and high-Eddington-ratio regime. The higher Eddington ratio of the QOP=4 quasars could be partly due to the systematic uncertainty of BH mass estimates with the ${\rm C_{\Rmnum{4}}}$ emission line. For example, the ${\rm C_{\Rmnum{4}}}$ blueshift can result in an under-estimated BH mass and thus over-estimated Eddington ratio for the less-luminous quasars (for detailed discussions see section~\ref{sec:s5_CIV_blueshift}). Regardless of the systematic effects, for the $z\sim3.5$ quasars whose BH masses are estimated with the reliable H$\beta$ or ${\rm Mg_{\Rmnum{2}}}$ emission line, they also show a sudden decrease towards $\log \lambda_{\rm Edd}<-1$, similar to the QOP=4 quasars. 

\section{Broad-line AGN BHMF and ERDF}
\label{sec:section4}

\subsection{The luminous and less-luminous quasar samples at $z\sim4$}
\label{sec:s4twoQSOsamp}

In order to determine the $z=4$ BHMF and ERDF, we limit our sample to the 52 that meet the c1 criteria (the primary selection window of the $z\sim4$ quasars in section~\ref{sec:s2sample}), since the selection function of the c1 criteria is well evaluated as a function of magnitude and redshift \citep[][]{akiyama2018}. We refer the sample as the less-luminous quasar sample hereafter. The less-luminous quasar sample covers the redshift range of $3.5< z_{\rm spec}<4.25$, with a median and standard deviation at 3.69 and 0.18, respectively. The quasar sample spans in the BH mass range of $7.44< \log M_{\rm BH}/M_{\odot}<9.73$, with a median of $\log M_{\rm BH}/M_{\odot}=8.34$; and their Eddington ratios cover the range of $-1.38<\log \lambda_{\rm Edd}<0.46$, with a median of $\log \lambda_{\rm Edd}=-0.43$. Distributions of the BH mass, Eddington ratio, and luminosity of the less-luminous quasar sample are plotted by red filled histograms in Figure~\ref{fig:mbh-hist} and red filled stars in Figure~\ref{fig:mbh-bol}. 

In addition, we construct a luminous quasar sample in the same redshift range from the SDSS DR7 quasar catalog. In the SDSS legacy survey, \citet{Richards02} select quasar candidates by stellar morphology and multi-color criteria, and apply a uniform target selection for spectroscopy. Its selection function is evaluated as a function of redshift and magnitude by \citet{Richards06b}. At $z\sim4$, the selection efficiency is close to 100\% down to $i_{\rm SDSS}=20.2$. The SDSS DR7 catalog contains the full quasar sample \citep{Schneider10}. As described in section~\ref{sec:s3_comp}, the virial BH masses of quasars in the catalog are estimated using the same ${\rm C_{\Rmnum{4}}}$ calibration in this work (i.e., equation~\ref{eq:mbh_civ_fwhm}) at $z>2$ \citep{shen2011}.

For the luminous quasar sample, firstly, we only consider quasars with ``uniform flag'' of 1, which means the quasars are uniformly selected by the final quasar selection algorithm in \citet{Richards02}. Then, we limit the redshift range of quasars to be $3.5<z<4.25$ to match that of the less-luminous quasar sample. There are 2,144 quasars meeting the criteria. Finally, to prevent overlaps in the selection function on the UV absolute magnitude $M_{\rm 1450}$ vs. redshift plane with that of the less-luminous quasar sample, we only select quasars above the magnitude limit of $M_{\rm 1450}=-25.75$, which is determined by the upper limit of the selection function of the less-luminous quasar sample (for details see section~\ref{sec:s4_bin}). Here, we evaluate $M_{\rm 1450}$ of SDSS quasars by their virial BH masses and Eddington ratios provided by \citet{shen2011}. We firstly derive the bolometric luminosity through $L_{\rm bol}=\lambda_{\rm Edd}L_{\rm Edd}$. Then, we obtain $M_{\rm 1450}$ of the SDSS quasars by adopting the bolometric correction in \citet{Runnoe12} as follows:
\begin{equation}\label{eq:bolCor_r12}
\log L_{\rm bol}=4.74+0.91\log \lambda L_{\rm 1450}.
\end{equation}
In total, we select 1,462 quasars from the SDSS DR7 quasar catalog as the luminous quasar sample. 

The luminous quasar sample covers the BH mass range of $8.1<\log M_{\rm BH}/M_{\odot}<10.67$, with a median of $\log M_{\rm BH}/M_{\odot}=9.49$; and their Eddington ratios distribute in the range of $-1.71<\log \lambda_{\rm Edd}<0.81$, with a median of $\log \lambda_{\rm Edd}=-0.49$. Distributions of the BH mass, Eddington ratio and luminosity of the luminous quasar sample can be seen by grey filled histograms in Figure~\ref{fig:mbh-hist} and grey dots in Figure~\ref{fig:mbh-bol}. We confirm the distributions of BH mass and Eddington ratio of the luminous sample are consistent with those of the entire SDSS DR7 quasars at $3.5<z<4.25$.

\subsection{Broad-line AGN BHMF and ERDF with the $V_{\rm max}$ method}
\label{sec:s4_bin}

We firstly determine the broad-line AGN BHMF and ERDF at $3.5<z<4.25$ using the $V_{\rm max}$ method \citep[the \textit{binned} BHMFs and ERDFs hereafter;][]{Avni1980}, which is applied to determine the AGN luminosity function in previous studies. The mass range of $7.5\leq\log M_{\rm BH} / M_{\odot}<11$ is divided into 12 bins with a bin width of $\Delta {\rm log} M_{\rm BH}=0.3$ dex. The number density in the bin between $(\log M_{\rm BH}-\Delta \log M_{\rm BH}/2)\sim(\log M_{\rm BH}+\Delta \log M_{\rm BH}/2)$ is calculated with
\begin{equation} \label{eq:binBHMF}
\Phi(M_{\rm BH})\Delta {\rm log} M_{\rm BH}=\sum_{i=1}^{n}\frac{1}{V_{i}},
\end{equation}
where $i$ is the index of each quasar, and $n$ is the total number of quasars in the mass bin. Here, $V_{i}$ is the comoving effective survey volume of the $i$-th quasar in the mass bin, and $1/V_{i}$ represents the contribution of the $i$-th quasar to the number density of the mass bin. The effective survey volume between $z_{\rm min}$ and $z_{\rm max}$ is calculated with 
\begin{equation} \label{eq:V}
V_{i}=\int_{z_{\rm min}}^{z_{\rm max}} \Omega(M_{\mathrm{1450},i},z)~(1+z)^{3} ~d^{2}_{A}(z) ~c ~\frac{\mathrm{d} \tau}{\mathrm{d} z}~\mathrm{d}z,
\end{equation} 
where $d_{A}$ is the angular diameter distance, $c$ is the speed of light, and ${\mathrm{d} \tau}/{\mathrm{d} z}=1/\left[H_{0}(1+z)E(z)\right]$ is the look-back time at $z$. Here, $E^{2}(z)=\Omega_{m}\times (1+z)^3+\Omega_{\Lambda}$. $\Omega(M_{\mathrm{1450},i},z)$ represents the effective survey area of the quasar with $M_{\mathrm{1450},i}$ and $z$. {The $1\sigma$ uncertainty} of the number density is calculated following the Poisson statistics by
\begin{equation} \label{eq:Poisson_BHMF}
\sigma=\left[\sum_{i}^{n}\left(\frac{1}{V_{i} {\Delta {\rm log} M_{\rm BH}}}\right)^2\right]^{1/2}.
\end{equation} 

\begin{figure*} [!ht]
\centering
\includegraphics[keepaspectratio,width=0.9\textwidth]{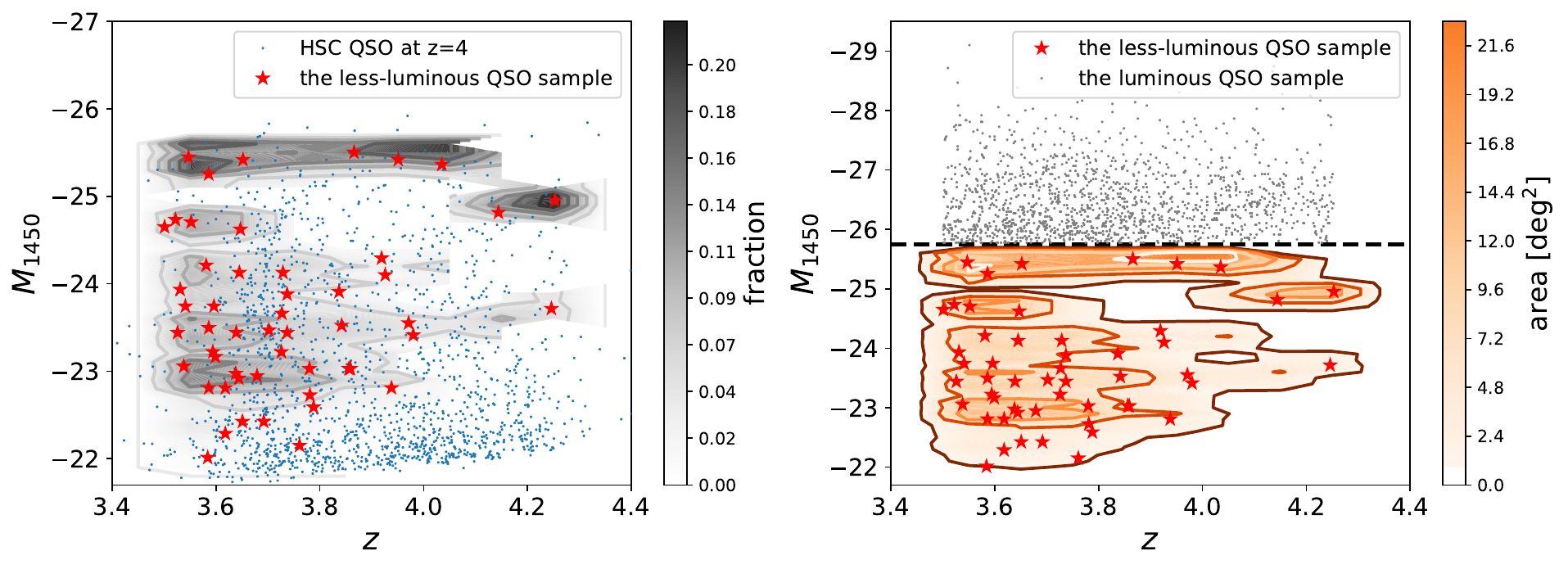}
\caption{{Left) fraction of the less-luminous quasars among the entire c1 quasar sample.  Red stars and blue dots represent the less-luminous and the entire c1 quasar samples, respectively. Grey contours show the fraction of number counts between the two samples with the color code displayed on the right. Right)} effective survey area $\Omega(M_{\mathrm{1450}},z)$ of the less-luminous quasar sample at $z\sim4$. Red stars and grey dots represent the less-luminous and luminous quasar samples used in the determination of the $z=4$ BHMF in this work, respectively. Orange contours show the effective survey area of the less-luminous quasar sample with the color code displayed on the right. Black dashed line indicates the absolute magnitude cut of the luminous quasar sample. }
\label{fig:sel_fun}
\end{figure*}

In order to evaluate the effective survey area $\Omega(M_{\mathrm{1450},i},z)$ of the less-luminous quasar sample, we adopt the effective survey area of the c1 quasars determined as a function of absolute magnitude and redshift in \citet{akiyama2018} (see Fig. 16 therein). The less-luminous quasars are not uniformly selected for spectroscopy on the absolute magnitude vs. redshift plane. {We thus estimate their survey area by multiplying the fraction of the spectroscopically-identified quasars among the entire c1 quasars in each absolute magnitude and redshift bin. We divide the absolute magnitude range of $-27 \leq M_{\rm 1450} \leq -20$ into 70 bins with a bin width of $\Delta M_{\rm 1450}=0.1$, and the redshift range of $3.05 \leq z \leq 4.85$ into 18 bins with a bin width of $\Delta z=0.1$. In each bin ($M_{\mathrm{1450},i}, ~z_{j}$), we evaluate the fraction considering a range of $(M_{\mathrm{1450},i}-0.19) \sim (M_{\mathrm{1450},i}+0.19)$ and $(z_{j}-0.19) \sim (z_{j}+0.19)$. The range is chosen to ensure a sufficient number of c1 quasars in each bin while avoiding excessive smoothing. We use spectroscopic redshifts for spectroscopically-identified quasars, and photometric redshifts for the remaining targets. We do not correct for the fraction of the identified contaminants in each bin since it is negligibly small as described in section~\ref{sec:s2iden}. The resulting effective survey area of the less-luminous quasar sample is shown with orange contours in the right panel of Figure~\ref{fig:sel_fun}.}
For the luminous quasar sample, we directly adopt the effective survey area of 6,248 deg$^2$ estimated by \citet{SK2012}. 

The comoving effective survey volume for each quasar is obtained by integrating equation~\ref{eq:V} over the redshift range between $z_{\rm min}=3.5$ and $z_{\rm max}=4.25$. While it is suggested the observed number density of AGNs rapidly evolves at $z<3$ \citep[e.g.,][]{Hasinger}, the evolution becomes milder at $z>3$ \citep[e.g.,][]{Ueda14}. Since our quasar samples cover a narrow redshift range ($3.5<z<4.25$), we do not consider the redshift dependence in the integration for simplicity. We confirm the change would be negligibly small even if we introduce the redshift dependence. 

\begin{figure*} [!ht]
\centering
\includegraphics[keepaspectratio,width=\textwidth]{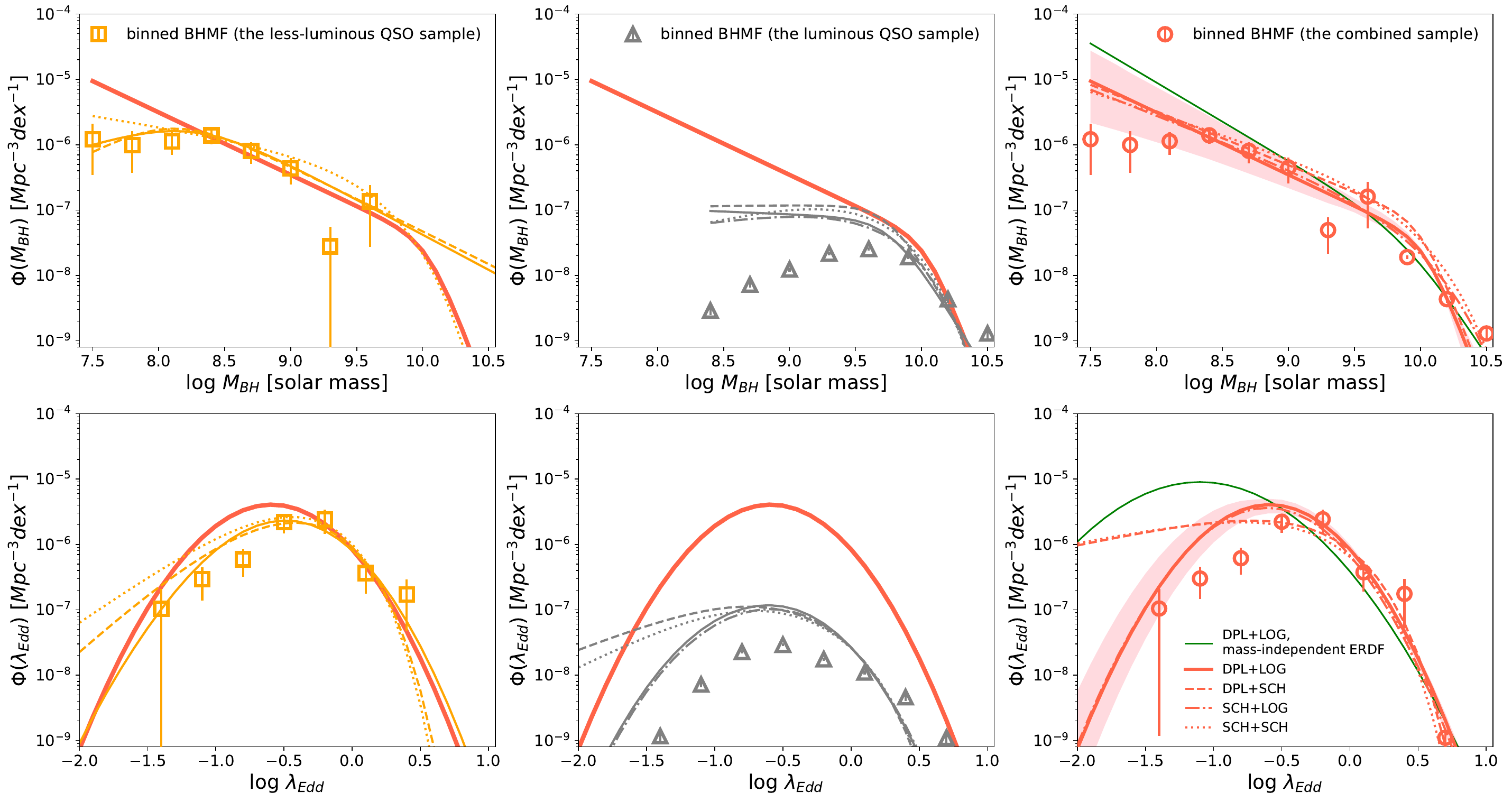}
\caption{$z=4$ broad-line AGN BHMFs (upper panels) and ERDFs (bottom panels) determined by the less-luminous (left panels), luminous (middle panels) and combined (right panels) quasar samples. Data points indicate the \textit{binned} results obtained with the $V_{\rm max}$ method, and the lines show the \textit{intrinsic} results derived with the maximum likelihood method. There are four different functional forms adopted in the fitting: solid lines are for the double power-law BHMF and the log-normal ERDF, dashed lines are for the double power-law BHMF and the Schechter ERDF, dashed-dotted lines are for the Schechter BHMF and the log-normal ERDF, and dotted lines are for the Schechter BHMF and the Schechter ERDF. The best-fit results are shown by orange, grey and red lines for the less-luminous, luminous and combined quasar samples, respectively. In the right panels, pink shaded areas indicate the {1$\sigma$} uncertainty of the best-fit model shown by the red solid lines. Green solid lines represent the best-fit results by adopting the log-normal ERDF with the double power-law BHMF, but excluding the mass dependence of ERDF. Red solid lines in the right panels are also plotted in the left and middle panels for comparison. }
\label{fig:BHMF}
\end{figure*}

The resulting \textit{binned} broad-line AGN BHMFs determined by the less-luminous, luminous and combined quasar samples are plotted by orange open squares, grey open triangles and red circles in the upper left, upper middle and upper right panels of Figure~\ref{fig:BHMF}, respectively. 
Tabel~\ref{tab:bin_BHMF} summarizes the results. 
\begin{table*}[!ht]
\caption{{\textit{Binned} broad-line AGN BHMFs at $z=4$}}
\label{tab:bin_BHMF}
\begin{adjustbox}{width=0.8\textwidth,center=16cm}
\centering
\begin{tabular}{ccccccc}
\hline\hline
$\log M_{\rm BH}/M_{\odot}$   & N$_{\rm HSC}$  & $\Phi_{\rm HSC}(M_{\rm BH})$  & N$_{\rm SDSS}$ &$\Phi_{\rm SDSS}(M_{\rm BH})$ &  N$_{\rm HSC+SDSS}$ & $\Phi_{\rm HSC+SDSS}(M_{\rm BH})$\\
 &   & $10^{-7}$ Mpc$^{-3}$ dex$^{-1}$ &  &$10^{-7}$ Mpc$^{-3}$ dex$^{-1}$   & &$10^{-7}$ Mpc$^{-3}$ dex$^{-1}$\\
\hline
7.50 & 5 & 12.14 $\pm$ 8.70 & 0 & 0.00 $\pm$ 0.00 & 5 & 12.14 $\pm$ 8.70 \\
7.80 & 4 & 9.92 $\pm$ 6.24 & 0 & 0.00 $\pm$ 0.00 & 4 & 9.92 $\pm$ 6.24 \\
8.10 & 9 & 11.30 $\pm$ 4.34 & 6 & 0.00 $\pm$ 0.00 & 15 & 11.30 $\pm$ 4.34 \\
8.40 & 16 & 13.90 $\pm$ 3.68 & 45 & 0.03 $\pm$ 0.00 & 61 & 13.93 $\pm$ 3.68 \\
8.70 & 8 & 7.99 $\pm$ 2.87 & 112 & 0.07 $\pm$ 0.01 & 120 & 8.07 $\pm$ 2.87 \\
9.00 & 7 & 4.34 $\pm$ 1.89 & 191 & 0.12 $\pm$ 0.01 & 198 & 4.46 $\pm$ 1.89 \\
9.30 & 1 & 0.28 $\pm$ 0.28 & 333 & 0.22 $\pm$ 0.01 & 334 & 0.49 $\pm$ 0.28 \\
9.60 & 2 & 1.36 $\pm$ 1.08 & 392 & 0.25 $\pm$ 0.01 & 394 & 1.61 $\pm$ 1.08 \\
9.90 & 0 & 0.00 $\pm$ 0.00 & 295 & 0.19 $\pm$ 0.01 & 295 & 0.19 $\pm$ 0.01 \\
10.20 & 0 & 0.00 $\pm$ 0.00 & 67 & 0.04 $\pm$ 0.01 & 67 & 0.04 $\pm$ 0.01 \\
10.50 & 0 & 0.00 $\pm$ 0.00 & 20 & 0.01 $\pm$ 0.00 & 20 & 0.01 $\pm$ 0.00 \\
10.80 & 0 & 0.00 $\pm$ 0.00 & 1 & 0.00 $\pm$ 0.00 & 1 & 0.00 $\pm$ 0.00 \\
\hline\\
\end{tabular}
\end{adjustbox}
\end{table*}
The BHMF of the luminous quasar sample peaks at $\log M_{\rm BH}/M_{\odot} \sim 9.6$, but the turnover disappears in the BHMF of the combined quasar sample. Thus, the peak is likely due to incompleteness of the luminous quasar sample in the less-massive bins. The BHMF of the combined sample covers a wide mass range of $7.5\leq\log M_{\rm BH}/M_{\odot}\leq10.5$ {with the massive end dominated by the luminous sample and the less-massive end dominated by the less-luminous sample. There is} a sharp decline in the high mass end above $\log M_{\rm BH}/M_{\odot}>9.6$ as well as a mild decline in the low mass end below $\log M_{\rm BH}/M_{\odot}<8.5$. The decline in the low mass end suggests the BHMF in the less-massive bins may be still affected by incompleteness even after combining the less-luminous quasar sample.

The \textit{binned} broad-line AGN ERDFs are evaluated in the same way as for the \textit{binned} broad-line AGN BHMFs. We divide the Eddington ratio range of $-2\leq{\rm log}\lambda_{\rm Edd}<1$ into 11 bins with an interval of $\Delta \log \lambda_{\rm Edd}=0.3$ dex. The number density in each bin is determined over the range of $(\log \lambda_{\rm Edd}-\Delta \log \lambda_{\rm Edd}/2)\sim (\log \lambda_{\rm Edd}+\Delta \log \lambda_{\rm Edd}/2)$. {The $1\sigma$ uncertainty} of the number density is calculated following the Poisson statistics.

The resulting \textit{binned} broad-line AGN ERDFs determined by the less-luminous, luminous and combined quasar samples are plotted by orange open squares, grey open triangles and red circles in the lower left, lower middle and lower right panels of Figure~\ref{fig:BHMF}, respectively. The results are summarized in Tabel~\ref{tab:bin_ERDF}.  
\begin{table*}[!ht]
\caption{{\textit{Binned} broad-line AGN ERDFs at $z=4$}}
\label{tab:bin_ERDF}
\begin{adjustbox}{width=0.8\textwidth,center=16cm}
\centering
\begin{tabular}{ccccccc}
\hline\hline
$\log \lambda_{\rm Edd}$   & N$_{\rm HSC}$  & $\Phi_{\rm HSC}(\lambda_{\rm Edd})$  & N$_{\rm SDSS}$ &$\Phi_{\rm SDSS}(\lambda_{\rm Edd})$ &  N$_{\rm HSC+SDSS}$ & $\Phi_{\rm HSC+SDSS}(\lambda_{\rm Edd})$\\
 &   & $10^{-7}$ Mpc$^{-3}$ dex$^{-1}$ &  &$10^{-7}$ Mpc$^{-3}$ dex$^{-1}$   & &$10^{-7}$ Mpc$^{-3}$ dex$^{-1}$\\
\hline
$-2.00$ & 0 & 0.00 $\pm$ 0.00 & 0 & 0.00 $\pm$ 0.00 & 0 & 0.00 $\pm$ 0.00 \\
$-1.70$ & 0 & 0.00 $\pm$ 0.00 & 5 & 0.00 $\pm$ 0.00 & 5 & 0.00 $\pm$ 0.00 \\
$-1.40$ & 1 & 1.03 $\pm$ 1.03 & 18 & 0.01 $\pm$ 0.00 & 19 & 1.04 $\pm$ 1.03 \\
$-1.10$ & 4 & 2.93 $\pm$ 1.56 & 111 & 0.07 $\pm$ 0.01 & 115 & 3.00 $\pm$ 1.56 \\
$-0.80$ & 6 & 5.91 $\pm$ 2.71 & 350 & 0.23 $\pm$ 0.01 & 356 & 6.14 $\pm$ 2.71 \\
$-0.50$ & 19 & 21.94 $\pm$ 7.13 & 450 & 0.29 $\pm$ 0.01 & 469 & 22.23 $\pm$ 7.13 \\
$-0.20$ & 16 & 24.07 $\pm$ 9.67 & 270 & 0.17 $\pm$ 0.01 & 286 & 24.24 $\pm$ 9.67 \\
0.10 & 4 & 3.64 $\pm$ 1.86 & 170 & 0.11 $\pm$ 0.01 & 174 & 3.75 $\pm$ 1.86 \\
0.40 & 2 & 1.71 $\pm$ 1.22 & 71 & 0.05 $\pm$ 0.01 & 73 & 1.75 $\pm$ 1.22 \\
0.70 & 0 & 0.00 $\pm$ 0.00 & 17 & 0.01 $\pm$ 0.00 & 17 & 0.01 $\pm$ 0.00 \\
1.00 & 0 & 0.00 $\pm$ 0.00 & 0 & 0.00 $\pm$ 0.00 & 0 & 0.00 $\pm$ 0.00 \\

\hline\\
\end{tabular}
\end{adjustbox}
\end{table*}
The ERDFs of the luminous and less-luminous quasar sample {cover the similar range of $-1.5\leq\log \lambda_{\rm Edd}\leq0.5$. Both of them have similar shape with a turnover around $\log \lambda_{\rm Edd}=-0.5$. Since the number density of less-luminous quasars is much higher than that of luminous quasars, the ERDF of the combined sample is dominated by the less-luminous quasars.}

\subsection{Broad-line AGN BHMF and ERDF with the Maximum Likelihood Method}
\label{sec:s4_correct_method}

Though we consider the effective survey area in determining the \textit{binned} broad-line AGN BHMFs and ERDFs, the results can still be affected by the selection incompleteness caused by the flux limit of the quasar samples. As can be seen in Figure~\ref{fig:mbh-bol}, with a certain flux limit, only quasars with sufficiently high Eddington ratios can be observable in the less-massive bins, and only quasars with large enough BH mass can be detected in the low-Eddington-ratio bins. Quasars with low BH masses and low Eddington-ratios can be missed as they are below the flux limit. The \textit{binned} BHMFs and ERDFs constrained by using the flux-limited quasar samples can thus be incomplete in the less-massive and lower-Eddington-ratio ranges.

The BHMFs and ERDFs corrected for the flux limit (the \textit{intrinsic} BHMFs and ERDFs hereafter) can be derived statistically by assuming multiple functional shapes for the intrinsic BHMF and ERDF \citep[e.g.,][]{AS2010,nobuta,SK2012,ks2013,AS2015}. In this work, we follow the maximum likelihood method introduced in \citet{AS2015}. The method enables to combine multiple samples with different selection functions, and it considers the correction for uncertainties in the virial BH mass estimates {and the bolometric correction}. The mass dependence of ERDF and/or the redshift evolution in the BHMF and ERDF can also be flexibly taken into account. We determine the \textit{intrinsic} BHMFs and ERDFs with the luminous, less-luminous and combined quasar samples separately.

In \citet{AS2015}, the \textit{intrinsic} broad-line AGN BHMF and ERDF are determined jointly through the \textit{intrinsic} bivariate distribution function $\Psi(M_{\rm BH}, \lambda_{\rm Edd}, z)$, which can be treated as the multiplication of BHMF and ERDF component, i.e., $\Psi(M_{\rm BH}, \lambda_{\rm Edd}, z)=\phi^*\times\psi_{\rm Edd} \times \psi_{\rm BH}$, where $\phi^*$ is the normalization. Here, a mass dependence in the ERDF component is allowed, i.e., $\psi_{\rm Edd}=\psi_{\rm Edd}(M_{\rm BH},\lambda_{\rm Edd},z)$. The \textit{intrinsic} ERDF $\phi_{\rm Edd}$ is then given by integrating the bivariate distribution function over $\log M_{\rm BH}$ between $\log M_{\rm BH,min}/M_\odot=7.5$ and $\log M_{\rm BH,max}/M_\odot=11$, i.e.,
\begin{equation} \label{eq:intERDF}
{\phi_{\lambda_{\rm Edd}}=\int \Psi(M_{\rm BH}, \lambda_{\rm Edd}, z) ~\mathrm{d} \log M_{\rm BH}.}
\end{equation}
The \textit{intrinsic} BHMF $\phi_{\rm BH}$ is also given by the same integration over $\log \lambda_{\rm Edd}$ between $\log \lambda_{\rm Edd,min}=-2$ and $\log \lambda_{\rm Edd,max}=1$.

We assume the BHMF can be described either with a double power-law function
\begin{equation} \label{eq:DPL_BHMF}
\psi_{\rm BH}=\frac{1}{\log e}\frac{1}{(M_{\rm BH}/M^*)^{-\alpha-1} + (M_{\rm BH}/M^*)^{-\beta-1}},
\end{equation}
or a Schechter function  
\begin{equation} \label{eq:Sch_BHMF}
\psi_{\rm BH}=\frac{1}{\log e}\left(\frac{M_{\rm BH}}{M^*}\right)^{\alpha+1} {\rm exp}\left(-\frac{M_{\rm BH}}{M^*}\right),
\end{equation}
since the double power-law function can reproduce the observed AGN luminosity functions up to $z\sim6$, and the Schechter function can represent the observed stellar mass functions of galaxies. 

We assume the ERDF follows a log-normal function
\begin{align}
\label{eq:LOG_ERDF}
\psi_{\rm Edd}=\frac{1}{\log e\sqrt{2\pi}\sigma_{\lambda}}{\rm exp}\left(-\frac{(\log \lambda_{\rm Edd}-\log \lambda_{\rm Edd}^\prime)^2}{2\sigma_{\lambda}^2}\right),
\end{align}
as the observed ERDFs at $z>3$ in literature show clear turnover, similar to the log-normal distribution \citep[e.g.,][]{Kollmeier06,willott,SK2012,shen2019,Farina22}. Meanwhile, there are also studies indicating the turnover can be caused by shallow detection limits of quasar surveys, and the Schechter function may be more reliable in describing the \textit{intrinsic} ERDF \citep[e.g.,][]{SK2012,AS2015,Li23}. Thus, we also adopt a Schechter function for the \textit{intrinsic} ERDF
\begin{equation} \label{eq:Sch_ERDF}
\psi_{\rm Edd}=\frac{1}{\log e}\left(\frac{\lambda_{\rm Edd}}{\lambda_{\rm Edd}^\prime}\right)^{\alpha_{\lambda}+1} {\rm exp}\left(-\frac{\lambda_{\rm Edd}}{\lambda_{\rm Edd}^\prime}\right).
\end{equation}

The BH mass dependence of ERDF is taken into account by assuming the $\log \lambda_{\rm Edd}^\prime$ term varies with BH mass in a linear form of $\log \lambda_{\rm Edd}^\prime=\log \lambda^*_{\rm Edd}+ k_{\lambda}(\log M_{\rm BH}-\log M_{\textrm{BH},l})$, where $\log M_{\textrm{BH},l}=7$ is {fixed}. A second-order polynomial function is also considered in the mass dependence, but we {find} the best-fit second-order coefficient is consistent with zero. We thus only consider the linear term. The redshift dependence of BHMF and ERDF is not included in the fitting since the quasar samples cover a narrow redshift range of $3.5<z<4.25$.  

There are two effects affecting the intrinsic distribution $\Psi(M_{\rm BH}, \lambda_{\rm Edd}, z)$ to the observed one $\Psi_{o}(M_{\rm BH}, \lambda_{\rm Edd}, z)$: one is the uncertainties associated with the virial BH mass estimates {and the bolometric correction}, and another one is the selection function of quasar samples over the $M_{\rm BH}$-$\lambda_{\rm Edd}$ plane. The former effect can be considered by convolving the intrinsic bivariate distribution function $\Psi(M_{\rm BH}, \lambda_{\rm Edd}, z)$ with the uncertainties of virial BH mass estimates $\sigma_{\rm VM}$ {and of bolometric correction $\sigma_{\rm BC}$} by
\begin{align} \label{eq:psi_c}
&{\Psi_{c}(M_{\textrm{BH}, c}, \lambda_{\textrm{Edd}, c}, z)  \notag }\\
&= \int g(M_{\textrm{BH},c}, \lambda_{\textrm{Edd}, c} \mid M_{\rm BH}, \lambda_{\rm Edd} ) \notag \\
& \times \Psi(M_{\rm BH}, \lambda_{\rm Edd}, z) \mathrm{~d} \log M_{\rm BH}\mathrm{~d} \log \lambda_{\rm Edd},
\end{align}
where the uncertainties $\sigma_{\rm VM}$ {and $\sigma_{\rm BC}$ are} assumed to follow a log-normal function as
\begin{align} \label{eq:g}
&{g(M_{\textrm{BH},c}, \lambda_{\textrm{Edd}, c} \mid M_{\rm BH}, \lambda_{\rm Edd}) \notag }\\
&=\frac{1}{2\pi \sigma_{\rm VM} \sigma_{\rm BC}} \notag \\
& \times \exp \left[-\frac{(\log M_{\textrm{BH}, c}-\log M_{\textrm{BH}})^2}{2\sigma^2_{\rm VM}}-\frac{(\log L_{\textrm{bol}, c}-\log L_{\textrm{bol}})^2}{2\sigma^2_{\rm BC}}\right].
\end{align}
$M_{\textrm{BH},c}$ and $M_{\rm BH}$ are the virial and true BH mass, respectively. {$L_{\textrm{bol},c}$ and $L_{\rm bol}$ are the converted and true bolometric luminosity, respectively. For the uncertainty in the virial BH mass estimates, we adopt $\sigma_{\rm VM}=0.2$ dex  following \citet{AS2015}; for the uncertainty in the bolometric correction, we assume $\sigma_{\rm BC}=0.1$ dex \citep{Richards06,shen2008}.}

The latter effect can be evaluated by directly multiplying the convolved bivariate distribution function {$\Psi_{c}(M_{\textrm{BH}, c}, \lambda_{\textrm{Edd}, c}, z)$} with the effective survey area $\Omega(M_{\rm BH},\lambda_{\rm Edd},z)$, i.e.,
\begin{align} \label{eq:psi_o}
 \Psi_{o}(M_{\textrm{BH}, c}, \lambda_{\textrm{Edd},c}, z) &=\Omega(M_{\textrm{BH}, c},\lambda_{\textrm{,c}Edd},z) \notag \\
 &\times \Psi_{c}(M_{\textrm{BH}, c}, \lambda_{\textrm{ Edd},c}, z).
\end{align}
The effective survey area of the luminous and less-luminous quasar samples is described as a function of absolute magnitude and redshift in section~\ref{sec:s4_bin}. To obtain the effective survey area as a function of BH mass and Eddington ratio, we convert $\Omega(M_\textrm{1450},z)$ to $\Omega(M_\textrm{BH},\lambda_{\rm Edd},z)$. We follow the method in \citet{ks2013}. Firstly we calculate a bolometric luminosity $L_{\rm bol}$ for any pair of BH mass $M_{\rm BH}$ and Eddington ratio $\lambda_{\rm Edd}$, and then transform it to the monochronic luminosity at 1450{\AA} through the bolometric correction in \citet{Runnoe12} (i.e., equation~\ref{eq:bolCor_r12}). We confirm the converted monochronic luminosity at 1450{\AA} is consistent with that directly measured from the spectrum. The effective survey area at the corresponding absolute magnitude $M_\textrm{1450}$ and redshift $z$ is then adopted as the effective survey area at that BH mass, Eddington ratio and redshift. 

\begin{figure}[!ht] 
\centering
\includegraphics[keepaspectratio,width=0.45\textwidth]{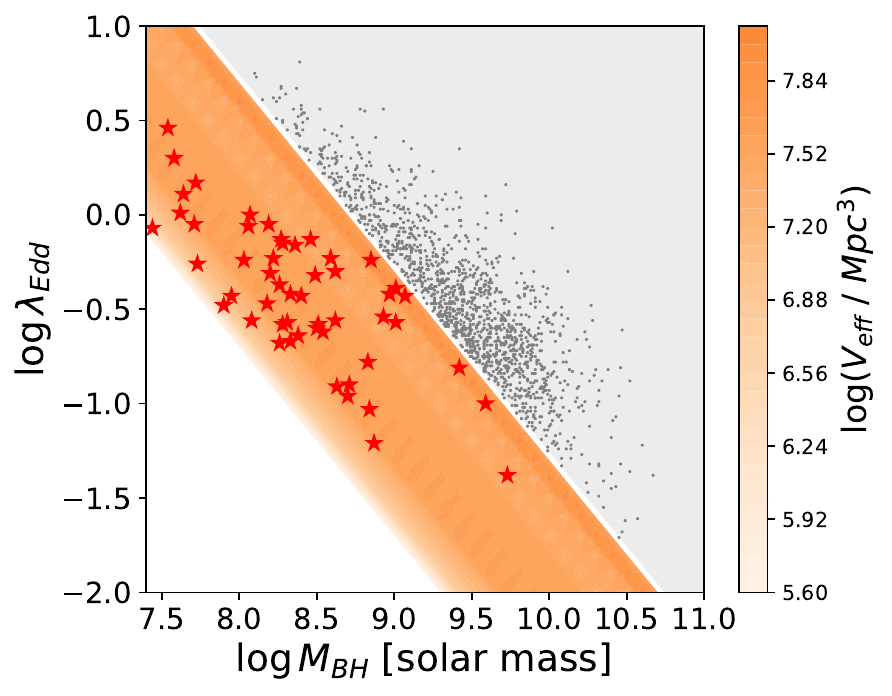}
\caption{Comoving effective survey volume, $V_{\rm eff}$, in the $\lambda_{\rm Edd}$ vs. $M_{\rm BH}$ plane for the luminous (grey shaded area) and less-luminous quasar samples (orange contour). Symbols have the same meanings as Figure~\ref{fig:sel_fun}. 
}
\label{fig:sel_fun1}
\end{figure}

For multiple quasar samples, if the respective survey area is independent with each other, their effective survey areas can be combined to one, i.e., $\Omega(M_\textrm{BH},\lambda_{\rm Edd},z)=\sum_{j} \Omega_{j}(M_\textrm{BH},\lambda_{\rm Edd},z)$, where $j$ is the index of the respective quasar sample \citep{AS2015}. As mentioned in section~\ref{sec:s4twoQSOsamp}, the luminous quasar sample is carefully selected not to overlap with the less-luminous quasar sample in absolute magnitudes. The respective survey area of the two quasar samples can thus be treated as independent area, and can be directly summed up for the combined quasar sample. In Figure~\ref{fig:sel_fun1}, the comoving effective survey volume $V_{\rm eff}$ of the two quasar samples is plotted against BH mass and Eddington ratio.   

With the observed bivariate distribution function of BH mass and Eddington ratio $\Psi_{o}(M_{\rm BH}, \lambda_{\rm Edd}, z)$, the probability of detecting the $i$-th observed quasar at its observed BH mass, Eddington ratio, and redshift can be written with 
\begin{equation} \label{eq:p}
p_{i}(M_{\rm BH},\lambda_{{\rm Edd}}, z)=\frac{1}{N}\Psi_{o}(M_{\textrm{BH},i}, \lambda_{\textrm{Edd},i}, z_{i})  \frac{\textrm{d}V}{\textrm{d}z},
\end{equation}
where 
\begin{equation} \label{eq:N}
N=\iiint \Psi_{o} (M_{\rm BH}, \lambda_{\rm Edd}, z) \frac{\textrm{d}V}{\textrm{d} z} \textrm{~d} \log M_{\rm BH} \textrm{~d} \log \lambda_{\rm Edd} \textrm{~d} z
\end{equation}
is the normalization. We then apply the maximum likelihood method to minimize the likelihood function $S=-2 \ln \mathcal{L}$, where $\mathcal{L}=\prod_{i}^{n} p_{i}$ is the likelihood of detecting the observed quasar sample \citep{Marshall}, and $n$ is the total number of quasars. The likelihood function can be written by 
\begin{equation} \label{eq:S}
S=- 2 \sum_{i}^{n} \ln \left[\Psi_{o}(M_{\textrm{BH},i}, \lambda_{\textrm{Edd},i}, z_{i})  \frac{\textrm{d}V}{\textrm{d}z}\right] +
 2\sum_{i}^{n}\ln N.
\end{equation}
The latter term can be simplified to $2n\ln N$ if $\Psi_{o}$ is constant for all of the quasars. 

We minimize the likelihood function $S$ by applying the downhill simplex algorithm \citep{NM}. The free parameters are $M^*$, $\alpha$ and $\beta$ for the \textit{intrinsic} BHMF; and $\lambda_{\rm Edd}^*$, $\alpha_{\lambda}$/$\sigma_{\lambda}$ and $k_{\lambda}$ for the \textit{intrinsic} ERDF. $\phi^*$ is the normalization of the best-fit models, and it is derived by scaling the number count of quasars predicted by the best-fit models to that of the observed quasars, i.e., $\phi^*=N_{\rm obs}/N_{\rm model}$, where $N_{\rm obs}=$52, 1462 and 1514 for the less-luminous, luminous and combined quasar samples, respectively. $N_{\rm model}$ can be obtained by integrating equation~\ref{eq:N} over BH mass, Eddington ratio and redshift. The integration ranges are $7.5<{\rm log}M_{\rm BH}/M_{\odot}<11$, $-2<{\rm log}\lambda_{\rm Edd}<1$, and $3.5<z<4.3$. For the luminous quasar sample, we further limit the integration range of BH mass to {$8.5<{\rm log}M_{\rm BH}/M_{\odot}<11$}.

Uncertainty of the best-fit parameters is determined following the descriptions in \citet{Pawitan01}. For the maximum likelihood estimates, the inverse of the observed Fisher information matrix can be the asymptotic covariance matrix, and the 1$\sigma$ uncertainties of the best-fit parameters are then the square roots of the diagonal elements of the covariance matrix. Here, the observed Fisher information matrix is the second-order derivative of the log-likelihood evaluated at the maximum likelihood estimates, i.e., 
\begin{equation} \label{eq:I}
I(\theta_{\rm ML})=-\frac{\partial^2}{\partial\theta_{i}\partial\theta_{j}}\ln \mathcal{L} (\theta_{\rm ML}),
\end{equation}
where $\theta_{\rm ML}$ are the best-fit parameters, and $1\leq i,j\leq n_{\rm para}$ are the index of parameters. Uncertainty of the $k$-th parameter ($1\leq k\leq n_{\rm para}$) can then be calculated by 
\begin{equation} \label{eq:uncertainty_MLE}
\sigma_{k}(\theta_{\rm ML})=\frac{1}{\sqrt{I_{kk}(\theta_{\rm ML})}}.
\end{equation}

The Fisher information matrix does not directly provide uncertainty of the entire best-fit models. Here, we determine that uncertainty following the $\chi^2$ assumption. If the uncertainty of maximum likelihood estimates follows a $\chi^2$ distribution, the {$1\sigma$} uncertainty of the best-fit models can be determined when the likelihood function $S$ increases from its best-fit value by 1. We then search for models having the change in $S$ by {one} from its best-fit value. For each parameter, we randomly change it from its best-fit value and fix it. We then re-minimize the likelihood function $S$ with the remaining parameters, and measure how much the minimum of $S$ changes from its best-fit value. The same procedure is repeated for 50 times for each parameter, and all the models having the minimum of $S$ within {one} from its best-fit value are adopted. The upper and lower boundaries of the adopted models are regarded as the {1$\sigma$ uncertainty}. We confirm the measured $1\sigma$ uncertainties are consistent with those evaluated by the Fisher information matrix.

\subsection{The \textit{intrinsic} broad-line AGN BHMFs and ERDFs}
\label{sec:s4_correct_result}

The \textit{intrinsic} broad-line AGN BHMFs and ERDFs derived with the maximum likelihood method are shown in Figure~\ref{fig:BHMF}. The functions determined with the less-luminous, luminous and combined quasar samples are plotted by orange, grey and red lines in the left, middle and right panels, respectively. Table~\ref{tab:intrinsicBHMF} summarizes the best-fit parameters with the $1\sigma$ uncertainty of the respective model.
\begin{table*}[!ht]
\caption{{\textit{Intrinsic} broad-line AGN BHMFs and ERDFs determined with the combined quasar sample at $z=4$}}
\label{tab:intrinsicBHMF}
\begin{adjustbox}{width=1.1\textwidth,center=16cm}
\centering
\begin{tabular}{ccccccccccccc}
\hline\hline
 BH estimates & BHMF & ERDF  & $\log \phi^*$ & $\log M_{\rm BH}^{*}$ & $\alpha$ & $\beta$ & $\lambda_{\rm Edd}^*$ & $\alpha_{\lambda}$ & $\sigma_{\lambda}$ & $k_{\lambda}$ & $\Delta$AIC & 2DKS  \\
&  & & Mpc$^{-3}$ dex$^{-1}$ & $M_\odot$ &  & &  &  &  &  & & \% \\
 \hline
 & \textbf{DPL} & \textbf{LOG} &\textbf{$-$8.19} & \textbf{10.05 $\pm$ 0.03} & \textbf{$-$1.96 $\pm$ 0.05} & \textbf{$-$6.38 $\pm$ 0.85} & \textbf{0.40 $\pm$ 0.07} & -&\textbf{0.32 $\pm$ 0.01} & \textbf{$-$0.19 $\pm$ 0.02} &0 & 53 \\
\citet{VP06} & DPL & SCH & $-7.64$& 10.02 $\pm$ 0.02 & $-$1.79 $\pm$ 0.05 & $-6.32\pm 0.73$& 0.86 $\pm$ 0.08 & $-0.56\pm0.13$&- & $-0.26$ $\pm$ 0.02& 21.4 & 36\\
 & SCH & LOG & $-7.79$& 9.91 $\pm$ 0.04 & $-$1.79 $\pm$ 0.05 &- &0.39 $\pm$ 0.05 & -&0.32 $\pm$ 0.01 & $-0.19$ $\pm$ 0.02& 11.2 & 49\\
  & SCH & SCH & $-7.24$& 9.84 $\pm$ 0.03 & $-1.62\pm0.06$  & -&0.79 $\pm$ 0.08 & $-0.59$ $\pm$ 0.13 &-& $-0.25$ $\pm$ 0.02& 40.3 & 31\\
   \hline
\citet{park17}& DPL & LOG &$-6.72$ & 8.71 $\pm$ 0.04 & $-1.71\pm 0.12$  & $-5.54\pm 0.36$ & 0.03 $\pm$ 0.10 & -&0.15 $\pm$ 0.02 & 0.73 $\pm$ 0.01 &- &47\\
\hline
\end{tabular}
\end{adjustbox}
\end{table*}

All of the four parametric models can get converged with the combined quasar sample, and they yield broadly consistent BHMFs. As plotted by lines in the left and middle panels of Figure~\ref{fig:BHMF}, large discrepancies among the models only appear at the high-mass end for the less-luminous quasar sample, and at the low-mass end for the luminous quasar sample. In those BH mass ranges, the number of the observed quasars in the respective sample is limited, resulting in large scatter in different models. After combining the two quasar samples, that scatter disappears. For the ERDFs, there is still a difference among parametric models, especially between the log-normal and Schechter models, even after combining the quasar samples. The difference may be caused by the turnover functional shape of the log-normal function. The pink shaded areas in the figure represent the {1$\sigma$} uncertainties of the double-power-law BHMF and log-normal ERDF model. These uncertainties only become significant at the low-mass and low-Eddington-ratio ends.

We compare the \textit{intrinsic} BHMFs with the \textit{binned} ones. The \textit{intrinsic} BHMFs of the less-luminous and luminous quasar sample show broadly consistent number densities with the \textit{binned} BHMFs over the mass ranges of {$7.5<\log M_{\rm BH}/M_{\odot}<9.5$} and $9.5<\log M_{\rm BH}/M_{\odot}<10.2$, respectively. In the higher mass ranges, the \textit{intrinsic} BHMFs decline more sharply than the \textit{binned} ones; while in the lower mass ranges, the \textit{intrinsic} BHMFs {show higher number densities} than the \textit{binned} ones. The former is caused by correcting for the uncertainty associated with the virial BH mass estimates {and the bolometric correction}, and the latter is the consequence of accounting for the incompleteness of quasar samples due to their flux-limit selections. After combining the two quasar samples, the \textit{intrinsic} and \textit{binned} BHMFs show basically consistent number densities in the range of $8<\log M_{\rm BH}/M_{\odot}<10.2$, and the flux-limit correction only becomes significant around $\log M_{\rm BH}/M_{\odot}\sim7.5$. For the ERDFs, in the combined quasar sample, the \textit{intrinsic} log-normal ERDFs well follow the \textit{binned} ones, while the \textit{intrinsic} Schechter ERDFs suggest a flattened low-Eddington-ratio end at $\log \lambda_{\rm Edd}<-1$ without clear turnover. 

\begin{figure*}[!ht]
\centering
\includegraphics[keepaspectratio,width=0.9\textwidth]{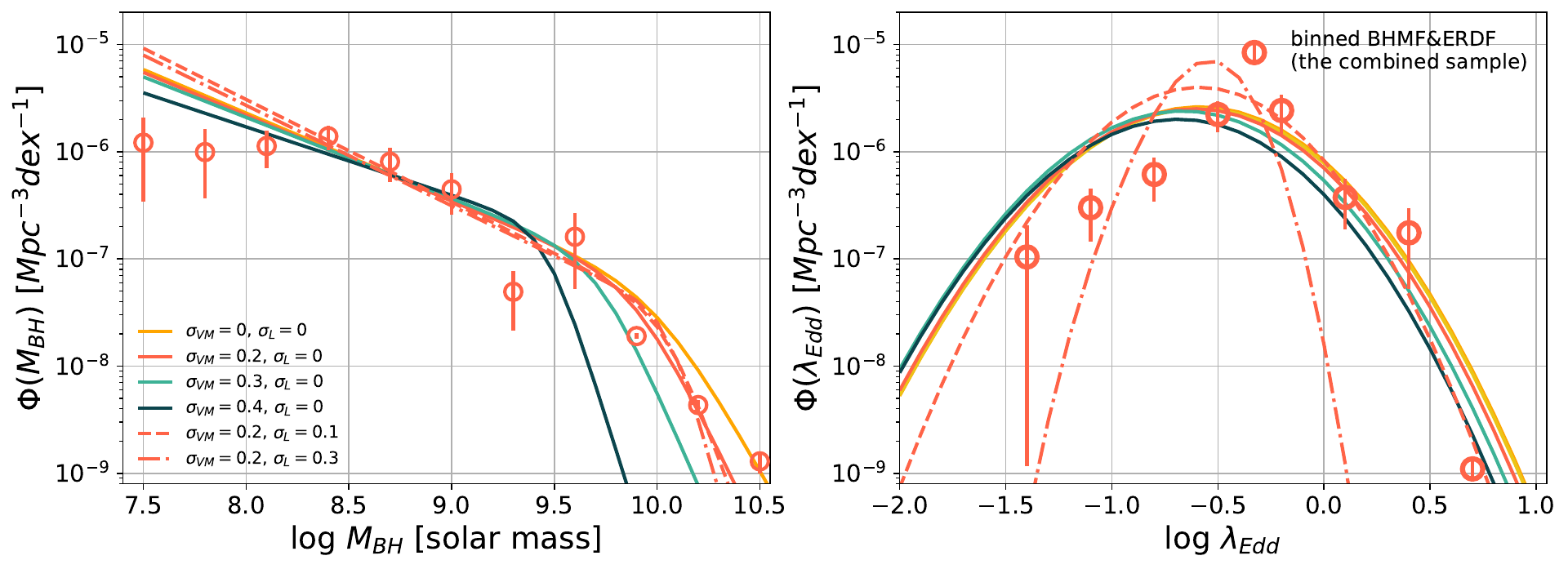}
\caption{$z=4$ broad-line AGN BHMFs (left) and ERDFs (right) by assuming the uncertainty in the virial BH mass estimates of $\sigma_{\rm VM}=$ 0 (orange line), 0.1 (yellow line), 0.2 (red line), 0.3 (green line) and 0.4 dex (dark green line), {and that in the bolometric correction of $\sigma_{\rm BC}=$ 0 (solid line), 0.1 (dashed line), and 0.3 (dashed-dotted line)}. Symbols have the same meanings as Figure~\ref{fig:BHMF}.
}
\label{fig:BHMF_sig}
\end{figure*}

As described in section~\ref{sec:s4_correct_method}, in the maximum likelihood fitting, we consider the uncertainty in the virial BH mass estimates by assuming $\sigma_{\rm VM}=0.2$ dex {and that in the bolometric correction by adopting $\sigma_{\rm BC}=0.1$ dex}. It is suggested that the $\sigma_{\rm VM}$ can reach $\sim$0.4 dex for an individual BH mass estimate with the ${\rm C_{\Rmnum{4}}}$-based calibration \citep[e.g.,][]{VP06,Coatman16}. Here, we examine how {these two uncertainties} affect the determination of BHMFs by assuming multiple uncertainties of $\sigma_{\rm VM}=$ 0, 0.1, 0.2, 0.3 and 0.4 dex, {and $\sigma_{\rm BC}=$ 0, 0.1, 0.3 dex}. The resulting BHMFs and ERDFs are plotted by lines in Figure~\ref{fig:BHMF_sig}. 

{Compared to the \textit{binned} BHMF, which does not consider the uncertainty, assuming a $\sigma_{\rm VM}$ uncertainty results in a steepened massive end of the \textit{intrinsic} BHMF. The high-Eddington-ratio end of \textit{intrinsic} ERDF also becomes sharper, while the change is relatively small. On the other hand, assuming a $\sigma_{\rm BC}$ uncertainty leads to narrow \textit{intrinsic} ERDF. The less-massive end of \textit{intrinsic} BHMF also becomes slightly steepened. Correcting for larger uncertainties can yield a more steepened high-mass (high-Eddington-ratio) end in the \textit{intrinsic} BHMF (ERDF), while the less-massive end does not show significant change. In this study, since we focus on how much the uncertainties can affect the distributions of the statistical sample instead of an individual estimate, we fix $\sigma_{\rm VM}=0.2$ dex and $\sigma_{\rm BC}=0.1$ dex hereafter, consistent with previous work \citep[e.g.,][]{ks2013,AS2015}.}

\begin{figure*} 
\centering
\includegraphics[keepaspectratio,width=0.9\textwidth]{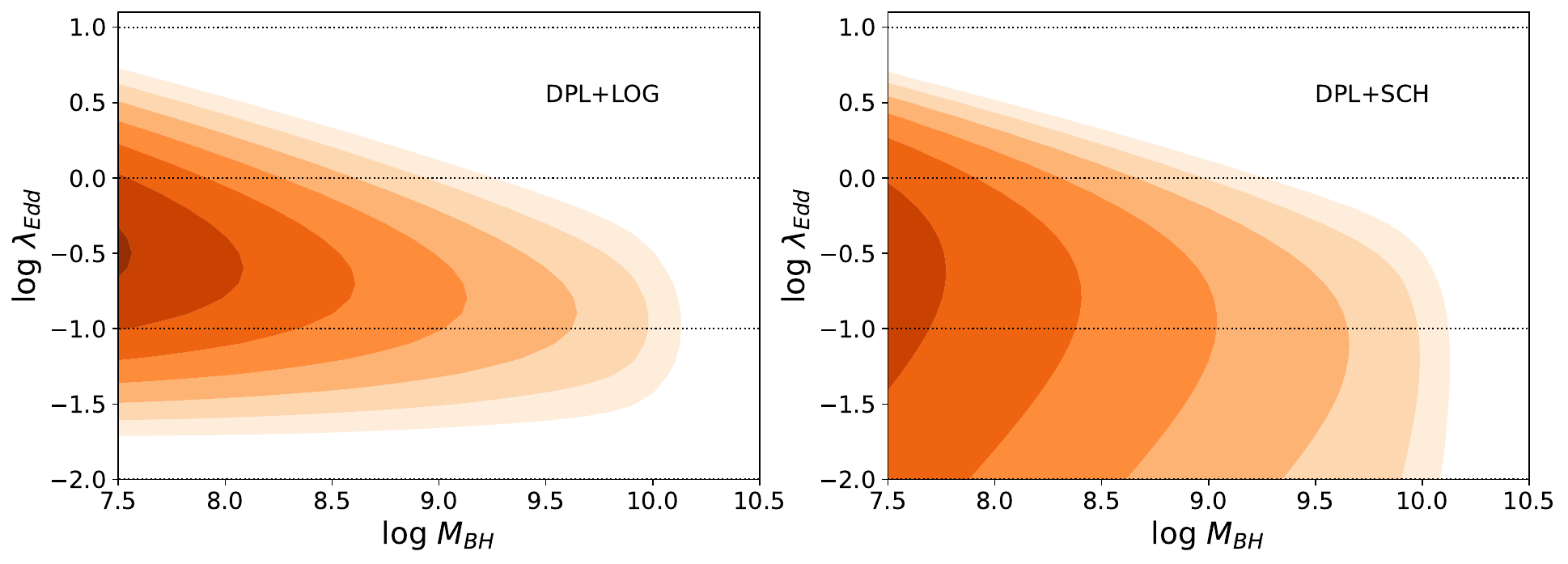}
\caption{\textit{Intrinsic} bivariate distribution function $\Psi(M_{\rm BH}, \lambda_{\rm Edd})$ of the best-fit double-power-law BHMF with the log-normal (left) and Schechter ERDF (right). Constant number densities of $\log \Psi(M_{\rm BH}, \lambda_{\rm Edd})=-8\sim-5.5$, with an interval of 0.5, are displayed by orange contours from light to dark.}
\label{fig:mbh_edd_rep}
\end{figure*}

Thanks to the wide BH mass ranges covered by the combined quasar sample, we are able to investigate the mass dependence of ERDFs at $z=4$. In Figure~\ref{fig:mbh_edd_rep}, we plot the \textit{intrinsic} bivariate distribution function of BH mass and Eddington ratio. No matter the Schechter or log-normal model is applied for ERDF, the mass dependence of ERDF is visible. Massive quasars are likely to have lower Eddington ratios, while the less-massive ones are likely to have higher Eddington ratios. 

In the parametric model, the mass dependence of ERDFs is described with the parameter $k_{\lambda}$. As can be seen from the best-fit models in Table~\ref{tab:intrinsicBHMF}, no matter which parametric model combination is adopted, $k_{\lambda}$ keeps negative even with the uncertainty, suggesting a mass dependence of ERDFs is required. For comparison, we applied the maximum likelihood method with $k_{\lambda}$ fixed to zero, and the resulting double-power-law BHMF and log-normal ERDF are plotted by green solid lines in the right panels of Figure~\ref{fig:BHMF}. Much larger corrections for the incompleteness at the less-massive and low-Eddington-ratio ends are required in this case. Considering those large corrections, the number density in the low-mass and low-Eddington-ratio ranges of $\log M_{\rm BH}/M_{\odot}<8.5$ and $\log \lambda_{\rm Edd}<-1$, which corresponds to the absolute magnitude range of $M_{\rm 1450}>-22$, is predicted to exceed the observation constraints of $\sim 10^{-6}$ Mpc$^{-3}$ \citep[e.g.,][]{akiyama2018}. We {confirm} that the predicted number density will not be largely changed if the Schechter function is applied to ERDF. 

\begin{figure}
\centering
\includegraphics[keepaspectratio,width=0.45\textwidth]{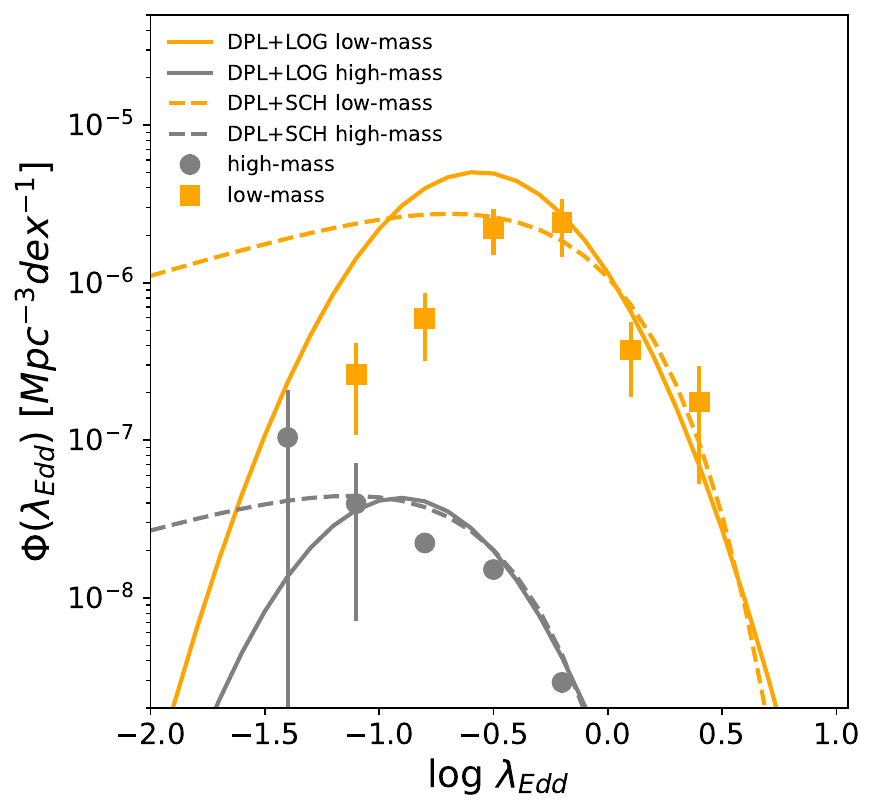}
\caption{$z=4$ broad-line AGN ERDFs in the two mass bins. Grey dots and orange squares show the \textit{binned} ERDFs in the high mass bin of $\log M_{\rm BH} / M_{\odot}>9.5$ and the low mass bin of $\log M_{\rm BH} / M_{\odot}<9.5$, respectively. Grey and orange lines indicate the \textit{intrinsic} ERDFs in the high and low mass bins, respectively. Types of lines have the same meanings as Figure~\ref{fig:BHMF}.
}
\label{fig:mbh_depend}
\end{figure}

In order to further examine the mass dependence of ERDFs, we divide the combined quasar sample into the high mass bin of $\log M_{\rm BH} / M_{\odot}>9.5$ and the low mass bin of $\log M_{\rm BH} / M_{\odot}<9.5$, where $\log M_{\rm BH} / M_{\odot}=9.5$ is the median BH mass of the combined quasar sample. In the respective mass bins, the \textit{binned} ERDFs are evaluated by the same $V_{\rm max}$ method described in section~\ref{sec:s4_bin}. Additionally, we acquire the \textit{intrinsic} ERDFs in the two mass bins by integrating equation~\ref{eq:intERDF} over the corresponding mass ranges. Since the Schechter and double-power-law BHMFs are roughly consistent, we focus on the latter model hereafter for simplicity. Results are plotted in Figure~\ref{fig:mbh_depend}. 
Except for the low-Eddington-ratio ends largely affected by the flux-limit corrections, the \textit{intrinsic} mass-dependent ERDFs well follow the \textit{binned} ones in both of the mass bins. By comparing the ERDFs in the two mass bins, the mass dependence of ERDFs can be clearly seen: the ERDFs in the high mass bin tend to peak at lower Eddington ratio of $\log \lambda_{\rm Edd}\sim-1$, while those in the low mass bin peak at higher Eddington ratio of $\log \lambda_{\rm Edd}\sim-0.5$. The semi-analytic model simulating the mass assembly of BHs at $z>4$ in \citet{Piana21} shows a similar trend that more massive BHs tend to accrete at lower Eddington ratios at $z\sim4-5$, as they switch to the gas-limited accretion phase at higher redshifts.

\subsection{Goodness of the Maximum Likelihood fitting}
\label{sec:s4_goodness_fit}

\begin{figure*} 
\centering
\includegraphics[keepaspectratio,width=.9\textwidth]{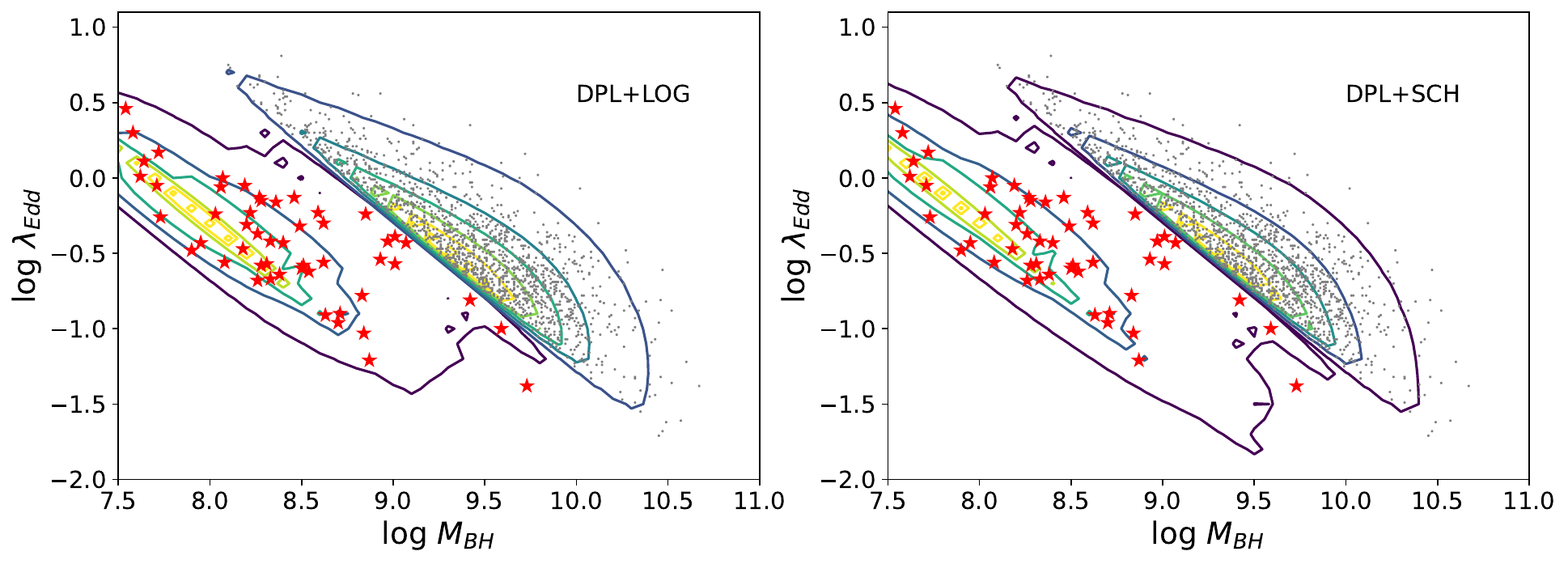}
\caption{Observed bivariate distribution function $\Psi_{o}(M_{\rm BH}, \lambda_{\rm Edd})$ of the best-fit double-power-law BHMF with the log-normal (left) and Schechter ERDF (right). Contours indicate the respective model distributions. Red stars and grey dots represent the less-luminous and luminous quasar samples, respectively.}
\label{fig:mbh_edd_rep1}
\end{figure*}

Maximum likelihood fitting does not directly provide the goodness of fit. Thus, we evaluate the goodness of fit by comparing the best-fit bivariate distribution functions above the detection limits with the distributions of the two quasar samples on the $M_{\rm BH}$ and $\lambda_{\rm Edd}$ plane. As plotted in Figure~\ref{fig:mbh_edd_rep1}, for the luminous quasar sample, we see good agreements between the observed and model distributions with both of the ERDF models; and for the less-luminous quasar sample, the observed and model distributions are still consistent with each other, but the model distribution with the log-normal ERDF follows better with the observed distributions than that with the Schechter ERDF. We quantitatively evaluate the goodness of fit by the two-dimensional Kolmogorov-Smirnov (2DKS) test on the $M_{\rm BH}$ and $\lambda_{\rm Edd}$ plane \citep{2dks}. The resulting probabilities exceed {5\%} for both of the models, i.e., we can not reject the statement that the observed and model distributions are drawn from the same parent distribution.

The quasar luminosity functions, which can be obtained by convolving the \textit{intrinsic} BHMFs and ERDFs, is another test of the goodness of fit. For the two quasar samples, using the same $V_{\rm max}$ method mentioned in section~\ref{sec:s4_bin}, we evaluate their \textit{binned} luminosity functions in the range of $-30<M_{\rm 1450}<-20$, with a bin width of 0.5. We then convolve the best-fit \textit{intrinsic} BHMFs and ERDFs of the combined quasar sample to derive the \textit{intrinsic} luminosity functions of quasars over the same luminosity range. 

\begin{figure*} [!ht]
\centering
\includegraphics[keepaspectratio,width=.9\textwidth]{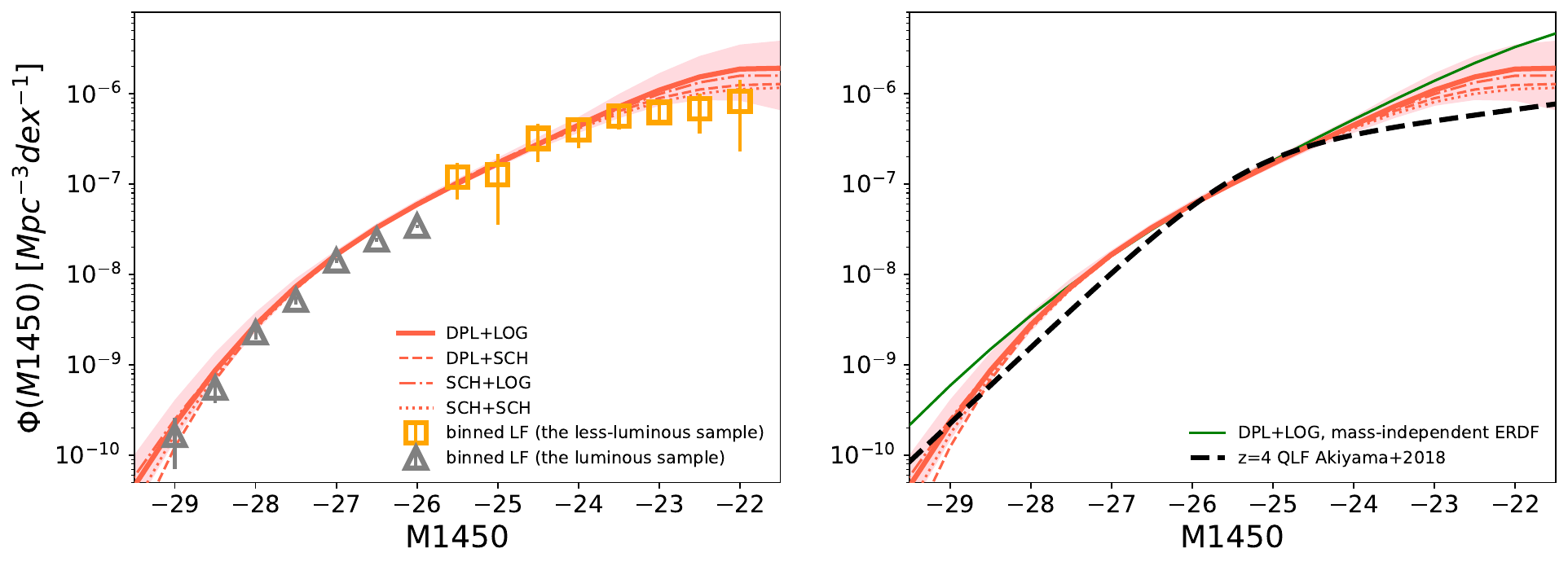}
\caption{$z=4$ quasar luminosity functions. Orange open squares and grey open triangles show the \textit{binned} luminosity functions estimated with the less-luminous and luminous quasar samples, respectively. Red lines plot the \textit{intrinsic} luminosity functions convolved by the \textit{intrinsic} BHMFs and ERDFs of the combined quasar sample. Types of lines are the same with Figure~\ref{fig:BHMF}. Pink shaded areas indicate the 3$\sigma$ uncertainties of the best-fit model shown by the red solid line. Green solid line displays the \textit{intrinsic} luminosity function of the best-fit model without the mass dependence of ERDF. Black solid line shows the $z=4$ quasar luminosity function constrained with the HSC imaging dataset \citep{akiyama2018}.}
\label{fig:LF}
\end{figure*}

The resulting \textit{binned} and \textit{intrinsic} luminosity functions of quasars are plotted in the left panel of Figure~\ref{fig:LF}. The observed and model luminosity functions show good agreements with each other. 
For the four parametric models, they show almost the same number densities over the luminosity ranges of $-28<M_{\rm 1450}<-23$. Discrepancies are seen only in the faintest bins with poor statistics. \citet{akiyama2018} determine the luminosity function of quasars at $z=4$ using photometric quasar candidates selected by the same c1 criteria from the HSC-SSP S16A Wide2 imagings, in combination with the SDSS DR7 quasars. The best-fit model in their work is plotted by black solid line in the right panel of Figure~\ref{fig:LF}. Overall, the \textit{intrinsic} luminosity functions follow the observed one in \citet{akiyama2018}, both showing the double-power-law shape with flattened faint-end. 

In order to examine the effect of the mass dependence of ERDF, we derive the \textit{intrinsic} luminosity function of the best-fit model with $k_{\lambda}=0$. Result is plotted by the green solid line in the right panel of Figure~\ref{fig:LF}. Due to the large correction at the less-massive and low-Eddington-ratio ends, the \textit{intrinsic} luminosity function can not reproduce the flattened faint-end found in \citet{akiyama2018}, supporting the necessity of the mass dependence of ERDF. 

Furthermore, following \citet{Bongiorno16}, we use the Akaike information criterion \citep[AIC;][]{AIC} to compare the fitting quality of the respective parametric model. It is defined as AIC$=S+2K+2K(N/(N-K-1))$, where $S$ is the minimum likelihood of fitting, $K$ is the number of free parameters in model, and $N$ is the size of sample. We list the resulting relative AIC for each parametric model in Table~\ref{tab:intrinsicBHMF}. 

For fitting the combined quasar sample, BHMFs in the double power-law model get better quality than those in the Schechter model, and ERDFs in the log-normal model achieve smaller AIC value than those in the Schechter model. The parametric model with BHMF in the double power-law function and ERDF in the log-normal function has the minimum AIC value, i.e., it can be regarded as the best-fit model. We note that the relative AICs can be enhanced by $\gtrsim50$ if no mass dependence of ERDF is assumed in the fitting. 

In summary, for discussions hereafter, we focus on the best-fit double-power-law BHMF with the log-normal ERDF, determined by the combined quasar sample with the mass-dependent ERDF. We also put the best-fit double-power-law BHMF with the Schechter ERDF as a comparison.  

\section{Discussion}
\label{sec:dis}

\subsection{Comparison with previous work}
\label{sec:s5_correct_comp_z4}

\begin{figure*} [!ht]
\centering
\includegraphics[keepaspectratio,width=0.9\textwidth]{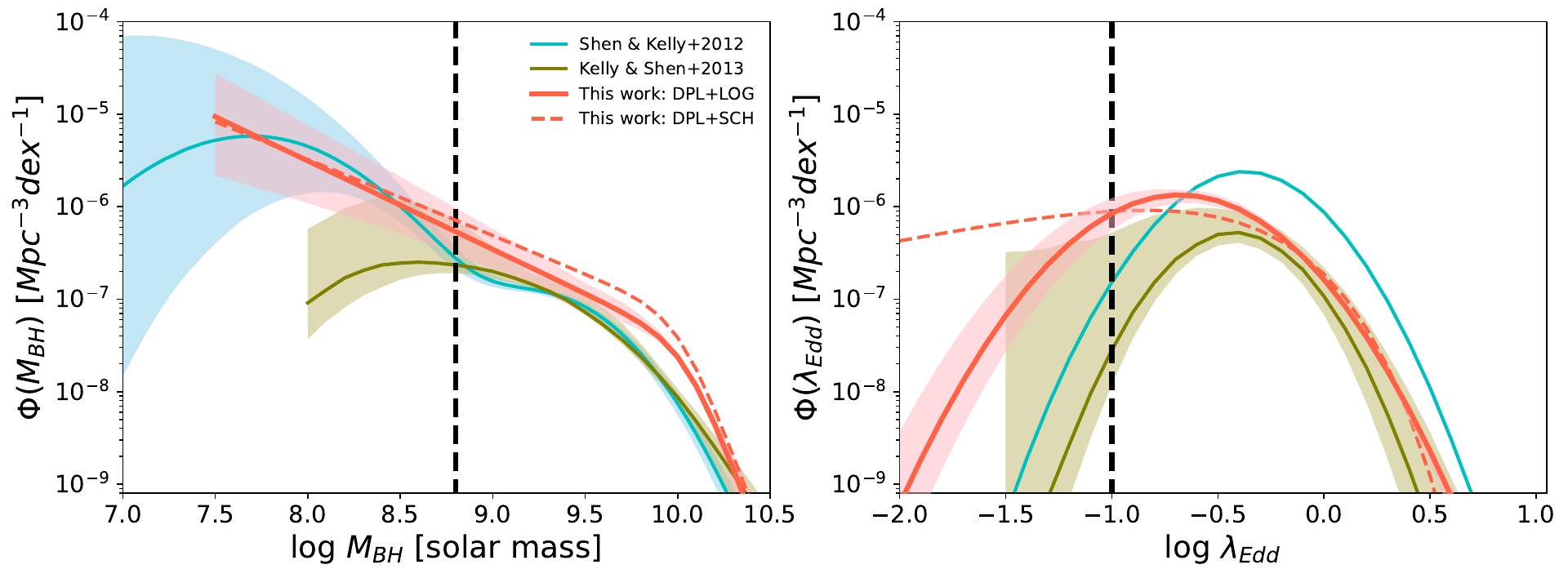}
\caption{Broad-line AGN BHMFs (left) and ERDFs (right) at $z\sim4$. Red lines show the best-fit results obtained by this work. Types of lines are the same with Figure~\ref{fig:BHMF}. Blue and olive lines display the $z=3.75$ BHMFs and ERDFs constrained by \citet{SK2012} and \citet{ks2013}, respectively. Shaded areas indicate the $1\sigma$ uncertainty of the respective constraints. Black vertical dashed lines show the {upper values} where the completeness of the SDSS quasar sample drops below $<10\%$, given by \citet{ks2013}. Normalization of the ERDFs are re-scaled by integrating the corresponding BHMFs over $8<\log M_{\rm BH}/M_{\odot}<11$.}
\label{fig:BHMF_compz4}
\end{figure*}

We compare the derived broad-line AGN BHMFs and ERDFs with those in literature. \citet{SK2012} and \citet{ks2013} examine the broad-line AGN BHMFs and ERDFs in the redshift range of $0.3<z<5$. They utilize a uniformly-selected sample of $\sim$58,000 SDSS DR7 quasars. At $z>1.9$, the virial BH masses of the quasars are estimated with the same ${\rm C_{\Rmnum{4}}}$-calibrated single-epoch estimator used in this work. In both of the studies, a Bayesian method with flexible models describing the underlying BHMFs and ERDFs is adopted to correct for the selection biases introduced by the flux limits, and luminosity-dependent scatter in the virial BH mass estimate is taken into account. The resulting BHMFs and ERDFs at $z=3.75$ in \citet{SK2012} and \citet{ks2013} are shown by sky-blue and olive lines in Figure~\ref{fig:BHMF_compz4}, respectively, together with those derived in this work.

In the left panel of the figure, we find a {broad} agreement between BHMFs in this work and theirs in the high-mass range of $\log M_{\rm BH}/M_{\odot}\gtrsim 9$. The luminous quasars mainly determine the constraints in that mass range, and the luminous quasar sample used in this work is constructed from the same SDSS DR7 quasar catalog. Although different methods, i.e., the Bayesian and Maximum likelihood methods, and the functional models are adopted, the consistent \textit{intrinsic} BHMFs can be obtained. Discrepancy in BHMFs appears in the low-mass range of $\log M_{\rm BH}/M_{\odot}\lesssim 8.5$, where the completeness of the SDSS quasar sample drops below 10\% \citep{ks2013}. The discrepancy can thus be caused by the uncertainty introduced by the large incompleteness correction of the SDSS quasar sample. In this work, by combining the luminous and less-luminous quasar samples, we can constrain the low-mass end of the $z=4$ BHMF down to $\log M_{\rm BH}/M_{\odot}\sim7.5$ with the completeness more than 10\%. 

In the right panel of Figure~\ref{fig:BHMF_compz4}, the ERDFs derived in this work are broadly consistent with those in \citet{SK2012} and \citet{ks2013}, especially at the high-Eddington-ratio end. The ERDFs show discrepancy at $\log \lambda_{\rm Edd}<-1$, which can also be caused by the incompleteness of the SDSS quasar sample in this range as indicated by the black dashed line.

\subsection{Cosmological evolution of the broad-line AGN BHMFs and ERDFs}
\label{sec:s5_correct_comp}

In order to clarify what drives the ``down-sizing'' trend seen in the AGN luminosity functions, we compare the $z\sim4$ broad-line AGN BHMFs and ERDFs with those at $0<z<6$ in literature: \citet{SK2012} and \citet{ks2013} covering $0.3<z<5$ as described above; \citet{AS2010} using 329 $z<0.3$ quasars and Seyfert-1 galaxies drawn from the Hamburg/ESO Survey; \citet{AS2015} with the $1.1<z<2.1$ quasars from the VVDS epoch-2 AGN, the zCOSMOS 20k AGN, and the SDSS DR7 QSO samples; \citet{nobuta} with the X-ray-selected $1.18<z<1.68$ quasars in the Subaru XMM-Newton Deep Survey (SXDS); \citet{willott} using 17 $z\gtrsim6$ quasars from the Canada–France High-z Quasar Survey (CFHQS); and \citet{wu2022} with 47 quasars at $5.7\leq z\leq6.5$ from SDSS. 

We note that different single-epoch virial BH mass estimators are adopted in each work: \citet{SK2012} and \citet{ks2013} use H$\beta$, ${\rm Mg_{\Rmnum{2}}}$ and ${\rm C_{\Rmnum{4}}}$ emission lines at $z<0.7$, $0.7<z<1.9$ and $z>1.9$, respectively; \citet{AS2010} use H$\beta$ emission line; and other studies mostly rely on ${\rm Mg_{\Rmnum{2}}}$ emission line. In addition, the $z=6$ BHMF in \citet{willott} already corrects for the obscured fraction of AGNs, including the Compton-thick ones, and the active fraction. It thus should be treated as an upper limit in the comparison with other broad-line AGN BHMFs, especially in the low-mass range as the obscuration correction becomes significant \citep[e.g.,][]{Merloni}. 
  
\begin{figure*} [!ht]
\centering
\includegraphics[keepaspectratio,width=\textwidth]{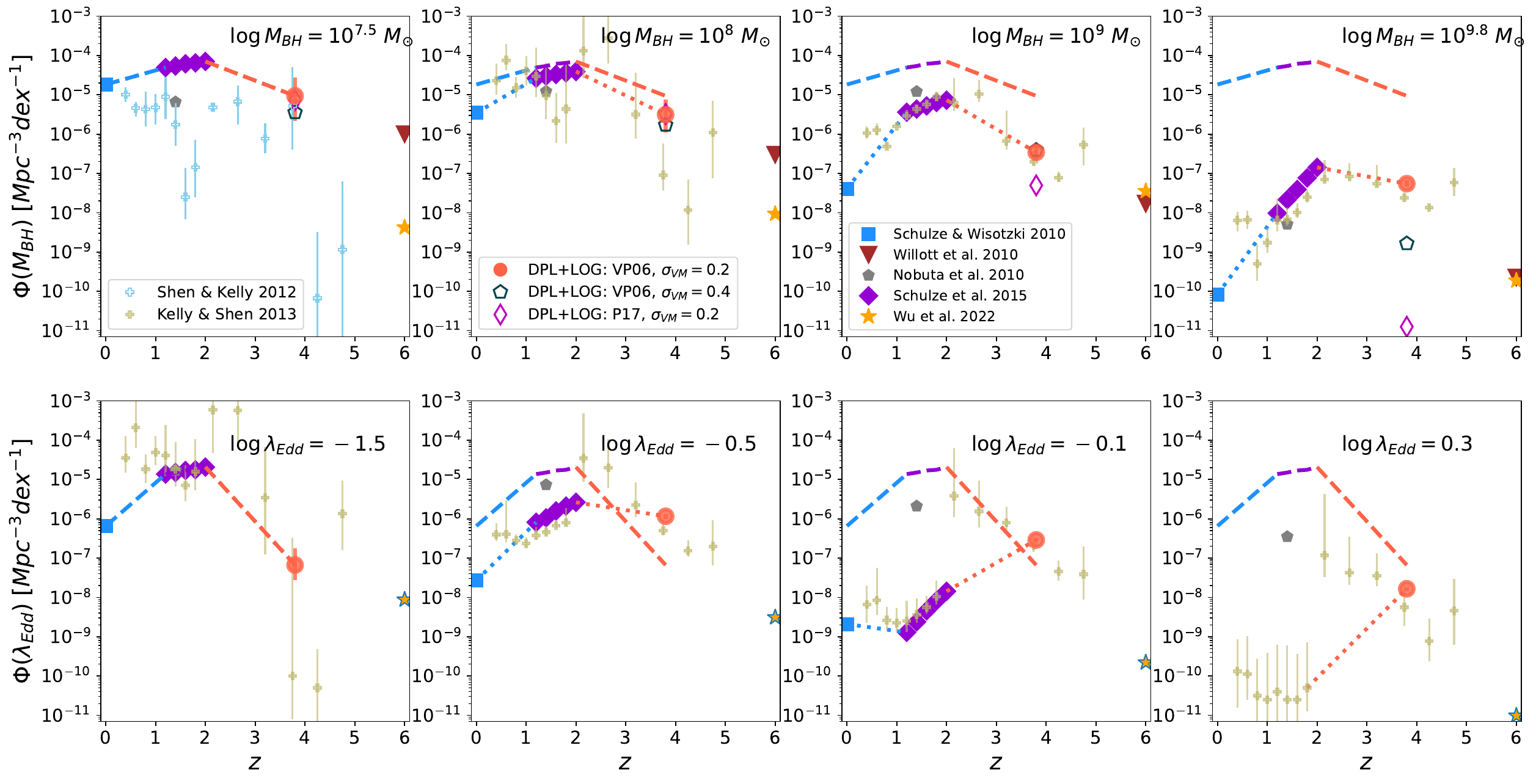}
\caption{Cosmic evolution of the number density of broad-line AGNs at a constant BH mass (top) and Eddington ratio (bottom) at $0<z<6$. Red dots, blue squares, purple diamonds, brown inverted triangles, light-blue open pluses, olive pluses, gray pentagons and orange stars show the results obtained in this work, \citet{AS2010}, \citet{AS2015}, \citet{willott}, \citet{SK2012}, \citet{ks2013}, \citet{nobuta} and \citet{wu2022}, respectively. Green open pentagons represent the $z\sim4$ results by adopting $\sigma_{\rm VM}=$ 0.4 dex, and magenta open diamonds show those derived with the single-epoch virial BH mass calibration in \citet{park17}. Dashed lines connect the blue squares at $z=0$, purple diamonds at $z=1-2$ and red dots at $z\sim4$ in the lowest BH mass and Eddington ratio bins; and the dotted lines connect those in the higher bins. All the ERDFs are re-scaled to match the number density of quasars in the mass range of $8<\log M_{\rm BH}/M_{\odot}<11$ before the comparison.}  
\label{fig:evo}
\end{figure*}

In Figure~\ref{fig:evo}, the cosmic evolution of the number density of the broad-line AGNs at a constant BH mass and Eddington ratio is shown. 
To guide the comparison, we connect the results at $z\sim0$ in \citet{AS2010}, at $z\sim1-2$ in \citet{AS2015} and at $z\sim4$ in this work by dotted lines. Results from other studies broadly follow the lines. There is a large scatter in the lowest BH mass and Eddington ratio bins, which can be caused by large incompleteness corrections of the SDSS quasars in the ranges \citep{SK2012,ks2013}. In addition, we see the number density at constant Eddington ratios constrained by \citet{ks2013} suddenly increases at $z\sim2$. Since the virial BH mass estimator is changed from ${\rm Mg_{\Rmnum{2}}}$ to ${\rm C_{\Rmnum{4}}}$ at $z\sim2$ in their work, that sharp increase can be a consequence of the inconsistency between different virial BH mass estimators. Furthermore, results in \citet{nobuta} show large excesses in the high Eddington ratio bins. It should be noted that the quasar sample in their work {covers a relatively less-massive range of $\log M_{\rm BH}<10$, and they do not correct for the scatter induced by the virial BH mass estimate}. The resulting BHMF and ERDF in their work thus show flattened massive and high-Eddington-ratio ends. 

Following the dotted lines in the figure, from $z=2$ to $z=6$, the number density of massive SMBHs with $\log M_{\rm BH}/M_{\odot}=9.8$ keeps constant up to $z=4$ and then drops towards $z=6$, while that of less-massive SMBHs with $\log M_{\rm BH}/M_{\odot}\lesssim 9$ shows continuous declines towards $z=6$. The number density of rapidly-accreting SMBHs with $\log \lambda_{\rm Edd}\gtrsim -0.1$ significantly increases from $z\sim2$ to $z\sim4$, while that of SMBHs with lower Eddington ratios of $\log \lambda_{\rm Edd}\lesssim -0.5$ decreases from $z\sim2$ to 4. Both of the evolutionary trends suggest the abundance of more massive and rapidly-accreting SMBHs peaks at higher redshifts, consistent with the ``down-sizing'' scenario. There are also semi-analytical simulations suggesting the number density of rapidly-accreting SMBHs with $\log \lambda_{\rm Edd}\sim 0$ tends to peak at high redshifts \citep[e.g.,][]{shirakata19}. The possible cause of the trend may be due to the gas-rich environment of galaxies at high redshifts, making rapid accretion to easily occur.

In the less-massive range of $\log M_{\rm BH}/M_{\odot}\lesssim9$, we find the number density basically evolves {with a similar slope} from $z=2$ to 4. The ``up-sizing'' evolutionary trend, which indicates a milder decline in the number density of less-massive or lower-Eddington-ratio SMBHs after the peaks towards high redshifts, is not seen in the comparison. 

\subsection{Obscuration correction and the active fraction of SMBHs at $z\sim4$}
\label{sec:s5_duty}

The current $z\sim4$ quasar sample only covers unobscured AGNs, here we estimate the BHMFs and ERDFs for the total AGN population by correcting for the obscured AGNs. Studies on hard-X-ray (2-10 keV) selected AGNs suggest the fraction of obscured AGNs among the entire AGN population deceases with increasing luminosity \citep[e.g.,][]{Ueda03,Hasinger08,Merloni,Ueda14}. Physical origin of the dependence is not disclosed yet, and a possible scenario is that the inner edge of the dusty obscuring torus recedes with increasing luminosity \citep[e.g.,][]{Lawrence1991,simpson05,ricci13,Toba14,Vijarnwannaluk22}.

\citet{Merloni} determine the relationship between the X-ray luminosity at 2-10~keV and the fraction of obscured AGNs among the entire AGN population with a complete sample of 1310 X-ray-selected AGNs at $0.3<z<3.5$ in the XMM-COSMOS field. Here, we assume the relationship does not evolve with redshift, and adopt it as follows:
\begin{equation} \label{eq:correction}
f_{\rm obs}=0.56+\frac{1}{\pi} {\arctan} \left[\frac{43.89-{\log}L_{2-10 ~{\rm keV}}}{0.46}\right],
\end{equation}
Additionally, since the relationship is poorly constrained in the less-luminous range, we apply an upper limit of $f_{\rm obs}$ as $f_{\rm max}=0.985$, which is the obscured fraction at the luminosity of $L_{2-10 ~{\rm keV}}=10^{42}$ erg s$^{-1}$. It should be noted that the contribution of Compton-thick AGNs is not considered here since it is still highly uncertain. 

The correction for the obscured fraction is applied in each BH mass and Eddington ratio bin on the $M_{\rm BH}$ vs. $\lambda_{\rm Edd}$ plane. Firstly we calculate the bolometric luminosity $L_{\rm bol}$ of the combination of BH mass $M_{\rm BH}$ and Eddington ratio $\lambda_{\rm Edd}$, and then convert the luminosity to the X-ray luminosity in the 2-10 keV band using the bolometric correction of $L_{2-10 ~{\rm keV}}=L_{\rm bol}/\kappa_{x}$, where $\kappa_{x}=a\times[1+(\log L_{\rm bol}/b)^{c}]$, and $a=12.76$, $b=2.15$ and $c=18.78$ \citep{duras}. The \textit{intrinsic} number density in that bin is then divided by the un-obscured fraction $(1-f_{\rm obs})$ at luminosity $L_{2-10 ~{\rm keV}}$. The total active BHMFs and ERDFs are obtained by integrating the obscuration-corrected bivariate distribution against the Eddington ratio over $-2<\log \lambda_{\rm Edd}<1$ and the BH mass over $7.5<\log M_{\rm BH}/M_{\odot}<11$, respectively. 

\begin{figure} 
\centering
\includegraphics[keepaspectratio,width=.45\textwidth]{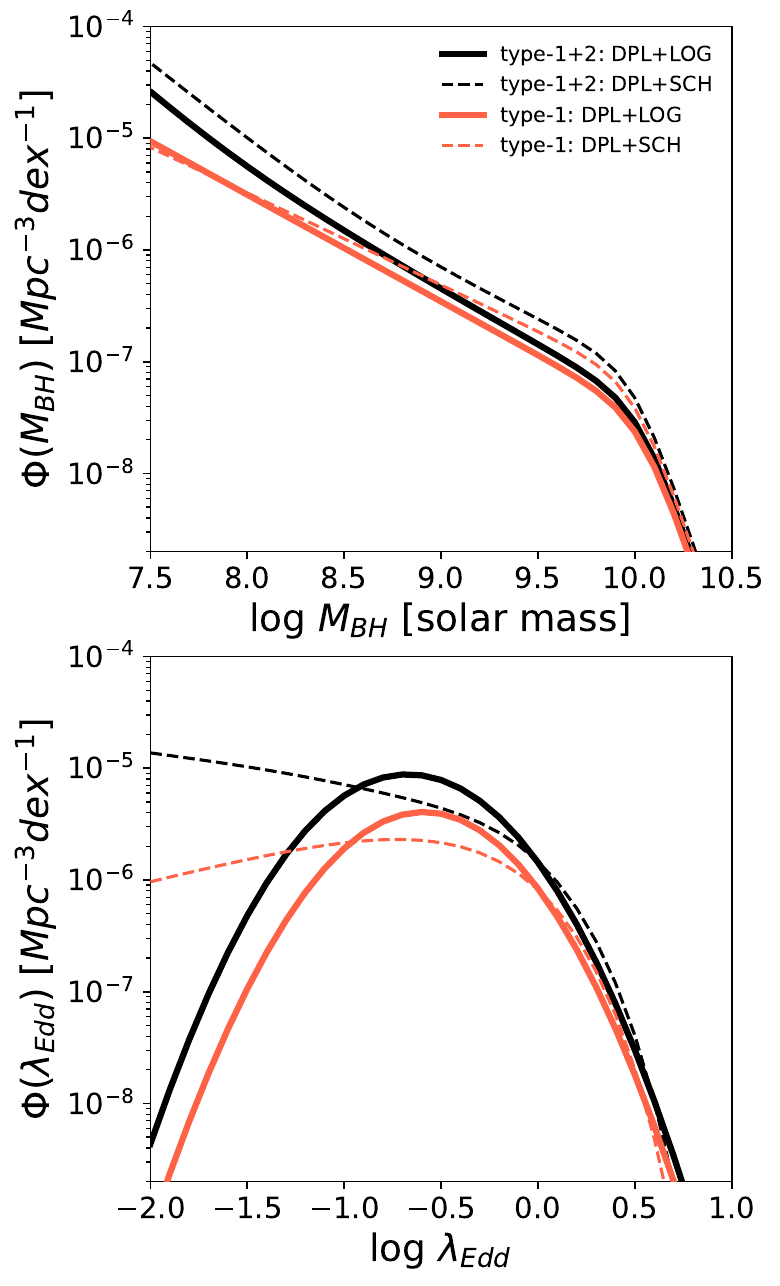}
\caption{Obscuration-corrected BHMFs (top) and ERDFs (bottom) at $z=4$. Black and red lines show the results with and without the correction for the fraction of obscured AGNs. Solid and dashed lines indicate the results given by the log-normal and Schechter ERDF, respectively.}
\label{fig:total_active}
\end{figure}

The resulting total active BHMFs and ERDFs are plotted with black lines in Figure~\ref{fig:total_active}. Solid and dashed lines show the results derived with the log-normal and Schechter ERDF, respectively.
As expressed in equation~\ref{eq:correction}, the obscured AGNs are assumed to dominate the less-luminous AGN population. We thus see significant correction for the obscured AGNs in the low-mass or low-Eddington-ratio ranges of the total active BHMFs and ERDFs. At the high-mass or high-Eddington-ratio ends, the contribution from the obscured AGNs can be negligibly small. 

We note that the correction in the model with log-normal ERDF is weaker than that with Schechter ERDF. The reason can be due to the sharp decline of the log-normal ERDF towards the low-Eddington-ratio end, where the obscured AGNs can dominate. Therefore, even after correcting for the obscured AGNs, the total spatial number density of active AGNs in the low-mass and low-Eddington-ratio regime predicted by the log-normal ERDF remains small.   

Utilizing the total active BHMFs, we evaluate the active fraction of SMBHs at $z=4$, which is defined as the fraction of active SMBHs with $\lambda_{\rm Edd}>0.01$ among the total SMBH population, by comparing the total active BHMFs to the total BHMFs including quiescent SMBHs with $\lambda_{\rm Edd}<0.01$. In order to derive the total BHMF, following \citet{AS2015}, we convolve the total stellar mass function of galaxies $\Psi{^*}$ with the $M_{\rm BH}-M_{\rm star}$ relation, i.e., 
\begin{align} \label{eq:totBHMF}
\Psi_{\rm tot}&(M_{\rm BH},z)=\\
&\frac{1}{\sqrt{2 \pi} \sigma} \int \exp \left\{-\frac{(\log M_{\rm BH}-\alpha-\beta s)^2}{2 \sigma^2}\right\} {\Psi{^*}(s)} \mathrm{d} s ,\\
\end{align}
where $s=\log M_{\rm star}-11$, and $\alpha$, $\beta$ and $\sigma$ are the the normalization, slope and intrinsic scatter of the $M_{\rm BH}-M_{\rm star}$ relation. 

To compare with \citet{AS2015}, we adopt the same $M_{\rm BH}-M_{\rm Bulge}$ relation in \citet{MM}, i.e., $\alpha=8.46$, $\beta=1.05$ and $\sigma=0.34$, and the total stellar mass function of galaxies at $z\sim4$ in \citet{Ilbert}. Here, as explained in \citet{AS2015}, due to multiple uncertainties, such as the ratio between the stellar and spheroid mass at high redshifts, we use total stellar mass as a proxy of spheroid mass. 

The $M_{\rm BH}-M_{\rm Bulge}$ relation given by \citet{MM} is examined in the local universe. Whether this relation evolves with redshift still remains controversial. In the high-redshift galaxies, the decomposition of their bulge components is not applicable. The stellar mass is then treated as a surrogate of bulge mass, and the $M_{\rm BH}-M_{\rm star}$ relation is {used} instead \citep[e.g.,][]{RV15,Shankar16,Suh2020}. There are studies suggesting a possible redshift evolution in the $M_{\rm BH}-M_{\rm star}$ relation \citep[e.g.,][]{McLure,Decarli}, while others suggest no evolution \citep[e.g.,][]{w2017,AS2014,Suh2020}. 

Considering the uncertainties, in addition to the $M_{\rm BH}-M_{\rm Bulge}$ relation given by \citet{MM}, we apply the $M_{\rm BH}-M_{\rm star}$ relations determined in \citet{Suh2020} and \citet{Shankar16}. \citet{Suh2020} determine the relation for a sample of 100 X-ray-selected moderate-luminosity broad-line AGNs out to $z\sim2.5$ in the Chandra-COSMOS Legacy Survey, with the local AGNs in \citet{RV15}. The best-fit relation shows a steeper decline towards the low-$M_{\rm star}$ end than that in \citet{MM}. Additionally, they also derive the relation with only the high-$z$ AGNs. The resulting relation is consistent with that in \citet{MM}, but with a larger intrinsic scatter of $\sim0.5$ dex. \citet{Shankar16} suggest the $M_{\rm BH}-M_{\rm star}$ relation can be heavily biased by up to 50 times at small stellar masses due to the selection effect. We use the de-biased relation provided by \citet{Shankar16}. It shows the sharpest decline in the low-mass range of $\log M_{\rm star}/M_\odot<11$. 

\begin{figure*} [!ht]
\centering
\includegraphics[keepaspectratio,width=\textwidth]{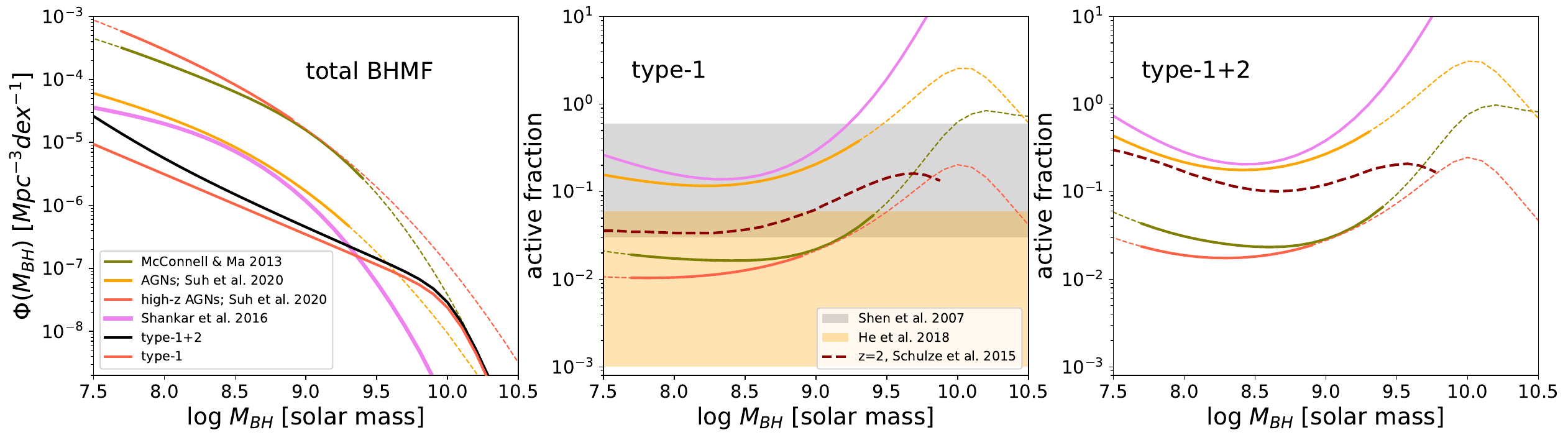}
\caption{Left) total BHMFs (thin lines), total active BHMF (black thick line), and broad-line AGN BHMF (red thick line) at $z=4$. Olive, orange, red and pink lines display the results derived with the $M_{\rm BH}-M_{\rm Bulge}$ or $M_{\rm BH}-M_{\rm star}$ relations in \citet{MM}, \citet[the total AGN sample;][]{Suh2020}, \citet[the high-$z$ sample;][]{Suh2020} and \citet{Shankar16}, respectively. The solid lines indicate the mass ranges where the $M_{\rm BH}-M_{\rm Bulge}$ or $M_{\rm BH}-M_{\rm star}$ relations are reliably constrained by the respective sample, while the dotted lines show the extrapolated ranges. Middle and right) active fraction of the broad-line AGNs (middle) and total AGNs (right) as a function of BH mass. \citet{AS2015} evaluate the same active fraction at $z=2$ (brown dashed lines). Grey and orange shaded areas present the active fractions of the luminous and less-luminous quasars among halos at $z=4$, constrained by the clustering analysis in \citet{shen2007} and \citet{He}, respectively. }
\label{fig:activeFrac_comp}
\end{figure*}

The resulting total BHMFs are plotted by thin lines in the left panel of Figure~\ref{fig:activeFrac_comp}.
Overall, the result obtained with the $M_{\rm BH}-M_{\rm star}$ relation in \citet[the high-$z$ sample;][]{Suh2020} is consistent with that derived with the $M_{\rm BH}-M_{\rm Bulge}$ relation in \citet{MM}. The other two results determined with the relations in \citet[the total AGN sample;][]{Suh2020} and \citet{Shankar16} show more flattened low-mass ends with the normalizations over one order of magnitude smaller than the former two results. 

The reason of this discrepancy can be caused by the sharp decline of the latter two relations in the low-mass range of $\log M_{\rm star}/M_\odot<11$. Meanwhile, the massive end of the resulting total BHMF shows strong decline if the adopted $M_{\rm BH}-M_{\rm Bulge}$ or $M_{\rm BH}-M_{\rm star}$ relation has small intrinsic scatter. The results given with the relations in \citet{Shankar16} and \citet{MM} thus have a high-mass end declining even sharper than the broad-line AGN BHMFs. 

With the total BHMFs, we evaluate the active fractions of broad-line AGNs as a function of BH mass. The results are presented by thin lines in the middle panel of Figure~\ref{fig:activeFrac_comp}. The active fraction of broad-line AGNs shows a positive dependence on the BH mass: the less-massive SMBHs with {$7.5<\log M_{\rm BH}/M_{\odot}<8.5$ show an average fraction of $0.01\sim0.3$, while the massive ones with $\log M_{\rm BH}/M_{\odot}>9$ occupy $0.04\sim0.8$}. Such BH mass dependence of the active fraction is also found in semi-analytical simulations of galaxy and BH evolution at $z>2$ \citep[e.g.,][]{shirakata20}. 

Beyond the peak around $\log M_{\rm BH}/M_{\odot}=9.5$, the active fractions start decreasing. It should be noted that most of the applied $M_{\rm BH}-M_{\rm Bulge}$ or $M_{\rm BH}-M_{\rm star}$ relations are extrapolated in the mass range of $\log M_{\rm BH}/M_{\odot}>9$. The converted total BHMF and the active fraction in that mass range thus become highly uncertain, and some resulting active fractions exceed unity. 

The fraction of dark matter halos with a broad-line AGN is also examined by the clustering analysis at $z\sim4$. The estimated active fractions are $0.001\sim0.06$ for the less-luminous quasars \citep{He}, and $0.03\sim0.6$ for the luminous quasars \citep{shen2007}. Although the methods to derive the fraction are different, we see broad agreements between the active fractions constrained in this study and those determined in the clustering analysis. The luminous SDSS quasars are expected to associate with massive SMBHs of $\log M_{\rm BH}/M_{\odot}>9$ with high active fractions, while the less-luminous quasars are more likely to represent smaller SMBHs with lower active fractions. 

With the correction for the obscured fraction, the active fractions of the total AGN population are plotted by thin lines in the right panel of Figure~\ref{fig:activeFrac_comp}. They show constant values with a mean at $0.01\sim0.15$ over a wide BH mass range of $7.5<\log M_{\rm BH}/M_{\odot}<9$, implying the phase of active accretion is occurring throughout the less-massive SMBH population without preferences on the BH mass. That active fractions are consistent with the values evaluated with the X-ray-selected AGNs in the deep survey fields \citep[e.g.,][]{Bongiorno12,Bongiorno16,Georgakakis2017}, and with the semi-analytical model for the BH formation and growth \citep[e.g.,][]{Li23}. Again, the increase beyond $\log M_{\rm BH}/M_{\odot}>9$ is highly uncertain because the total BHMFs are evaluated with the extrapolated $M_{\rm BH}-M_{\rm Bulge}$ or $M_{\rm BH}-M_{\rm star}$ relations. 

We investigate the cosmic evolution of the active fraction of the broad-line and total active AGNs by comparing the fractions at $z=4$ with that at $z=2$, which is estimated by \citet{AS2015} with the same method described above. Since only the $M_{\rm BH}-M_{\rm Bulge}$ relation in \citet{MM} is adopted in \citet{AS2015}, we consider the result derived with the same relation at $z=4$. While the uncertainty remains large, the massive SMBHs of $\log M_{\rm BH}/M_{\odot}>9.5$ are likely to keep a high active fraction of $\sim 10$\% since the cosmic noon. We note that the comparison highly depends on the stellar mass function of galaxies and $M_{\rm BH}-M_{\rm Bulge}$ relation at high redshifts. Future studies on examining the $M_{\rm BH}-M_{\rm Bulge}$ relation over wide BH mass and redshift ranges can be important to improve confining the active fractions.

\subsection{Time evolution of BHMFs at $z=4\sim6$}
\label{sec:s5_continuity}

We further examine the time evolution of BHMF between $z=4$ and 6 using the \textit{intrinsic} broad-line AGN BHMF and ERDF at $z=4$ determined in section~\ref{sec:s4_correct_result} with the total BHMF at $z=6$ given by \citet{willott}. We try to reproduce the observed broad-line AGN luminosity functions at $z=4$, 5, and 6, by optimizing the time evolution model of the broad-line AGN BHMF. We assume the evolution is driven only by the mass accretion during the AGN phase, and the evolution is described by the continuity equation \citep[e.g.,][]{YT2002,Shankar13}:
\begin{equation} \label{eq:continuity1}
\frac{\partial \Phi(M_{\rm BH}, t)}{\partial t}=-\frac{\partial \left[\langle\dot{M_{\rm BH}}\rangle \Phi(M_{\rm BH}, t)\right]}{\partial M_{\rm BH}}.
\end{equation}
Here, $\langle\dot{M}_{\rm BH}\rangle$ is the average accretion rate of SMBHs with BH mass $M_{\rm BH}$ at time $t$. {Assuming} the SMBH growth is driven only by the AGN activity, this rate corresponds to the total mass accreted onto the SMBHs during the AGN phase, i.e., $\dot{M_{\rm BH}}=(1-\epsilon)\dot{M_{\rm in}}$, where $\epsilon$ is the radiative efficiency and $\dot{M_{\rm in}}$ is the total mass inflow rate. The remaining inflow mass is converted into luminosity $L=\epsilon\dot{M_{\rm in}}c^2$. As described in section~\ref{sec:s3_mbh}, the Eddington ratio $ \lambda_{\rm Edd}$ is defined as the ratio between luminosity $L$ and the Eddington luminosity $L_{\rm Edd}=\ell M_{\rm BH}$, where $\ell=1.26\times10^{38} M^{-1}_{\odot}$ erg s$^{-1}$. Thus, the average accretion rate can be written as
\begin{equation} \label{eq:ave_accre}
\langle\dot{M_{\rm BH}}\rangle=\frac{(1-\epsilon) \ell}{\epsilon c^{2}} U M_{\rm BH}\langle\lambda_{\rm Edd}\rangle,
\end{equation}
where $U$ and $\langle\lambda_{\rm Edd}\rangle$ are the active fraction and the average Eddington ratio of SMBHs with BH mass $M_{\rm BH}$ at time $t$, respectively. 

By combining equation~\ref{eq:continuity1} and equation~\ref{eq:ave_accre}, the continuity equation can be modified to:
\begin{align} \label{eq:continuity2}
& \frac{\partial\Phi(M_{\rm BH},z)}{\partial z}\notag \\
&=-\frac{(1-\epsilon) \ell}{\epsilon c^{2}} \frac{{\rm d}t}{{\rm d}z}\frac{\partial}{\partial M_{\rm BH}} \left[M_{\rm BH}\langle\lambda_{\rm Edd}\rangle\Phi_{\rm AGN}(M_{\rm BH}, z)\right] .
\end{align}
Since solving the continuity equation backwards in time can easily result in unphysical BHMFs with negative number densities \citep[e.g.,][]{Shankar13}, we start from $z=6$, and then move forwards in time to $z=4$ with a time step of $\Delta z=0.1$ to solve equation~\ref{eq:continuity2}. 

We use the total BHMF at $z=6$ given by \citet{willott} as the initial condition. At each time step, we adopt the total BHMF $\Phi(M_{\rm BH},z)$ obtained from the previous step or from the assumed initial condition in the first time step. Then, the active fraction $U$ is derived as the ratio between the active and total BHMF at the time step, i.e., $U(M_{\rm BH}, z)=\Phi_{\rm AGN}(M_{\rm BH}, z)/\Phi(M_{\rm BH}, z)$. Here, we confirm the time step of $\Delta z=0.1$ is sufficiently small that smaller time steps yield consistent results. The total number of black holes between the two time steps is conserved.

At the time step, the active BHMF $\Phi_{\rm AGN}(M_{\rm BH}, z)$ can be estimated from the broad-line AGN BHMF $\phi_{\rm BH}(M_{\rm BH}, z)$ after correcting for the obscured fraction of AGNs, as described in section~\ref{sec:s5_duty}. We set the broad-line AGN BHMF $\phi_{\rm BH}$ to be free. Since the Schechter and double-power-law broad-line AGN BHMFs are consistent with each other at $z=4$, we only consider $\phi_{\rm BH}$ with the Schechter function (equation~\ref{eq:Sch_BHMF}) during $z=4\sim6$ to reduce the number of free parameters. We describe the time evolution of $\phi_{\rm BH}$ through the characteristic BH mass, i.e., $\log M^*_{\rm new}=\log M^*+k_{M_{\rm BH},1}(z-z_0)$; the slope, i.e., $\alpha_{\rm new}=\alpha+k_{M_{\rm BH},2}(z-z_0)$; and the normalization, i.e., $\log \phi_{\rm new}^*=\log \phi^*+k_{M_{\rm BH},3}(z-z_0)+k_{M_{\rm BH},4}(z-z_0)^2$, where $z_0=4$. There are six free parameters in total, i.e., $M^*$, $\alpha$, $k_{M_{\rm BH},1}$, $k_{M_{\rm BH},2}$, $k_{M_{\rm BH},3}$ and $k_{M_{\rm BH},4}$. $\phi^*$ is the normalization at $z=4$ evaluated in section~\ref{sec:s4_correct_result}.

Meanwhile, at the time step, we evaluate the average Eddington ratio in each BH mass and redshift through $\left\langle\lambda_{\rm Edd}\right\rangle=\int P(\lambda_{\rm Edd} \mid M_{\rm BH}, z) \lambda_{\rm Edd} \mathrm{d} \log \lambda_{\rm Edd} $, where $P(\lambda_{\rm Edd} \mid M_{\rm BH}, z)$ is the normalized active ERDF, i.e., $\int \mathrm{d} \log \lambda_{\rm Edd} P(\lambda_{\rm Edd} \mid M_{\rm BH}, z)=1$ over $-2<\log \lambda_{\rm Edd}<1$. We again determine the active ERDF from the broad-line AGN ERDF by correcting for the obscured fraction of AGNs mentioned in section~\ref{sec:s5_duty}. 

Based on the broad-line AGN ERDFs at $z=4$ derived in section~\ref{sec:s4_correct_result}, we assume two models: the first one assumes no redshift evolution of ERDF during $z=4\sim6$; and the second one allows for a time evolution of ERDF by adding an additional term to the characteristic Eddington ratio, i.e., $\log \lambda^*_{\rm Edd, new}=\log \lambda^*_{\rm Edd}+ k_{\lambda}(\log M_{\rm BH}-\log M_{\textrm{BH},c})+k_{\lambda,z}(z-z_0)$, where $z_0=4$. We only adopt the log-normal ERDF as it shows better fitting quality than the Schechter ERDF in Section~\ref{sec:s4_goodness_fit} (see Table~\ref{tab:intrinsicBHMF}), and the shape of the observed ERDFs at $z\sim6$ can be better described by the log-normal function \citep[e.g.,][]{willott,shen2019,Farina22}. The coefficient is set to $k_{\lambda,z}=0.11$ so that the broad-line AGN ERDF can peak at the Eddington limit $\log \lambda_{\rm Edd}\sim0$ at $z=6$, which is consistent with \citet{willott}. We confirm using the Schechter ERDF does not significantly change the resulting best-fit broad-line AGN BHMF.

Finally, with the radiative efficiency $\epsilon$ and the six free parameters of $\phi_{\rm BH}$ as inputs, we can estimate the BH mass growth during the time interval through $\langle\dot{M_{\rm BH}}\rangle$, and derive the total BHMF at the time step by integrating both sides of equation~\ref{eq:continuity2}. The broad-line AGN BHMF is also the output. It should be noted that we adopt the total BHMF to be the active BHMF in the mass ranges where the latter exceeds the former to ensure the active fraction to be less than one. The broad-line AGN BHMF over those mass ranges will also be modified accordingly by considering the obscured fraction of AGNs. Furthermore, we do not consider the time evolution of obscured fraction of AGNs during the period as it still remains uncertain. 

For simplicity, {we fix the radiative efficiency to $\epsilon=0.05$, 0.1 and 0.15.} The six free parameters of $\phi_{\rm BH}$ are then determined by fitting the convolved luminosity function of broad-line quasars, i.e., $\Phi_{\rm QSO}(L, z)=\int \mathrm{d} \log \lambda_{\rm Edd}P_{\rm QSO}(\lambda_{\rm Edd} \mid M_{\rm BH}, z) \phi_{\rm BH}(M_{\rm BH}, z)$, to the observations at $z=4$ from \citet{akiyama2018}, $z=5$ from \citet{niida2020} and $z=6$ from \citet{Matsuoka18}, respectively. These observed luminosity functions are evaluated over a wide luminosity range so that the shape of $\phi_{\rm BH}$ can be well determined over a wide mass range. The fitting procedure utilizes the least-square method, and the range of fitting is set to $-28.5<M_{\rm 1450}<-21.5$, where the observed luminosity functions are constrained with good statistics. 

\begin{figure*}[!bt]
\centering
\includegraphics[keepaspectratio,width=.8\textwidth]{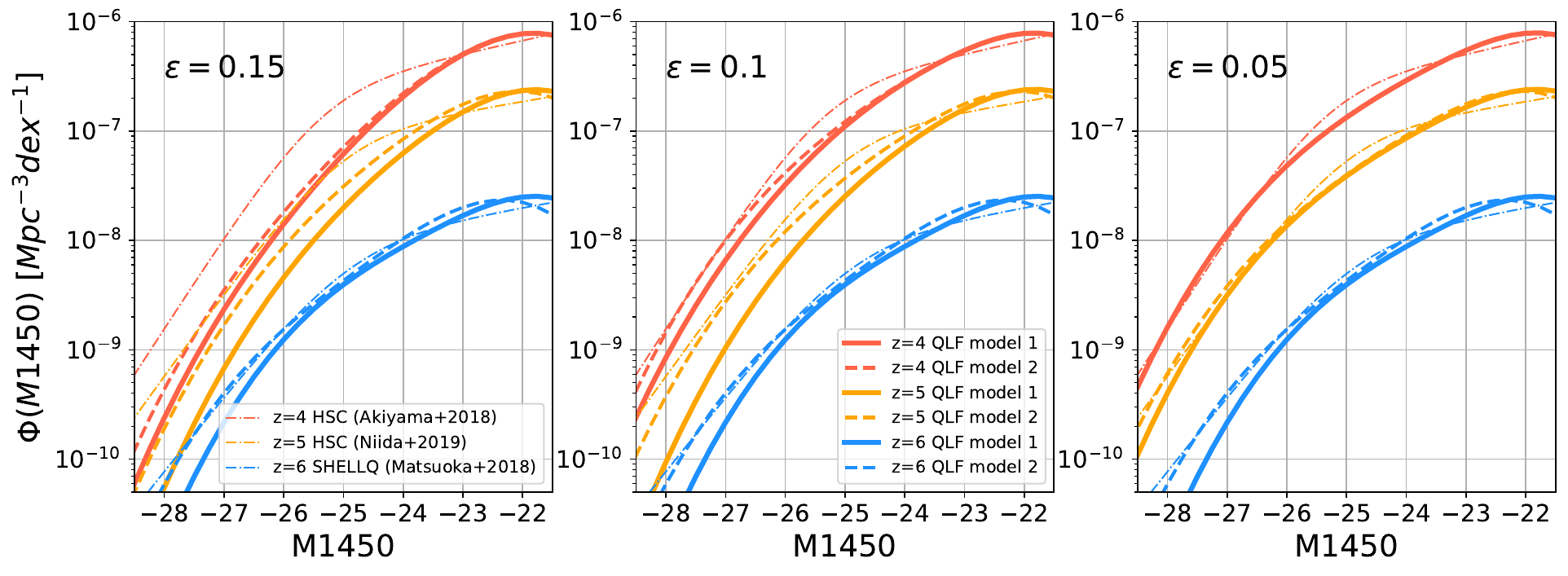}
\caption{Luminosity functions of quasars derived by solving the continuity equation from $z=6$ to 4. Results obtained with the radiative efficiency of 0.15, 0.1 and 0.05 are displayed from left to right. In each panel, red, orange and blue lines show results at $z=4$, 5 and 6, respectively. Solid and dashed lines plot results obtained with the constant and time-evolving ERDF models, respectively. For the comparison, the luminosity functions of quasars at $z=4$, 5 and 6 from \citet{akiyama2018}, \citet{niida2020} and \citet{Matsuoka18} are plotted by red, orange and blue thin dashed-dotted lines in each panel, respectively. }
\label{fig:z56QLF}
\end{figure*}

{With each of the radiative efficiency $\epsilon$ of 0.05, 0.1 and 0.15,} we adopt both of the ERDF models. The resulting luminosity functions of broad-line quasars at $z=4$, 5 and 6 are plotted by red, orange and blue lines in Figure~\ref{fig:z56QLF}, respectively. 
We see the time-evolving ERDF model can yield higher number densities than the constant ERDF model at the luminous ends. The discrepancy may be due to the higher average Eddington ratios allowed by the former model, and more luminous quasars are enabled to be produced. 

We compare the model luminosity functions with the observational ones at $z=4$, 5 and 6: the predicted number densities at the faint ends are consistent with observations at all redshifts and radiative efficiencies, while those at the luminous ends tend to be underestimated by over one order of magnitude compared to the observations at large radiative efficiencies $\epsilon\gtrsim0.1$. When adopting the radiative efficiency of $\epsilon=0.15$, the discrepancies between the model and observed luminosity functions at $z=4$ and 5 can be larger than three times at the luminous bins, even considering the time-evolving ERDF model. Thus, such large radiative efficiencies are likely to be discarded in our models. 

To fully reproduce the observed luminosity functions at $z=4$, 5 and 6 over $-28.5<M_{\rm 1450}<-21.5$, either a small radiative efficiency of $\epsilon\lesssim0.05$ with the constant ERDF model or a moderate radiative efficiency of $\epsilon\lesssim0.1$ with the time-evolving ERDF model is required. Such preference of small radiative efficiencies of $\sim0.05$ in the SMBH growth at $z\gtrsim4$ is also stated, and explained as the result of more common super-Eddington accretion at higher redshifts in semi-analytical simulations of galaxy and BH evolution \citep[e.g.,][]{shirakata20}. We note that this small radiative efficiency is close to the lower boundary for a geometrically thin disk \citep{SS1973}. It may indicate "radiatively inefficient accretion flows (RIAFs)" caused by low-Eddington ($\log \lambda_{\rm Edd}<-2$) or super-Eddington ($\log \lambda_{\rm Edd}>0$) accretion \citep[for review see][]{Inayoshi20}.

\begin{figure*} [!ht]
\centering
\includegraphics[keepaspectratio,width=0.8\textwidth]{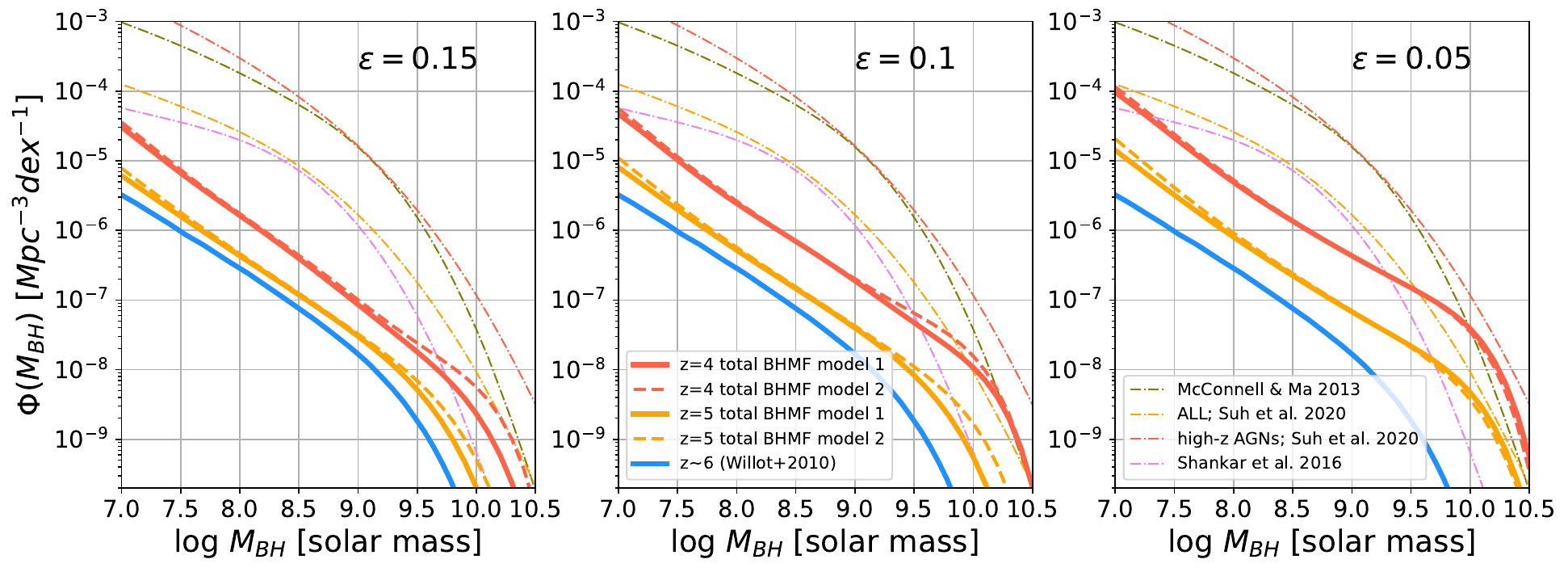}
\includegraphics[keepaspectratio,width=0.8\textwidth]{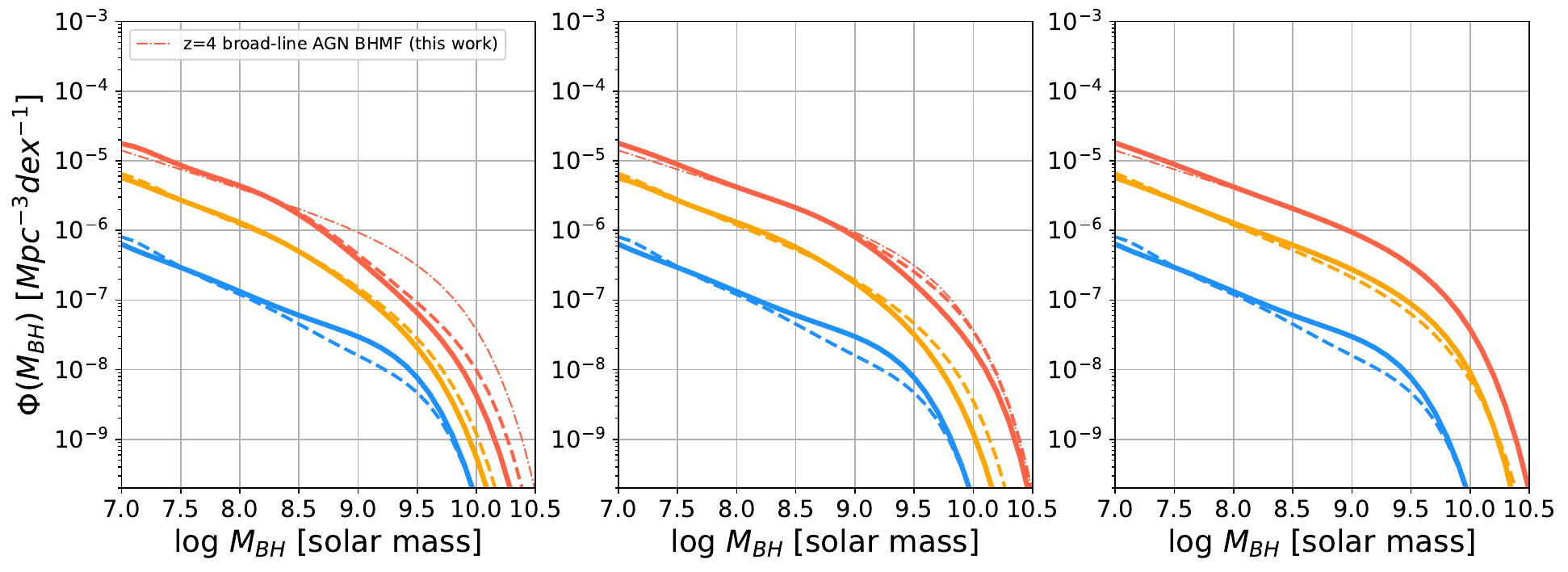}
\caption{Total BHMFs (top) and broad-line AGN BHMFs (bottom) derived by solving the continuity equation from $z=6$ to 4. Results obtained with the radiative efficiency of 0.15, 0.1 and 0.05 are plotted from left to right. Lines are the same with Figure~\ref{fig:z56QLF}. For the comparison, the total BHMFs convolved from the stellar mass function at $z=4$ are displayed by thin dashed-dotted lines in the top panels, and the \textit{intrinsic} broad-line AGN BHMF constrained in section~\ref{sec:s4_correct_result} is plotted by thin red dashed-dotted line in the middle panels. }
\label{fig:z56BHMF}
\end{figure*}

In Figure~\ref{fig:z56BHMF}, we plot the resulting total and broad-line AGN BHMFs in the top and bottom panels, respectively. The results at $z=4$, 5 and 6 are plotted by red, orange and blue lines, respectively. 
We see the results show a dependence on the radiative efficiency. With large radiative efficiencies $\epsilon\gtrsim0.1$, there is little evolution in the total BHMF from $z=6$. Thus, the massive ends of the broad-line AGN BHMFs at $z=4$ and 5 need to sharply decline so that they would not exceed the total BHMF at the respective redshift. The resulting low number density of massive SMBHs of $\log M_{\rm BH}/M_\odot>9$ causes the lack of luminous quasars, so the model luminosity functions can be highly underestimated at the luminous ends. 

With smaller radiative efficiencies $\epsilon<0.1$, the total BHMF can be significantly evolved from $z=6$, especially at the massive end of $\log M_{\rm BH}/M_\odot>9.5$. Compared with the total BHMFs convolved from the stellar mass function at $z=4$, which are obtained in section~\ref{sec:s5_duty}, the number densities at the massive end of $\log M_{\rm BH}/M_\odot>9.5$ can be well reproduced, while those at the less-massive end of $\log M_{\rm BH}/M_\odot<9$ require even smaller radiative efficiencies of $\epsilon\lesssim 0.05$. 

Under different small radiative efficiencies $\epsilon<0.1$, the resulting broad-line AGN BHMFs keep consistent at all redshifts, and the result at $z=4$ is in accordance with the \textit{intrinsic} broad-line AGN BHMF derived in section~\ref{sec:s4_correct_result}. We note that the time-evolving ERDF model can result in slightly higher BHMFs than the constant ERDF model due to the larger average Eddington ratios allowed by the former model. The best-fit parameters with the radiative efficiency of 0.05 are summarized in Table~\ref{tab:BHMF_ce}.
\begin{table}[!ht]
\caption{Best-fit parameters of the broad-line AGN BHMF.}\label{tab:BHMF_ce}
\begin{adjustbox}{width=.6\textwidth,center=7cm}
\centering
\begin{tabular}{ccccccccc}
\hline\hline
$\epsilon$ & ERDF model  & $\log \phi^*$ &$\log M^*$& $\alpha$& $k_{M_{\rm BH},1}$& $k_{M_{\rm BH},2}$& $k_{M_{\rm BH},3}$ & $k_{M_{\rm BH},4}$  \\
 &   & Mpc$^{-3}$ dex$^{-1}$ &$M_\odot$& & & & \\
\hline
 0.05 & time-evolving & $-7.65$ & 9.65  & $-1.44$ & 0.04 & $-0.14$ & $-0.55$  & $-0.24$ \\
 & constant    & $-7.66$ & 9.65  & $-1.45$ & 0.14 & $-0.05$ & $-0.49$  & $-0.23$ \\
\hline\\
\end{tabular}
\end{adjustbox}
\end{table}
 
As both the total and broad-line AGN BHMFs at $z=4$ can be best reproduced with the radiative efficiency of 0.05, we further examine the time evolution of broad-line AGN BHMF during $z=4\sim6$ under the radiative efficiency of 0.05. As shown in the bottom right panels of Figure~\ref{fig:z56BHMF}, overall, the predicted broad-line AGN BHMFs at $z=4$, 5 and 6 keep similar shapes. Only the normalization shows a major evolution. The {similar pure density evolution trend} can be also seen in the luminosity functions of quasars across $z=4-6$, which are displayed in Figure~\ref{fig:z56QLF}. We note that both the constant and time-evolving ERDF models can result in the same evolution trend.

\subsection{Effects of the ${\rm C_{\Rmnum{4}}}$-based BH mass estimator}
\label{sec:s5_CIV_blueshift}

As mentioned in section~\ref{sec:s4_correct_result}, the ${\rm C_{\Rmnum{4}}}$-based virial BH mass estimates can be under- or over-estimated due to the blueshifted component of the ${\rm C_{\Rmnum{4}}}$ emission line. Here, we evaluate its effect in estimating the virial BH masses using the relation between the ${\rm C_{\Rmnum{4}}}$ blueshift and the ratio of virial BH mass estimated by ${\rm C_{\Rmnum{4}}}$ and H${\alpha}$ lines, which is given by \citet{Coatman}. We set an upper limit of 4000 km s$^{-1}$ for the blueshift since the quasars used to determine the relation in \citet{Coatman} barely cover the range beyond the limit. 

The wavelength coverage of the quasar spectra obtained in this work does not cover other emission lines, such as ${\rm Mg_{\Rmnum{2}}}$, to determine the systematic redshift. We thus adopt the empirical relation between the ${\rm C_{\Rmnum{4}}}$ FWHM and its blueshift. The relation is determined by fitting a second-order polynomial to 19 luminous SDSS quasars at $2 < z < 2.7$ with both of the quantities available in \citet{Coatman16}. To prevent over-corrections, we limit the blueshift to be positive. Then, we derive the corrected BH mass with the ${\rm C_{\Rmnum{4}}}$ FWHM. 

After correcting for the effect caused by the ${\rm C_{\Rmnum{4}}}$ blueshifts, the luminous and less-luminous quasar samples have the median BH mass of $\log M_{\rm BH}/M_{\odot}=9.4$ and 8.7, respectively. Compared to the estimates without the blueshift correction, both of the quasar samples cover narrower mass range, and almost all massive SMBHs with $\log M_{\rm BH}/M_{\odot}\gtrsim9.6$ in the luminous quasar sample are corrected to have smaller mass around $\log M_{\rm BH}/M_{\odot}\sim9.5$. Meanwhile, the median Eddington ratio of the less-luminous quasar sample drops to $\log \lambda_{\rm Edd}=-0.7$, while that of the luminous quasar sample increases to $\log \lambda_{\rm Edd}=-0.4$. 

Considering the large uncertainty of the method described above, we further examine the effect by applying the calibration of the ${\rm C_{\Rmnum{4}}}$-based single-epoch virial BH mass estimate in \citet{park17}. It is suggested that the calibration can produce a similar BH mass distribution to that computed from the blueshift-corrected formula in \citet{Coatman} at large blueshifts of $>2000$ km s$^{-1}$. We derive the virial BH mass of both the luminous and less-luminous quasar samples with the calibration as follows:
\begin{align}
\log \left( \frac{M_{\rm BH, FWHM}}{M_{\odot}} \right)  =&  7.54 + \log \left[ \left( \frac{\lambda L_{\lambda} (1350)}{10^{44}\ {\rm erg\ s^{-1}}} \right)^{0.45} \right] \notag \\ 
 &+ \log \left[ \left( {\rm \frac{FWHM_{\rm C_{\Rmnum{4}}}}{1000 km\ s^{-1}}} \right)^{0.5} \right].
\label{eq:mbh_civ_fwhm_p}
\end{align}
The systematic uncertainty of the calibration is evaluated to be 0.37 dex \citep{park17}.

\begin{figure} [!ht]
\centering
\includegraphics[keepaspectratio,width=.45\textwidth]{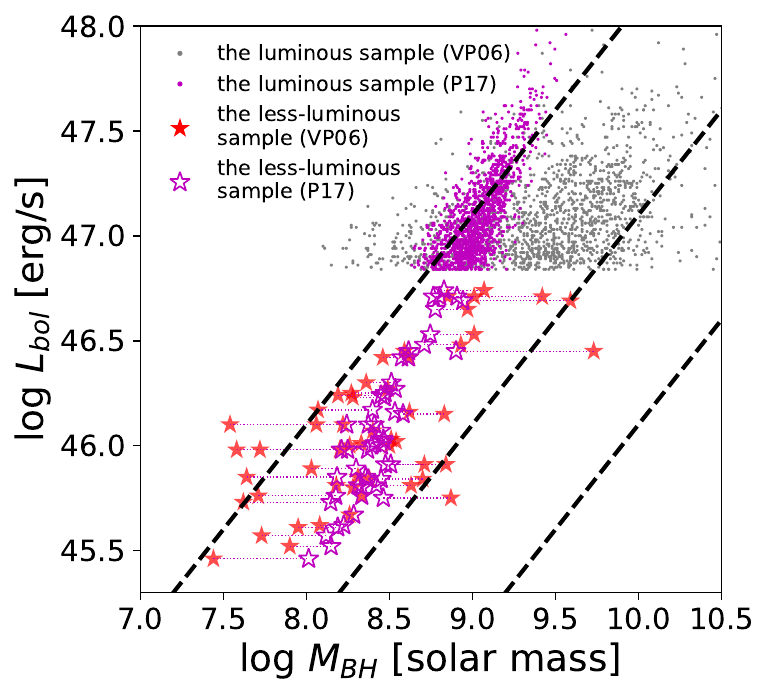}
\caption{Bivariate distributions of $z\sim4$ quasars on the $L_{\rm bol}$ vs. $M_{\rm BH}$ plane. Red filled stars and black dots represent the less-luminous and luminous quasar samples used to determine the $z=4$ BHMF in this work. Open stars and dots show the same less-luminous and luminous quasars, respectively, but with the virial BH mass estimated with the calibration in \citet{park17}. Black dashed lines denote constant Eddington ratios of 1, 0.1 and 0.01 from top to bottom. }
\label{fig:BHMF_ERDF_blueshift}
\end{figure}

In Figure~\ref{fig:BHMF_ERDF_blueshift}, we plot the luminous and less-luminous quasar samples with the virial BH mass estimated by the calibration in \citet{park17} by magenta dots and open stars, respectively. 
Compared to the results derived with the calibration in \citet{VP06}, we find that both of the quasar samples show much narrower mass ranges when adopting the calibration in \citet{park17}: the less-luminous quasar sample spans the range of $8.0<\log M_{\rm BH}/M_{\odot}<8.9$ with a median of $\log M_{\rm BH}/M_{\odot}=8.4$, and the luminous quasar sample covers the range of $8.6<\log M_{\rm BH}/M_{\odot}<9.6$ with a median of $\log M_{\rm BH}/M_{\odot}=9.0$. Meanwhile, the Eddington ratio range of the less-luminous quasar sample decreases to $-0.8<\log \lambda_{\rm Edd}<-0.2$ with a median of $\log \lambda_{\rm Edd}=-0.5$, while that of the luminous quasar sample increases to $-0.5<\log \lambda_{\rm Edd}<0.5$ with a median of $\log \lambda_{\rm Edd}=-0.0$. 

\begin{figure} [!ht]
\centering
\includegraphics[keepaspectratio,width=0.45\textwidth]{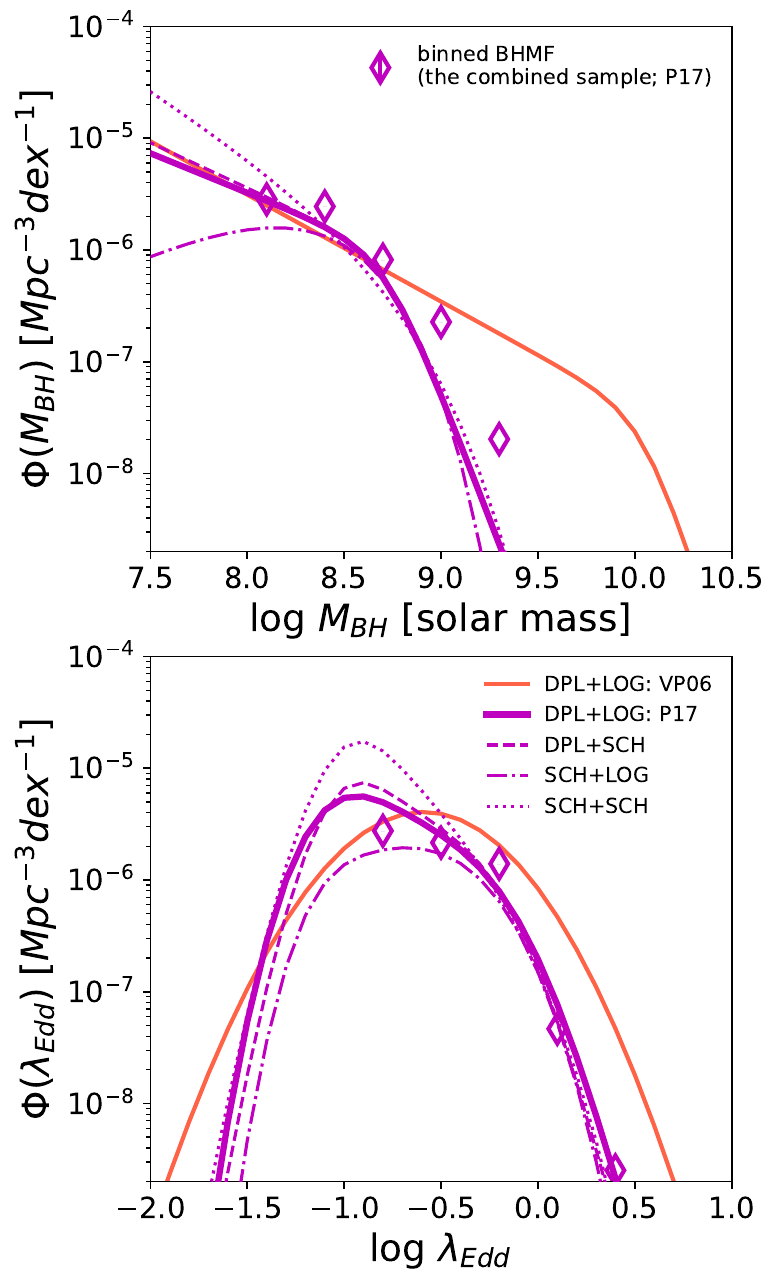}
\caption{\textit{Binned} (diamonds) and \textit{intrinsic} (lines) broad-line AGN BHMFs (top) and ERDFs (bottom) at $z=4$. Magenta color presents those with the virial BH mass re-calculated with the calibration provided in \citet{park17}. Types of lines have the same meaning with Figure~\ref{fig:BHMF}. As a comparison, the best-fit BHMF and ERDF model using the calibration in \citet{VP06} are plotted by red solid lines. }
\label{fig:BHMF_ERDF_p17}
\end{figure}

With the virial BH mass estimated with the calibration in \citet{park17}, we derive the \textit{binned} and \textit{intrinsic} BHMFs and ERDFs at $z=4$ for the combined sample with the same method introduced in section~\ref{sec:s4_correct_method}. The resulting BHMFs and ERDFs are plotted by magenta lines in the top and bottom panels of Figure~\ref{fig:BHMF_ERDF_p17}, respectively. The best-fit parameters are summarized in Table~\ref{tab:intrinsicBHMF}. 

The broad-line AGN BHMFs derived with the calibration in \citet{park17} show broad consistency with those obtained with the calibration in \citet{VP06} in the less-massive end of $\log M_{\rm BH}/M_{\odot}<8.5$. Significant discrepancies between the two occur at the massive end of $\log M_{\rm BH}/M_{\odot}>9$. The results derived with the calibration in \citet{park17} sharply decline towards the massive end as almost none of the luminous quasars are estimated to have BH masses greater than $\log M_{\rm BH}/M_{\odot}=9.5$ with that calibration. The rapid decline of the massive end suggests a different evolution trend in Figure~\ref{fig:evo}. As plotted by magenta open diamonds, when adopting the calibration in \citet{park17}, the number density of massive SMBHs with $\log M_{\rm BH}/M_{\odot}=9.8$ significantly drops from $z=2$ to 4, while that of less-massive SMBHs with $\log M_{\rm BH}/M_{\odot}\lesssim 9$ mildly declines. 

Additionally, when adopting the calibration in \citet{park17}, the mass dependence of ERDFs becomes positive with the mass-dependent term $k_{\lambda}$ significantly greater than zero, suggesting more massive SMBHs tend to accrete at higher Eddington ratios. The trend is consistent with the results determined with the ${\rm Mg_{\Rmnum{2}}}$-based single-epoch virial BH mass estimate at lower redshifts $z=1-2$ \citep[e.g.,][]{ks2013,AS2015}.

The main discrepancy between the calibrations in \citet{park17} and \citet{VP06} is caused by the relaxed constraints of slope parameters in \citet{park17}. \citet{park13} claim that treating the slope parameters in calibration as free parameters can mitigate the FWHM-dependent biases, which may be partly induced by the blueshifted component of the ${\rm C_{\Rmnum{4}}}$ emission line. However, we should also note that the relaxation of slope parameters in \citet{park17} may contradict the virial theorem (i.e., $M_{\rm BH}\propto {\rm FWHM}^{2}$). In addition, the AGN sample utilized to calibrate the virial BH mass estimates in their work does not cover massive SMBHs with $\log M_{\rm BH}/M_\odot>9$, which may also result in large uncertainties in the virial BH mass estimates of luminous quasars.

\section{Summary}
\label{sec:summary}

In this paper, we estimate the ${\rm C_{\Rmnum{4}}}$-based virial BH mass of less-luminous quasars at $z\sim4$. The quasar candidates are selected down to $i=23.2$ with the $g$-band dropout colors and stellar morphology from the HSC-SSP S16A-Wide2 dataset. In total, we identify 151 quasars at $z\sim4$, and estimate the virial BH mass of 80 less-luminous quasars at $3.04\leq z \leq 4.44$. Our main results can be summarized as follows:
\begin{enumerate}
\item The less-luminous quasars span a BH mass range of $7.4<\log M_{\rm BH}/M_\odot<9.7$ with a median of $\log M_{\rm BH}/M_{\odot}=8.5$, which is around one order of magnitude less-massive than the luminous SDSS DR7 quasars in the same redshift range. 
\item Both the luminous and less-luminous quasars have the Eddington ratio similarly distributed around $\log \lambda_{\rm Edd}\sim-0.5$, but the latter ones slightly extend towards the high Eddington ratios.
\end{enumerate}

Then, we determine the broad-line AGN BHMFs and ERDFs at $z=4$ with 52 less-luminous quasars covering the BH mass range of $7.4< \log M_{\rm BH}/M_{\odot}<9.7$ at $3.50\leq z\leq 4.25$. To improve the constraints over a wide BH mass range, we further construct a sample of 1,462 luminous quasars in the same redshift range from the SDSS DR7 quasar catalog. The combined sample can cover a BH mass range of $7.4< \log M_{\rm BH}/M_{\odot}<10.7$. We adopt both the $V_{\rm max}$ and maximum likelihood method. The former method yields the \textit{binned} results that only quasars above the flux limit are considered, while the latter one can derive the \textit{intrinsic} results by accounting for the incompleteness in the low-mass and/or low-Eddington-ratio ranges caused by the flux limit. Our main results can be summarized as follows:
\begin{enumerate}
\item For the \textit{intrinsic} broad-line AGN BHMFs and ERDFs, the parametric model with BHMF in double power-law function and ERDF in log-normal function shows the best-fit result. In the case, the \textit{intrinsic} and \textit{binned} BHMFs and ERDFs are broadly consistent with each other. Large discrepancies between the two only appear in the less-massive range of $\log M_{\rm BH}/M_{\odot}\sim7.5$, where the completeness of the combined quasar sample drops below 50\%. 
\item A negative mass dependence of ERDF, i.e., the massive SMBHs accrete with low Eddington ratios, is suggested by the best-fit model. Without that mass dependence, the flattened faint-end of the observed luminosity function of quasars at $z=4$ can not be reproduced.
\item The \textit{intrinsic} broad-line AGN BHMF and ERDF follow the ``down-sizing'' evolutionary trend in comparison with those at the cosmic noon $z\sim2$. The number density of the most massive SMBHs of $\log M_{\rm BH}/M_{\odot}\sim10$ roughly keeps constant from $z=2$ to 4, while that of smaller SMBHs continuously declines during the period. Additionally, the number density of rapidly-accreting SMBHs of $\log \lambda_{\rm Edd}\gtrsim 0$ is significantly enhanced since $z\sim2$, while that of SMBHs with lower Eddington ratios continuously declines from $z=2$ to 4. The abundance of massive and/or rapidly-accreting SMBHs tends to peak at high redshifts.
\end{enumerate}

Furthermore, we correct for the fraction of the obscured AGNs to estimate the BHMF and ERDF for the total AGN population at $z=4$. We then compare the \textit{intrinsic} broad-line and total active AGN BHMFs to the total BHMF at $z=4$. The results correspond to the active fractions of broad-line and total active AGNs among the entire SMBH population as a function of BH mass at $z=4$. In order to obtain the total $z=4$ BHMF, we convolve the total stellar mass function of galaxies with the $M_{\rm BH}-M^{*}$ relation. The main findings can be summarized as follows:
\begin{enumerate}
\item While uncertainties remain large, the active fraction of broad-line AGNs suggests a positive dependence on the BH mass: the less-massive SMBHs with {$7.5<\log M_{\rm BH}/M_{\odot}<8.5$ show an average fraction of $0.01\sim0.3$, while the massive ones with $\log M_{\rm BH}/M_{\odot}>9$ occupy $0.04\sim0.8$.}
\item SMBHs, especially the massive ones, are likely to keep a high active fraction of $\sim 10$\% across $z=2\sim4$. 
\end{enumerate}

Finally, we examine the time evolution of broad-line AGN BHMF between $z=4$ and 6 through solving the continuity equation. The time evolution model of the broad-line AGN BHMF is optimized to reproduce the observed luminosity functions of quasars at $z=4$, 5, and 6. We reach the following results:
\begin{enumerate}
\item Small radiative efficiencies of $\epsilon\lesssim0.1$ are indicated to fully reproduce the observed luminosity functions of quasars at $z=4$ and 5. 
\item The best-fit time evolution model suggests the broad-line AGN BHMF basically evolves in parallel during $z=4\sim6$ without significant changes in the shape. Meanwhile, the average Eddington ratios at $z=5$ and 6 tend to increase compared to those at $z=4$, suggesting an even more vigorous growth in SMBH towards the high redshifts. 
\end{enumerate}

We caution that the ${\rm C_{\Rmnum{4}}}$-based virial BH mass estimates can be under- or over-estimated due to the blueshifted component of the ${\rm C_{\Rmnum{4}}}$ emission line, especially for luminous quasars with potentially large ${\rm C_{\Rmnum{4}}}$ blueshifts. The effects can affect the conclusions obtained above, mostly in the massive and/or high-Eddington-ratio ends. For example, if we estimate the virial BH mass by adopting the calibration in \citet{park17}, which can produce BH mass estimates similar to those with the blueshift effect corrected using the formula in \citet{Coatman}, the resulting \textit{intrinsic} BHMF shows sharp decline towards the massive end of $\log M_{\rm BH}/M_{\odot}>9.5$, and the large excess at the high-Eddington-ratio end of $\log \lambda_{\rm Edd}>0$ is also reduced. Here, since we do not have direct measurements on the ${\rm C_{\Rmnum{4}}}$ blueshifts for our sample, and the calibration in \citet{park17} also suffers from the incompleteness of high-mass SMBHs, we adopt the results obtained with the calibration in \citet{VP06} as the nominal results. Future studies on estimating the virial BH mass of our quasar samples with the H$\beta$ and/or ${\rm Mg_{\Rmnum{2}}}$ emission lines can be necessary to verify the determination of BHMF and ERDF at $z=4$ in this work.


\begin{acknowledgments}

We {appreciate for the valuable comments from the reviewer. We} are grateful to Lidman, C. for providing the \texttt{python} scripts to run the \texttt{v6.46 2dfdr} pipeline for reducing the data taken with AAT/AAOmega. Inayoshi, K. acknowledges support from the National Natural Science Foundation of China (12073003, 12003003, 11721303, 11991052, 11950410493, 1215041030, and 12150410307), and the China Manned Space Project Nos. CMS-CSST-2021-A04 and CMS-CSST-2021-A06. We appreciate for the support from MEXT (Ministry of Education, Culture, Sports, Science and Technology), Japan. This research is supported by Japan Society for the Promotion of Science (JSPS) through Grant-in-Aid for JSPS Fellows with the grant number 19J11013. 

The data in this research is collected at Subaru Telescope, which is operated by the National Astronomical Observatory of Japan (NAOJ), Anglo-Australian Telescope and Keck II telescope. We would give a big thank for the great supports from people in these observatories during our observations. 

The Hyper Suprime-Cam (HSC) collaboration includes the astronomical communities of Japan and Taiwan, and Princeton University. The HSC instrumentation and software were developed by NAOJ, the Kavli Institute for the Physics and Mathematics of the Universe (Kavli IPMU), the University of Tokyo, the High Energy Accelerator Research Organization (KEK), the Academia Sinica Institute for Astronomy and Astrophysics in Taiwan (ASIAA), and Princeton University. Funding was contributed by the FIRST program from Japanese Cabinet Office, MEXT, JSPS, Japan Science and Technology Agency (JST), the Toray Science Foundation, NAOJ, Kavli IPMU, KEK, ASIAA, and Princeton University. 

This paper is based [in part] on data collected at the Subaru Telescope and retrieved from the HSC data archive system, which is operated by Subaru Telescope and Astronomy Data Center (ADC) at NAOJ. Data analysis was in part carried out with the cooperation of Center for Computational Astrophysics (CfCA) at NAOJ. We are honored and grateful for the opportu- nity of observing the Universe from Maunakea, which has the cultural, historical and natural significance in Hawaii.

This paper makes use of software developed for Vera C. Rubin Observatory. We thank the Rubin Observa- tory for making their code available as free software at http://pipelines.lsst.io/.

The Pan-STARRS1 Surveys (PS1) and the PS1 public science archive have been made possible through contributions by the Institute for Astronomy, the University of Hawaii, the Pan-STARRS Project Office, the Max Planck Society and its participating institutes, the Max Planck Institute for Astronomy, Heidelberg, and the Max Planck Institute for Extraterrestrial Physics, Garching, The Johns Hopkins University, Durham University, the University of Edinburgh, the Queens University Belfast, the Harvard-Smithsonian Center for Astrophysics, the Las Cumbres Observatory Global Telescope Network Incorporated, the National Central University of Taiwan, the Space Telescope Science Institute, the National Aeronautics and Space Administration under grant No. NNX08AR22G issued through the Planetary Science Division of the NASA Science Mission Directorate, the National Science Foundation grant No. AST- 1238877, the University of Maryland, Eotvos Lorand University (ELTE), the Los Alamos National Laboratory, and the Gordon and Betty Moore Foundation.

Funding for SDSS and SDSS-II has been provided by the Alfred P. Sloan Foundation, the Participating Institutions, the National Science Foundation, the U.S. Department of Energy, the National Aeronautics and Space Administration, the Japanese Monbukagakusho, the Max Planck Society, and the Higher Education Funding Council for England. The SDSS Web Site is http://www.sdss.org/.

\end{acknowledgments}



\end{document}